\documentclass[12pt,titlepage]{article}
\usepackage{amsmath}
\usepackage{amssymb}
\usepackage{graphicx}
\usepackage{caption2}
\usepackage{amsfonts}
\usepackage{cite}
\usepackage{lineno}
\usepackage{color}
\usepackage{multirow}
\usepackage{subfigure}
\usepackage{bm}

\newcommand{\be}{\begin{equation}}
\newcommand{\ee}{\end{equation}}
\newcommand{\bea}{\begin{eqnarray}}
\newcommand{\eea}{\end{eqnarray}}
\newcommand{\bef}{\begin{figure}}
\newcommand{\ef}{\end{figure}}
\newcommand{\bt}{\begin{tabular}}
\newcommand{\et}{\end{tabular}}
\newcommand{\bno}{\begin{enumerate}}
\newcommand{\eno}{\end{enumerate}}

\allowdisplaybreaks[4]

\def\3{\ss}

\catcode`\"=\active
\def"{\accent'177}

\begin{document}

\begin{center}

{\Large\bf  Clean numerical simulation \\ of three-dimensional turbulent Kolmogorov flow}

\vspace{0.3cm}

Shijie Qin$^{2}$ and Shijun Liao$^{1,2}$ \footnote{Corresponding author, Email: sjliao@sjtu.edu.cn}  


\vspace{0.3cm}

$^{1}$ State Key Laboratory of Ocean Engineering,  Shanghai 200240, China\\

$^{2}$ School of Ocean and Civil Engineering, Shanghai Jiao Tong University, China\\

\end{center}

\hspace{-0.75cm}{\bf Abstract} 
{  
Turbulence  holds immense importance across various scientific and engineering disciplines.  
The direct numerical simulation (DNS) of turbulence proposed by Orszag in 1970 is a milestone in fluid mechanics, which began an era of numerical experiment for turbulence.  Many researchers have reported that  turbulence should be chaotic, since spatiotemporal trajectories are very sensitive to small disturbance.  Thus, due to the famous butterfly-effect of chaos, unavoidable numerical noises of DNS  
 might have great influence on spatiotemporal trajectories of turbulence.  This is indeed true for a two-dimensional (2D) Kolmogorov turbulent flow, as currently revealed by a much more accurate algorithm than DNS, namely the ``clean numerical simulation'' (CNS).  Different from DNS, CNS can greatly reduce both of truncation error and round-off error to any required small level so that numerical noise can be rigorously negligible throughout a time interval long enough for calculating statistics.   However,   In physics,  3D turbulent flow is more important than 2D turbulence.  Thus, for the first time,  we solve a 3D turbulent Kolmogorov flow by means of CNS  in this paper, and compare our CNS result with that given by DNS in details.  It is found that the spatial-temporal trajectories of the 3D Kolmogorov turbulent flow given by DNS are indeed badly polluted by numerical noise rather quickly, and  besides  the DNS result has significant deviations from the CNS benchmark solution not only in the spatial symmetry of flow field and the energy cascade but also even in statistics.       
}

\hspace{-0.75cm}{\bf Keywords} {Turbulent Kolmogorov flow, three-dimensional, clean numerical simulation}




\section{Motivation}

More than a half-century ago, Orszag \cite{Orszag1970} proposed the so-called ``direct numerical simulation'' (DNS),  a simulation in computational fluid dynamics (CFD) in which the NS equations are numerically solved without any turbulence models. This is a milestone in fluid mechanics, which began an era of the so-called ``numerical experiment'' of turbulence.  
With the increasing performances of supercomputer, numerical experiment based on DNS has become a very useful tool in fundamental research in turbulence \cite{Rogallo1981NASA, She1990Nature, Kuhnen2018,  Nelkin1992Science, FeracoScience2024, MoinARFM1998, Scardovelli1999ARFM, Coleman2010DNS, Huang2022JFM},  since it can extract  information that are difficult or impossible to obtain in laboratory for physical experiments.  
Nowadays, numerical experiment based on DNS \cite{Orszag1970, Rogallo1981NASA, She1990Nature, Nelkin1992Science, FeracoScience2024, MoinARFM1998, Scardovelli1999ARFM, Coleman2010DNS, Huang2022JFM} has been widely used to develop turbulence models, such as the sub-grid scale models for large eddy simulation (LES) and models related to the Reynolds-averaged Navier-Stokes equations (RANS), and so on. 

Chaos theory is regarded as one among the three greatest theories of physics in the 20th century, compared to the quantum mechanics and the general theory of relativity. 
A chaotic system has the famous ``butterfly-effect'', i.e. a tiny difference in initial condition expands exponentially and could lead to significant deviation of spatiotemporal trajectory as the time increases so that long-term prediction of chaos is impossible \cite{lorenz1963deterministic}. 
Note that the Lorenz equation \cite{lorenz1963deterministic} as a famous chaotic system is a simplified model derived from the NS equations.   
It has been reported that spatiotemporal trajectories of the NS equations are very sensitive (i.e. unstable) to initial conditions  \cite{Deissler1986PoF, boffetta2017chaos, berera2018chaotic}. In addition, Ge, Rolland and Vassilicos reported that the average uncertainty energy of three-dimensional (3D) NS turbulence grows exponentially \cite{Vassilicos2023JFM}.  All of these clearly indicate that the NS turbulence should be chaotic in essence.  Note that  Lorenz \cite{Lorenz2006Tellus} discovered in 2006  that  numerical noises, which are artificial, might greatly change the characteristics of a chaotic system.   Unfortunately, each numerical algorithm has more or less  numerical noise, caused by truncation error and round-off error. Thus, the above-mentioned facts logically lead to the conclusion that  tiny  numerical noises of DNS should be enlarged very quickly to macroscopic level, since the NS equations for turbulence are chaotic but artificial numerical noise is avoidable for DNS.   

To study the influence of numerical noises on chaotic system and turbulence,  Liao \cite{Liao2009} proposed the so-called ``clean numerical simulation'' (CNS) in 2009. Different from DNS, numerical noise of CNS \cite{Liao2009, Liao2023book, LIAO2014On, hu2020risks, LiaoNA2022, qin2020influence, qin_liao_2022, Qin2024JOES, Liao-2025-JFM-NEC, Liao-2025-JFM-PS, Zhang2025CPC, Xiaoming2025-3D-3body} can be reduced to any required small level so that it is rigorously negligible compared to the corresponding physical variables (such as velocity, pressure and so on) in a {\em finite} time interval\footnote{This is a great difference between CNS and other traditional numerical methods that always give simulations in an {\em arbitrarily} long interval of time without considering the exponential growth of numerical noise of chaotic system.} that is long enough for statistics.  Therefore, one can give accurate trajectory of chaotic systems in a long enough interval of time by means of the CNS. For details, please refer to Liao's book \cite{Liao2023book} about CNS.   

In other words,  one can use CNS to do ``clean'' numerical  experiment  in a finite but long enough interval of time.   
For example, in 2022 Qin and Liao \cite{qin_liao_2022} solved a two-dimensional (2D) turbulent Rayleigh-B\'{e}nard convection (using  {\em thermal fluctuation as its initial condition}) by means of CNS and DNS, separately, in the case of aspect ratio
$\Gamma = 2\sqrt{2}$, Prandtl number $Pr = 6.8$  (corresponding to water at room temperature), and Rayleigh number $Ra = 6.8 \times 10^{8}$ (corresponding to a turbulent state).  Through detailed comparisons, it was found \cite{qin_liao_2022} that the CNS benchmark solution always sustains the non-shearing vortical/roll-like convection during the whole process of simulation, however the DNS result agrees well with the CNS benchmark solution in $0\leq t \leq 50$ and sustains the non-shearing vortical/roll-like convection until $t\approx 188$ when its corresponding flow field turns to a kind of zonal flow thereafter. This provides us a rigorous evidence that numerical noise as a kind of small-scale artificial stochastic disturbances have quantitatively and qualitatively large-scale influences on a sustained turbulence  \cite{qin_liao_2022}. 

To confirm the correctness and generality of the above conclusion, Qin et al. \cite{Qin2024JOES} further solved the 2D turbulent Kolmogorov flow subject to an initial condition with {\em a spatial symmetry} by means of CNS and DNS, respectively, and found that the CNS benchmark solution always maintains the {\em same} spatial symmetry as the initial condition but the DNS result first agrees well with the CNS  benchmark solution until $t\approx 188$ when its corresponding flow field loses the spatial symmetry completely.  We emphasize here that numerical noises are random and thus has {\em no} spatial symmetry at all: thus, the loss of spatial symmetry is an obvious sign for random numerical noise to have been enlarged to the same order of magnitude as the true solution. This clearly confirms the correction and generality of our previous conclusion, i.e. numerical noise as a kind of artificial tiny disturbances can lead to obvious deviations at large scale on turbulence governed by the NS equations, not only quantitatively (even in statistics) but also qualitatively (such as spatial symmetry of flow) \cite{Qin2024JOES}.  
      
Furthermore, using CNS as a numerical tool, Liao and Qin \cite{Liao-2025-JFM-NEC} \footnote{The related code of CNS can be downloaded via
GitHub (https://github.com/sjtu-liao/2D-Kolmogorov-turbulence).} did an ideal (thought) numerical experiment about a 2D turbulent Kolmogorov flow, which discovers, for the first time, the so-called ``noise-expansion cascade'' phenomena: all micro-level noises at different level of magnitudes consistently increase to a macroscopic level, and each of them might completely changes the macro-level characteristics of turbulence such as flow type, flow symmetry, statistics, and so on. This new concept of ``noise-expansion cascade'' directly connects the micro-level physical randomness and the macro-level disorder of turbulence and thus clearly reveals an origin of randomness of turbulence.  Besides, it confirms once again that the NS equations for turbulence are indeed chaotic. In addition, by means of an ideal (thought) numerical experiment based on CNS for 2D turbulent Kolmogorov flow, Liao and Qin \cite{Liao-2025-JFM-PS} indicated that artificial numerical noises are equivalent to physical micro-level noises or environmental disturbances, which provides an positive meaning of artificial numerical noises, for the first time,  to the best of our knowledge.    

It has been reported \cite{qin_liao_2022, Qin2024JOES, Liao-2025-JFM-NEC, Liao-2025-JFM-PS} that the spatiotemporal trajectories of the 2D turbulent flow given by DNS are  badly  polluted by artificial numerical noises rather quickly.    However, one might argue that these 2D results might be possibly due to the ``inverse'' energy cascade of the 2D turbulence, i.e. kinetic energy transfers from small-scale to large-scale, so that artificial numerical noise might be enlarged with energy transfer, thus one possibly should not gain the same conclusion for a 3D turbulence since DNS results for 3D turbulent flows always give direct energy cascade, i.e. kinetic energy transfers from large-scale to small-scale.   In mathematics, the CNS algorithm for 2D turbulent flow is based on stream-function and thus is invalid for 3D turbulence.   In physics,  3D turbulent flow is more important than 2D turbulence, as pointed out by Nobel Prize winner T.D. Lee \cite{Lee1951}.  So,  in this paper, {\em for the first time},  we extend the CNS algorithm in \cite{Liao-2025-JFM-NEC, Liao-2025-JFM-PS}  to  a 3D turbulent Kolmogorov flow  and compare our CNS result  with that given by DNS in details.  Shortly speaking, we gain the similar conclusions even for the 3D turbulent Kolmogorov flow, as mentioned below in details.    

\section{Set up of the clean numerical experiment}

Let us consider a 3D incompressible flow in a cubic domain $[0,L]^3$ under the so-called Kolmogorov forcing $\textbf{f}$, which is stationary, monochromatic and sinusoidally varying in space, with an integer $n_K$ describing the forcing scale and $\chi$ representing the corresponding forcing amplitude per unit mass of fluid.
Using the length scale $L$, the velocity scale $U_0=\sqrt{\chi L}$ and thus the time scale $T_0=L/U_0=\sqrt{L/\chi}$, the non-dimensional governing equation for the velocity field $\textbf{u}(\textbf{x},t)$ of this 3D Kolmogorov flow \cite{wu2021quadratic} is the incompressible
Navier-Stokes equation
\begin{align}
\frac{\partial \textbf{u}}{\partial t}+(\textbf{u}\cdot\nabla)\textbf{u} = -\nabla p + \frac{1}{Re}\Delta\textbf{u} + \textbf{f},       \label{NS}
\end{align}
together with the  continuity equation $\nabla\cdot\textbf{u}=0$, where $t$ denotes the time, $\textbf{x}=(x, y, z)\in[0,1]^3$ is the Cartesian coordinate for the cubic flow field that corresponds to the three Cartesian components of the velocity \[ \textbf{u}=(u, v, w)=u \textbf{e}_x + v  \textbf{e}_y + w \textbf{e}_z,\] $\textbf{e}_x, \textbf{e}_y, \textbf{e}_z$ are the direction vectors of the Cartesian coordinate system, $\nabla$ is the Hamilton operator, $\Delta$ is the Laplace operator, $p$ denotes the hydrodynamic pressure, $Re=L\sqrt{\chi L}/\nu$ is the non-dimensional Reynolds number, $\nu$ denotes the kinematic viscosity, and the external force is in the form $\textbf{f} = \sin(2\pi n_K z)\,\textbf{e}_x$, respectively.   Without loss of generality, let us consider here $n_K=4$. 
Following Wu {\em et al.} \cite{wu2021quadratic}, we choose $Re=1211.5$ so as to have a relatively strong turbulent state of the flow. Besides, the periodic boundary conditions are adopted here. According to Coleman and Sandberg \cite{Coleman2010DNS}, one has ``the complete control of the initial and boundary conditions'' for the NS equations. Thus,   we have reasons to choose here the following initial condition
\begin{equation}
\left\{
\begin{array}{l}
u(x,y,z,0)=\cos(2\pi x)\cos(2\pi y)\sin(2\pi z)/2,    \\
v(x,y,z,0)=\sin(2\pi x)\sin(2\pi y) \sin(2\pi z),  \\
w(x,y,z,0)=\sin(2\pi x)\cos(2\pi y)\cos(2\pi z)/2,   \\
\end{array}
\right.  \label{initial_condition}
\end{equation}
which is spatially periodic and satisfies the continuity equation $\nabla\cdot\textbf{u}=0$. Note that the corresponding  initial vorticity field has the two kinds of  {\em spatial  symmetries}  
\begin{equation}
\left\{
\begin{array}{ll}
\mbox{rotation:} &   \big| \bm{\omega}(x,y,z,t) \big| = \big| \bm{\omega}(1-x,1-y,1-z,t) \big|, \\
\mbox{translation:} &   \big| \bm{\omega}(x,y,z,t) \big|  = \left| \bm{\omega}\left(x+\frac{1}{2}, y+\frac{1}{2}, z+\frac{1}{2},t\right) \right|. \\
\end{array}
\right. 
\label{spatial-symmetry}
\end{equation} 
Note that,  due to the mathematical symmetry of the governing equation (\ref{NS}) and the adopted periodic boundary condition, the spatial symmetry (\ref{spatial-symmetry}) should {\em always} remain for $t\geq 0$. Therefore, when a numerical simulation loses the spatial symmetry (\ref{spatial-symmetry}),  it must be far away from the ``exact'' solution $s_{exact}$ of Eqs. (\ref{NS}) and  (\ref{initial_condition}). This provides us a {\em criterion} or a {\em sign} to check whether or not a numerical simulation of the NS turbulence  deviates far away from its exact solution $s_{exact}$.  This is exactly the reason why the periodic initial condition (\ref{initial_condition}) with the spatial symmetry (\ref{spatial-symmetry}) is used here.

It should be emphasized that, as pointed out by Coleman and Sandberg \cite{Coleman2010DNS}, we have ``the complete control of the initial and boundary conditions'' for the NS equations. Besides, the turbulent Kolmogorov flow is primarily studied as a {\em theoretical and idealized model} to understand the fundamental aspects of turbulence. Therefore, in theory,  we certainly can choose the initial condition (\ref{initial_condition}) with the spatial symmetry (\ref{spatial-symmetry}), although in practice it might be very difficult to have such a kind of initial condition. So, in fact, we do an `ideal (thought) numerical experiment' in this paper. Note that thought experiment sometimes could reveal physical essence deeply, just like Einstein's famous thought experiment about a lift that leads to the theory of general relativity.         

Both of DNS and CNS are used to solve the {\em same} equations mentioned above with the {\em same} physical parameters, respectively, and their results are compared in details.  
On the one hand, we apply the CNS \cite{Liao2009, Liao2023book, hu2020risks, qin2020influence, qin_liao_2022,Qin2024JOES, Liao-2025-JFM-NEC, Liao-2025-JFM-PS} to greatly decrease the background numerical noise, i.e. truncation error and round-off error, to such a tiny level that the corresponding artificial numerical noise of the simulation is much smaller than, and thus rigorously negligible compared with, the exact  solution $s_{exact}$ of the Kolmogorov flow in a {\em finite} interval of time that is long enough for statistics. In this way, a ``convergent'' (reproducible) numerical simulation of spatiotemporal trajectory of the 3D turbulent  Kolmogorov flow is obtained by CNS, which is used here as the ``clean'' and ``reliable'' benchmark solution.
Briefly speaking, in the frame of CNS, to decrease the spatial truncation error of the simulation to a  required small enough level, we, {\em like} DNS, discretize the spatial domain of the flow field by a uniform mesh $N^3 = 128^3$, say, the mesh spacing $h=1/128$, and adopt the Fourier spectral method for spatial approximation with the $3/2$ rule for dealiasing.  As verified by Wu~{\em et~al.}~\cite{wu2021quadratic} in the same case, the corresponding spatial resolution is high enough for the considered Kolmogorov flow: the grid spacing is less than the Kolmogorov scale, as shown later in \S~3.4.  Besides,  in order to decrease the temporal truncation error to a  required small enough level, we, {\em unlike} DNS, use the 40th-order ($M = 40$) Taylor expansion for the temporal advancement with a time-step $\Delta t=10^{-3}$. Especially, in order to decrease the round-off error to a required  small enough level, we, {\em unlike} DNS, use the {\em multiple-precision} \cite{oyanarte1990mp} with the 100 significant digits ($N_s = 100$) for {\em all} physical/numerical variables and parameters.  In addition,  we similarly run another CNS algorithm using the {\em same} pseudo-spectral method with the {\em same} uniform spacing mesh but the {\em smaller} temporal truncation error  by means of an even higher-order (i.e. $M = 42$) Taylor expansion with the same time-step ($\Delta t=10^{-3}$) and the  {\em smaller} round-off error by an even higher multiple-precision with more significant digits (i.e. $N_s = 105$).
Comparing these two CNS results by means of the so-called ``spectrum-deviation'' $\delta_s(t)$ defined by (\ref{delta_s}) in Appendix~C, it is found that there are no distinct deviations between these two CNS results in an interval of time $0\leq t \leq 500$. This guarantees the convergence  and reliability of our former CNS result in the {\em whole} spatial domain $\textbf{x}\in[0,1]^{3}$ within the {\em whole} temporal interval $t\in[0,500]$, which is therefore regarded as a ``clean''   benchmark solution of the 3D turbulent Kolmogorov flow.  
About the CNS algorithm in details, please refer to Appendix~B. 

On the other hand, the same NS equations with the same physical parameters and the same initial/boundary condition are solved numerically in $t \in [0, 500]$ by the traditional DNS algorithm using the same mesh spacing $h=1/128$ but the fourth-order Runge-Kutta method with time-step $\Delta t=10^{-4}$ and the {\em double-precision} that is widely used in DNS. Obviously, the temporal truncation error and the round-off error of the DNS algorithm are {\em much} larger than those of the CNS. For details about the DNS, please refer to \cite{Rogallo1981NASA, wu2021quadratic}.   

Let us discuss here the relationship between the DNS and CNS for turbulence. The big difference is that the level of background truncation error and round-off error of CNS is much smaller than that of DNS. Let ${\cal E}_{0}$ denote the level of background truncation error and round-off error and $\lambda$ be the Lyapunov exponent of turbulent flow, respectively. Since turbulent flow is chaotic, its numerical noise ${\cal E}(t)$ increases exponentially, say,
\begin{equation}
{\cal E}(t) = {\cal E}_{0} \exp(\lambda t). \label{Tc-1}
\end{equation}
Let $T_{c}$ denote the time when numerical noise enlarges to macro-scale, i.e. ${\cal E}(T_{c})\sim1$, called the ``critical predictable time''.  Then, we have
\begin{equation}
T_{c}\approx -\frac{\ln {\cal E}_{0}}{\lambda}. \label{Tc-2}
\end{equation}
Since numerical noise is random and thus has no spatial symmetry, symmetry breaking is an obvious sign that random numerical noise is at a macroscopic level.  Obviously, a smaller ${\cal E}_{0}$ corresponds to a larger $T_{c}$. Since ${\cal E}_{0}$ of CNS is much smaller than that of DNS, certainly result of CNS has much larger $T_{c}$ than that of DNS. From this point of view, DNS can be regarded as a special case of CNS, but unfortunately its $T_{c}$ is often too short\footnote{For example, $T_{c}$ of the DNS result given in this paper is 90, as mentioned later.} to calculate statistics: this is exactly the reason why DNS result is badly polluted quickly by numerical noise \cite{qin_liao_2022,Qin2024JOES, Liao-2025-JFM-NEC, Liao-2025-JFM-PS}. 

In general, one can obtain the system's Lyapunov exponent through the linear stability analysis of the governing equations.  A more convenient way is to calculate the maximal Lyapunov exponent of the system through numerical experiment of tracking perturbation growth rate.  The initial error magnitude is determined by the maximum of truncation error and round-off error.  
  
\section{Results of clean numerical simulation}    \label{Influence}

\subsection{Spatial symmetry}

\begin{figure*}[!htb]
    \begin{center}
        \begin{tabular}{cc}
             \subfigure[]{\includegraphics[width=1.7in]{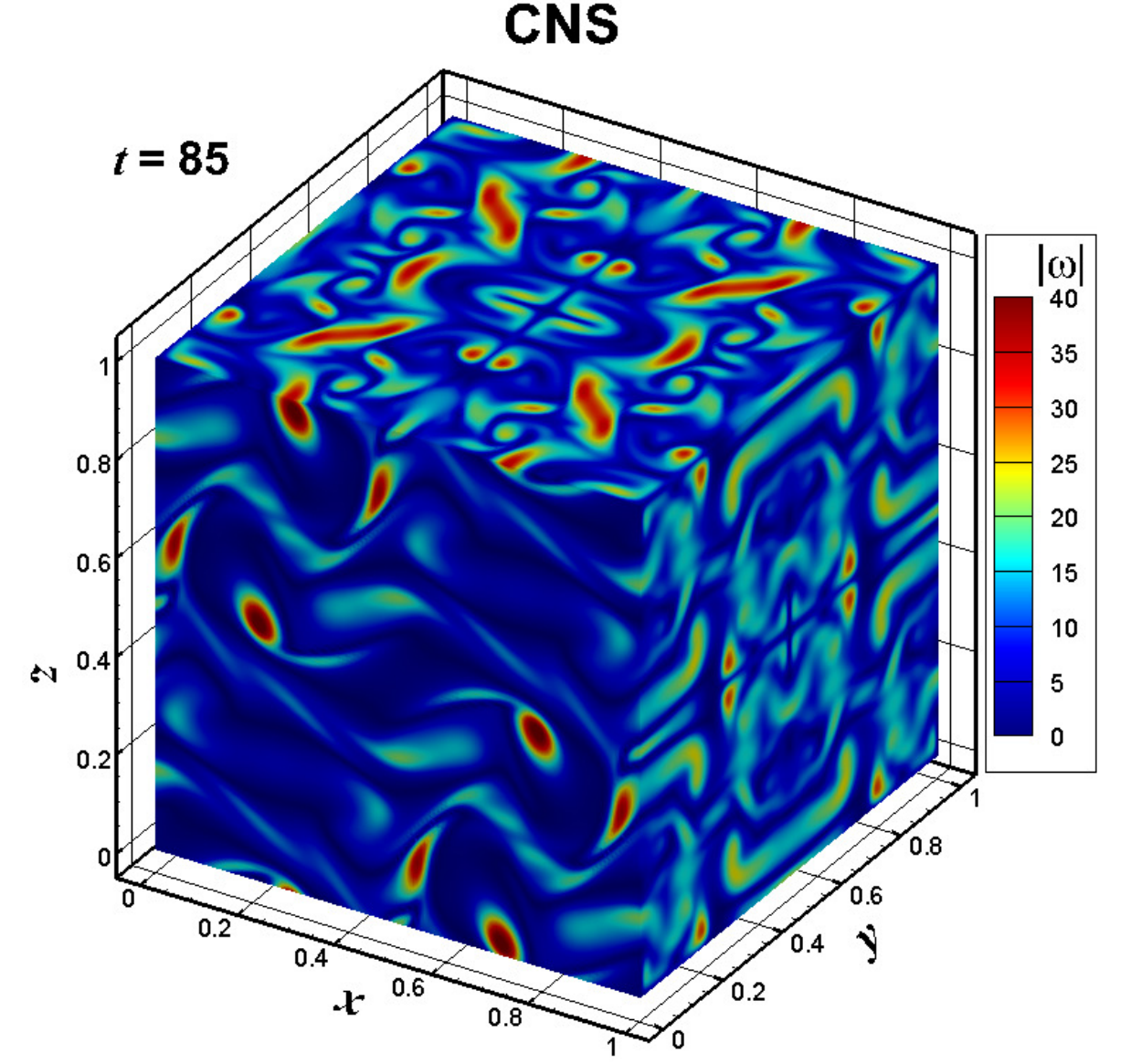}}
             \subfigure[]{\includegraphics[width=1.7in]{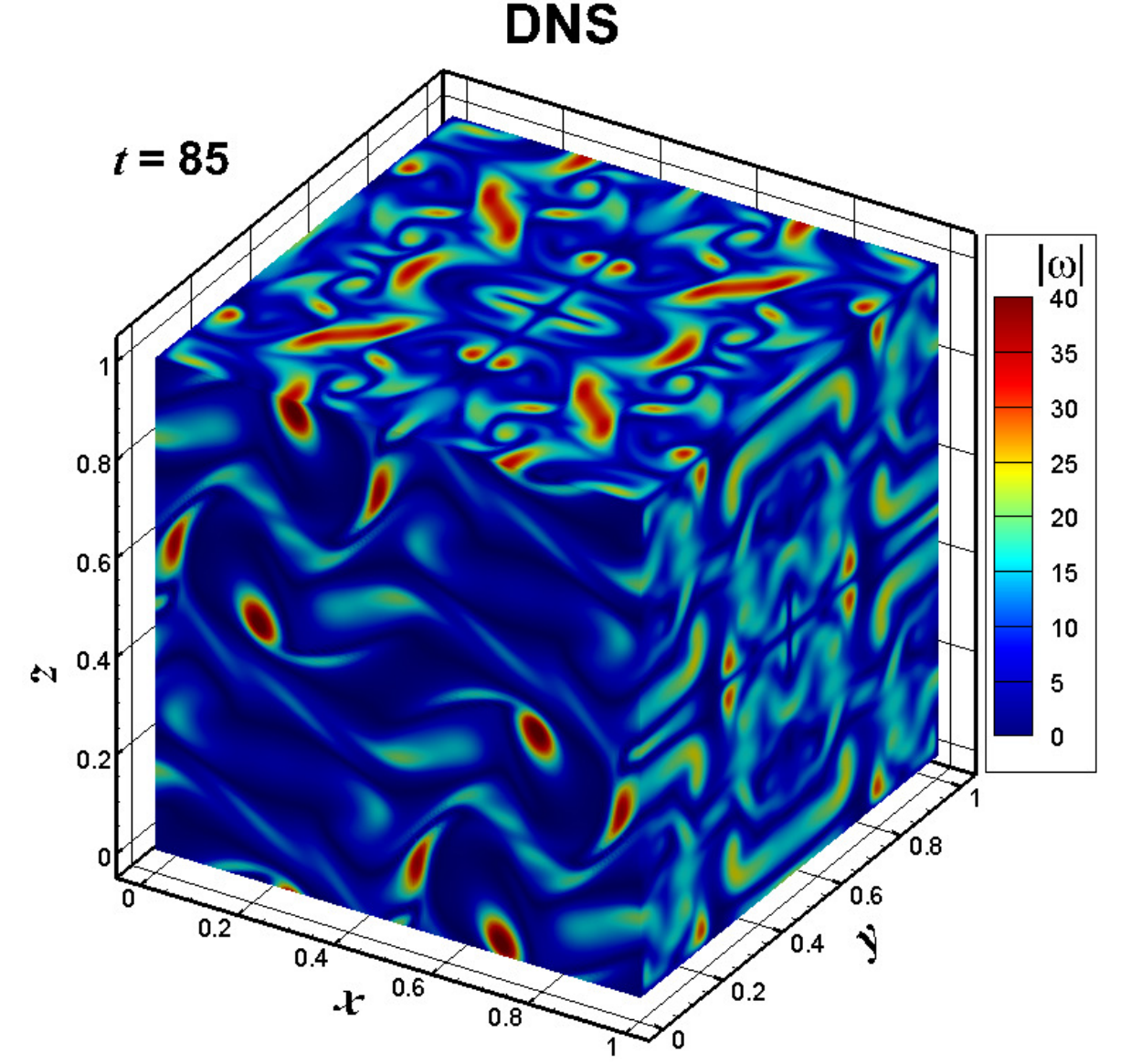}}\\
             \subfigure[]{\includegraphics[width=1.7in]{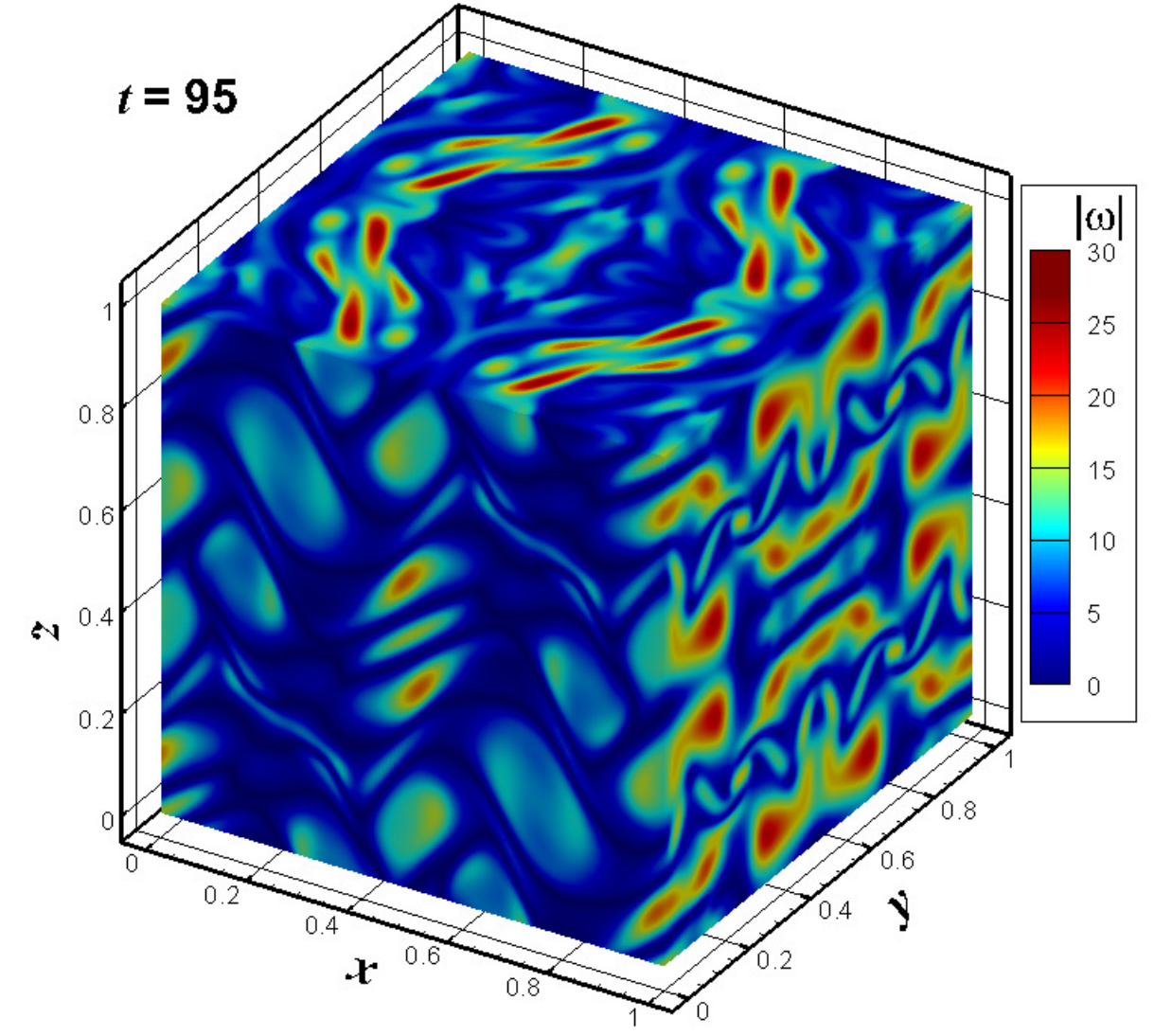}}
             \subfigure[]{\includegraphics[width=1.7in]{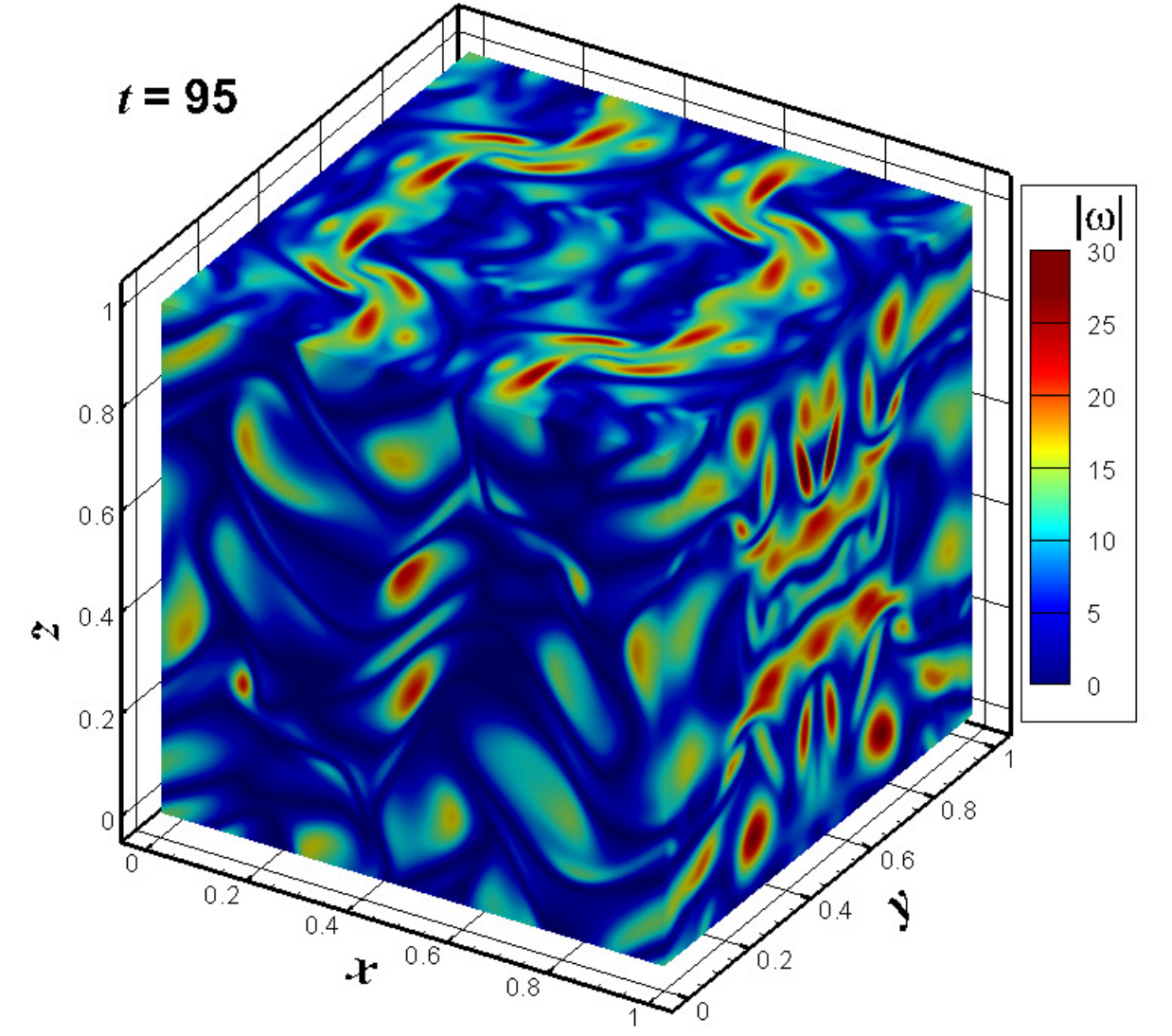}}\\
             \subfigure[]{\includegraphics[width=1.7in]{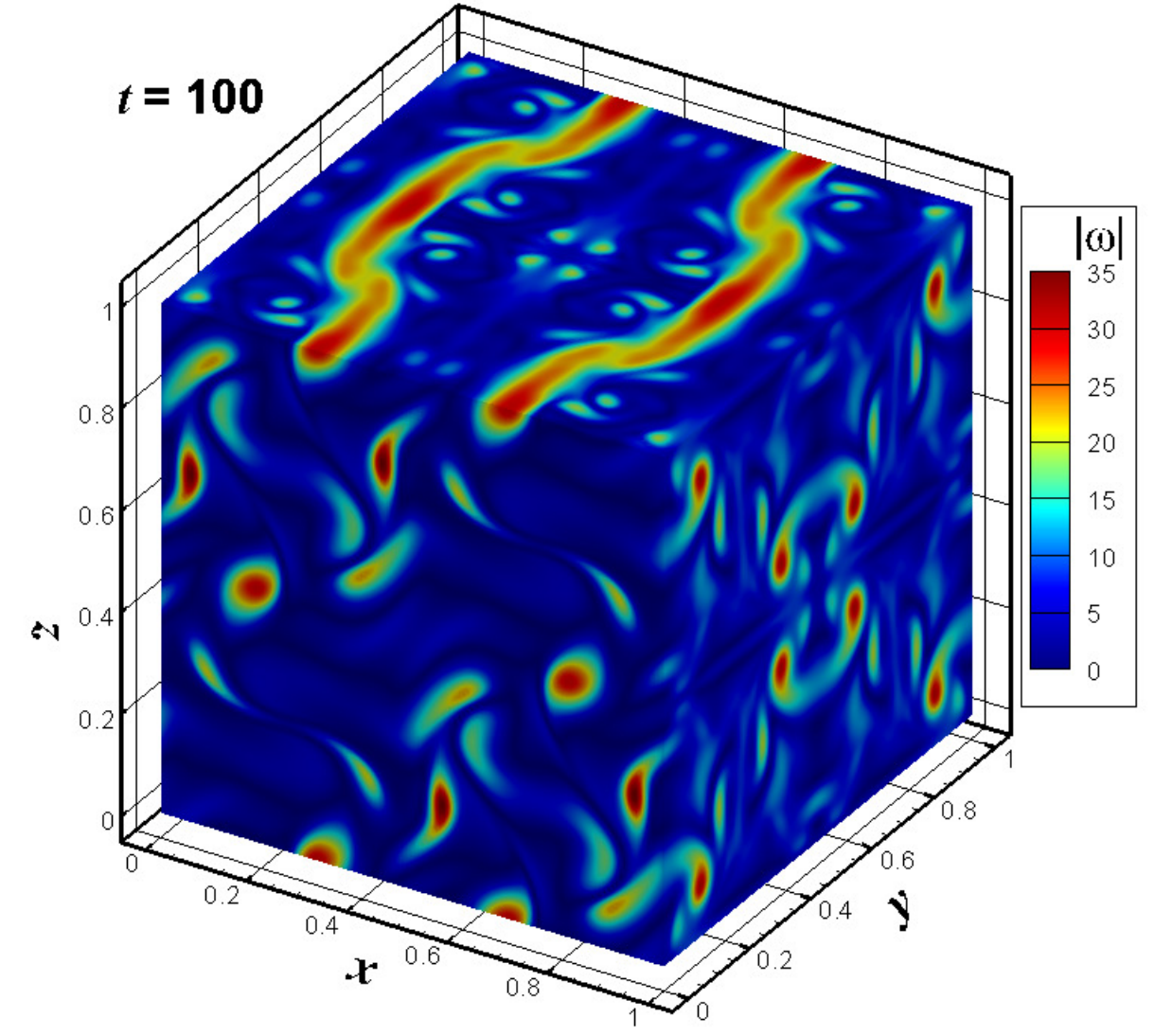}}
             \subfigure[]{\includegraphics[width=1.7in]{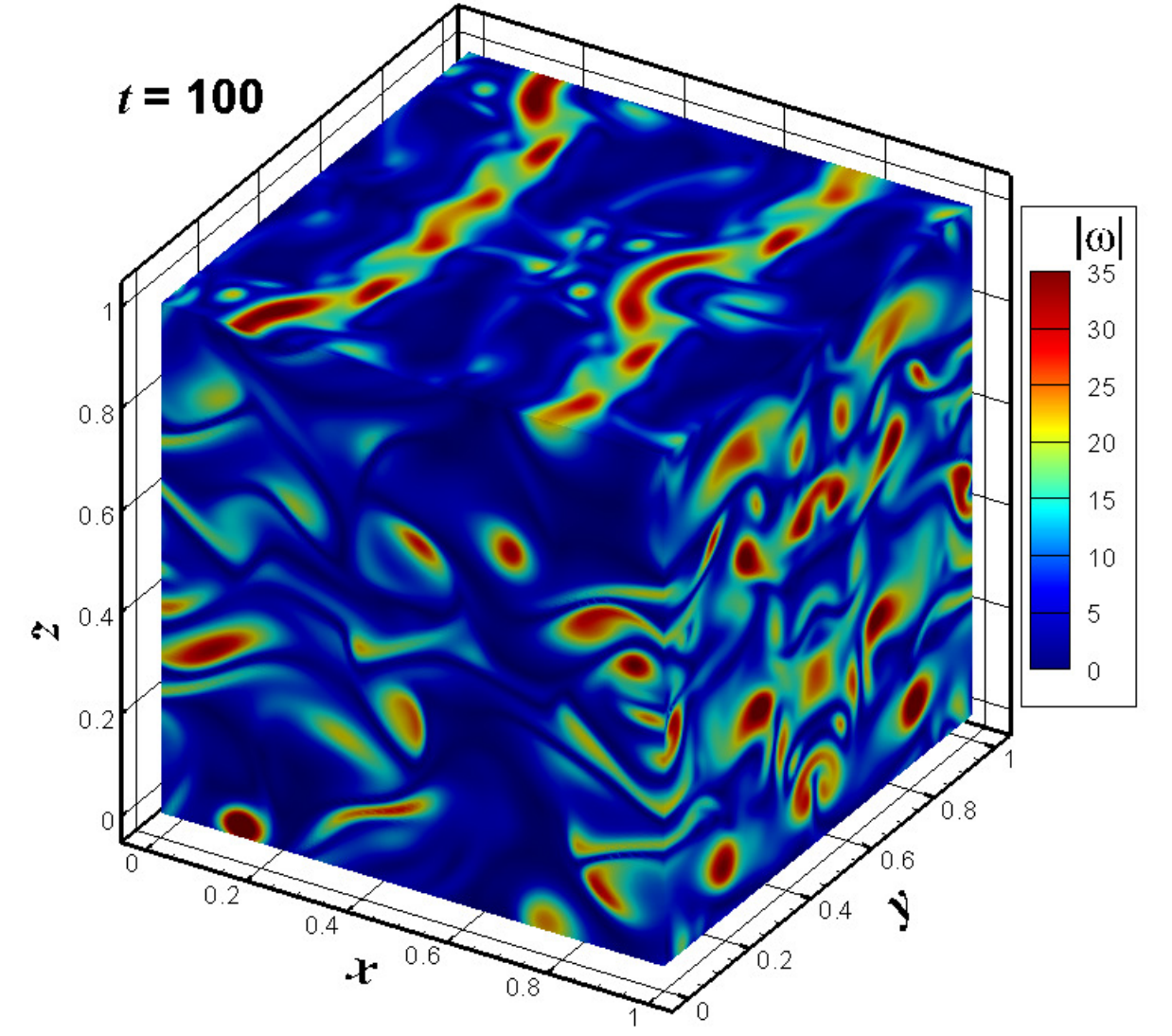}}\\
             \subfigure[]{\includegraphics[width=1.7in]{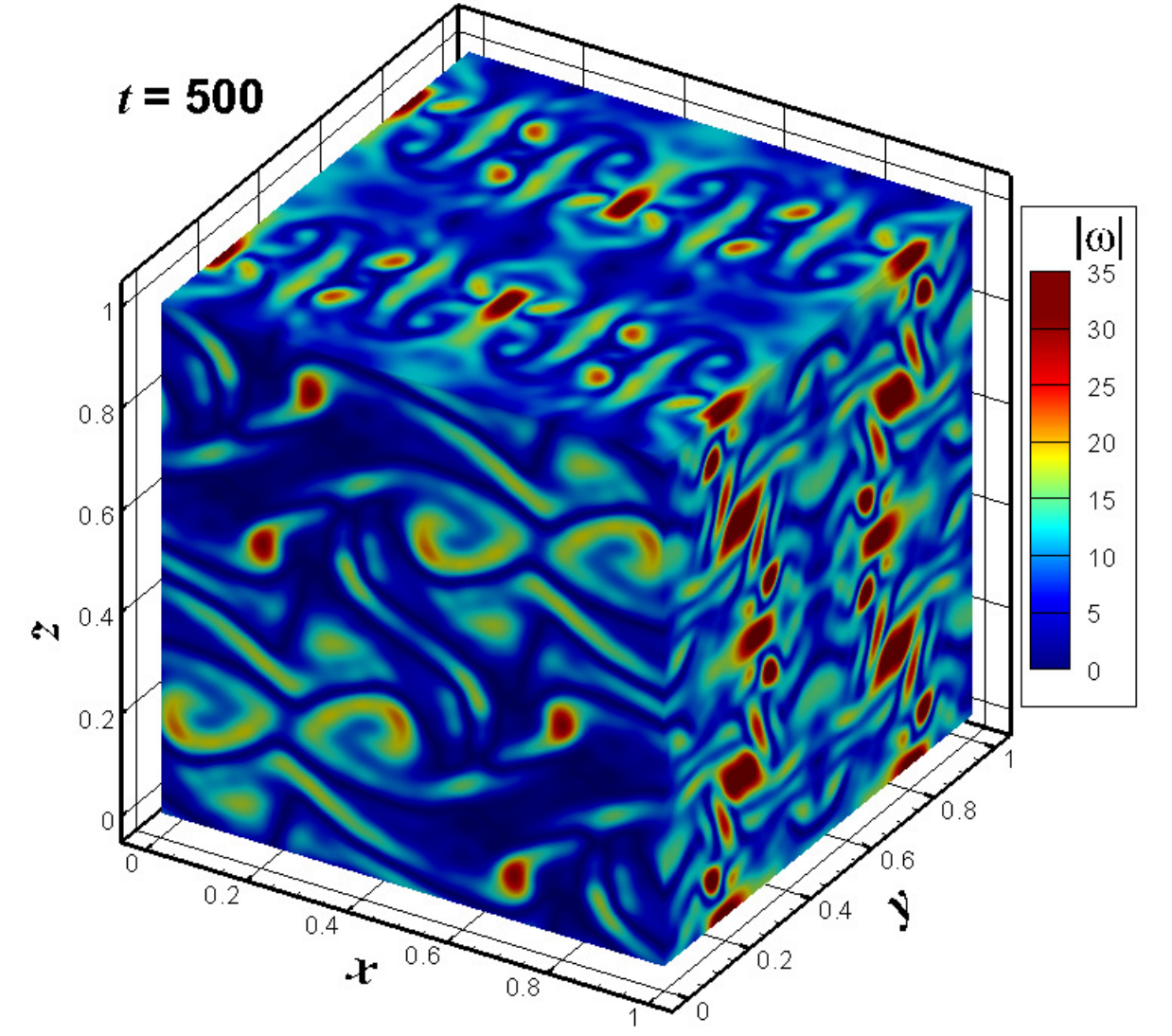}}
             \subfigure[]{\includegraphics[width=1.7in]{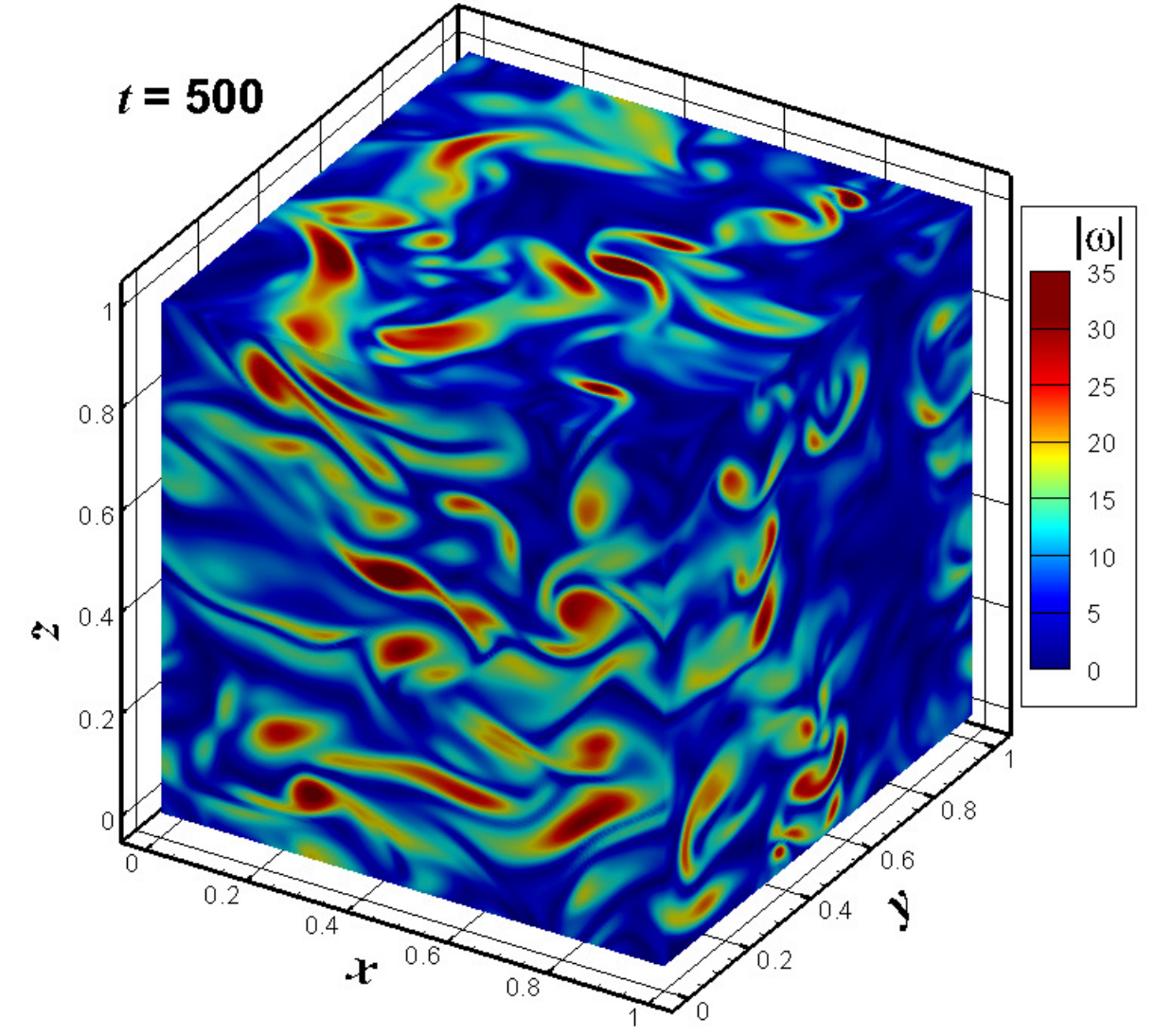}}\\
        \end{tabular}
    \caption{Distribution of  the vorticity modulus  $|\bm{\omega}(\textbf{x},t)|$ of the 3D turbulent Kolmogorov flow governed by (\ref{NS}) subject to the periodic boundary condition and the initial condition (\ref{initial_condition}) with the spatial symmetry (\ref{spatial-symmetry}) in the case of $n_K=4$ and $Re=1211.5$, given respectively by the CNS benchmark solution (left) and the DNS result (right) at some typical times, i.e. (a)-(b) $t=85$, (c)-(d) $t=95$, (e)-(f) $t=100$, and (g)-(h) $t=500$, respectively.
}     \label{Vor_Evolutions}
    \end{center}
\end{figure*}

\begin{figure*}[!htb]
    \begin{center}
        \begin{tabular}{cc}
             \subfigure[]{\includegraphics[width=1.7in]{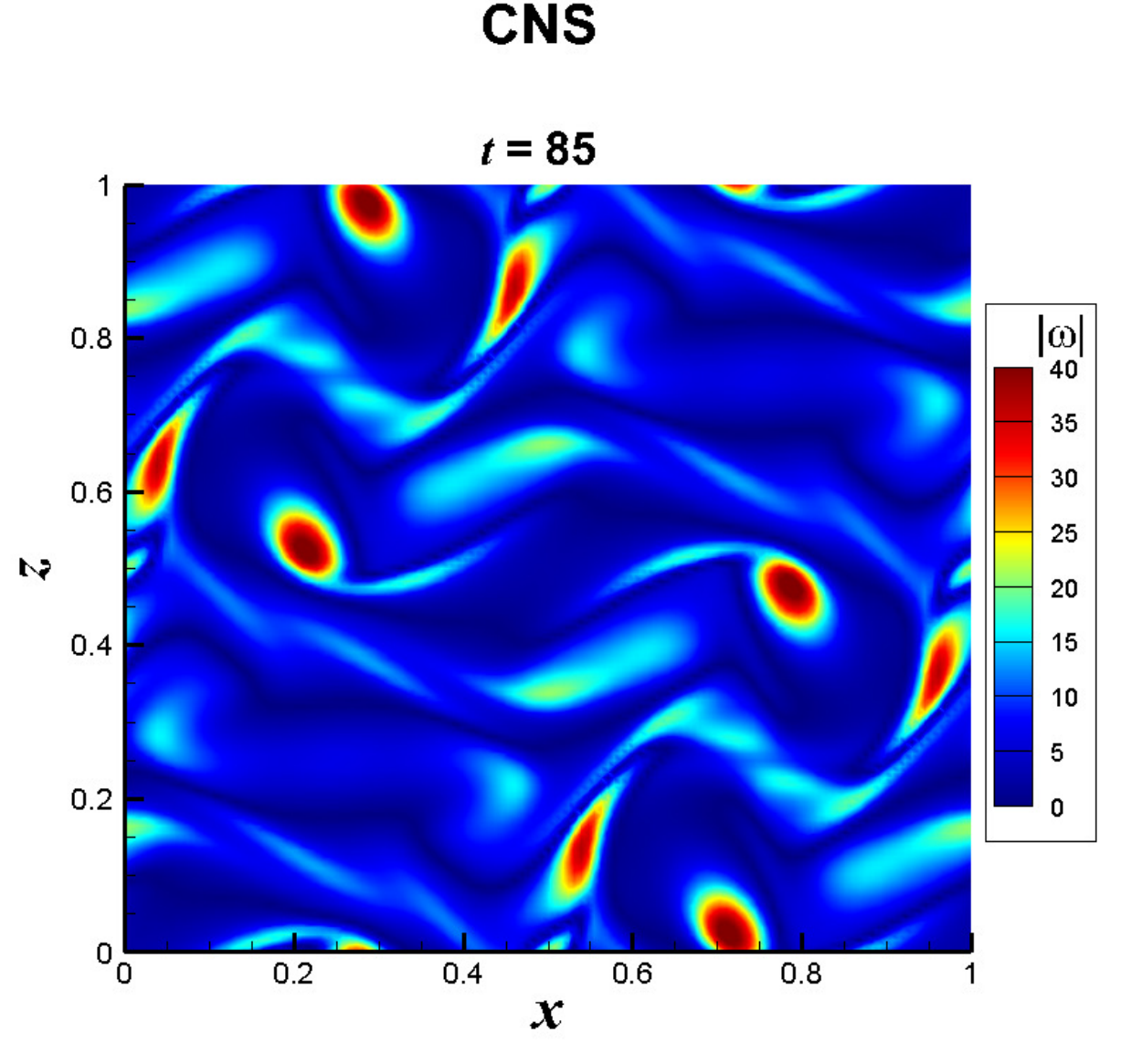}}
             \subfigure[]{\includegraphics[width=1.7in]{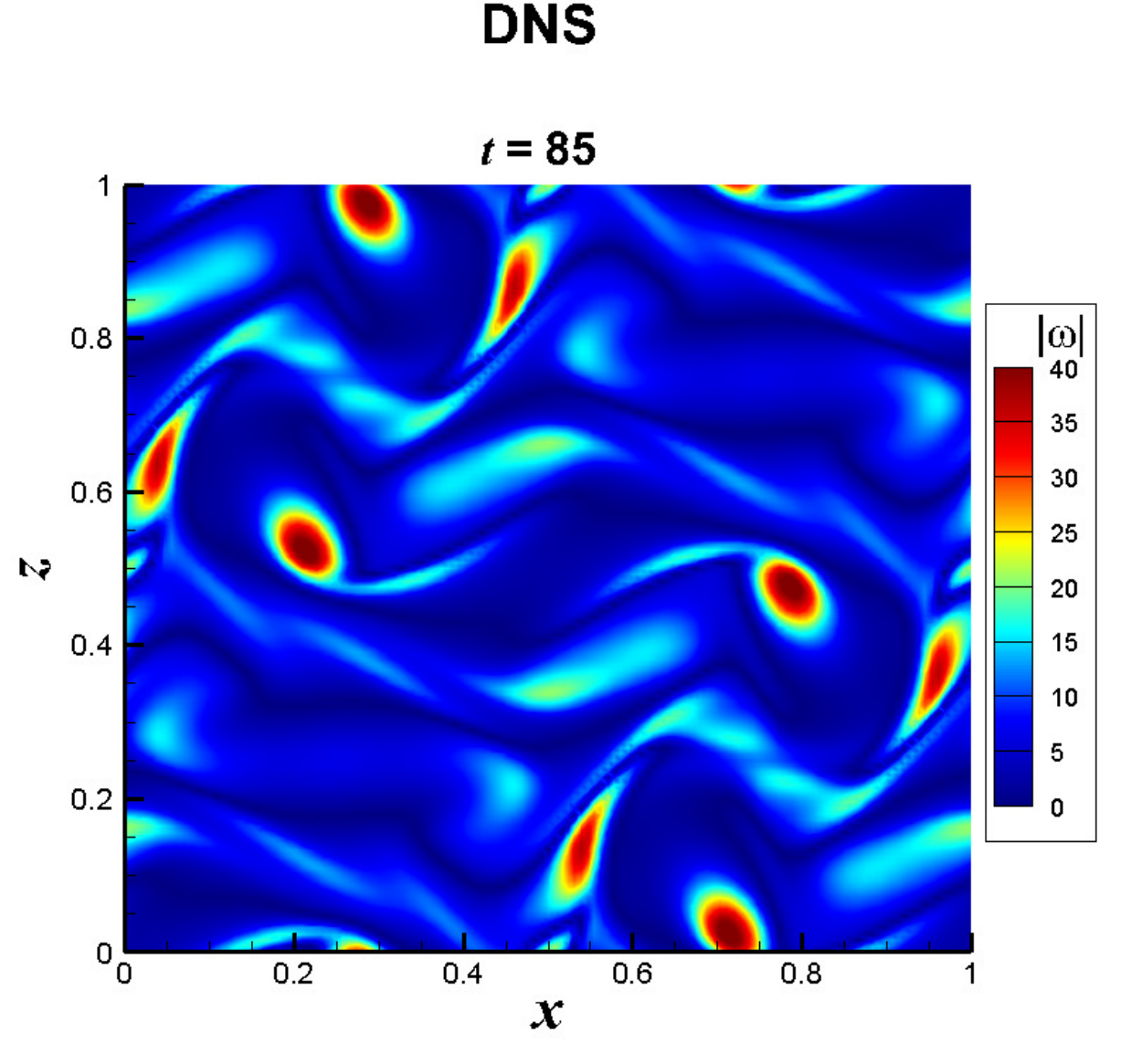}}\\
             \subfigure[]{\includegraphics[width=1.7in]{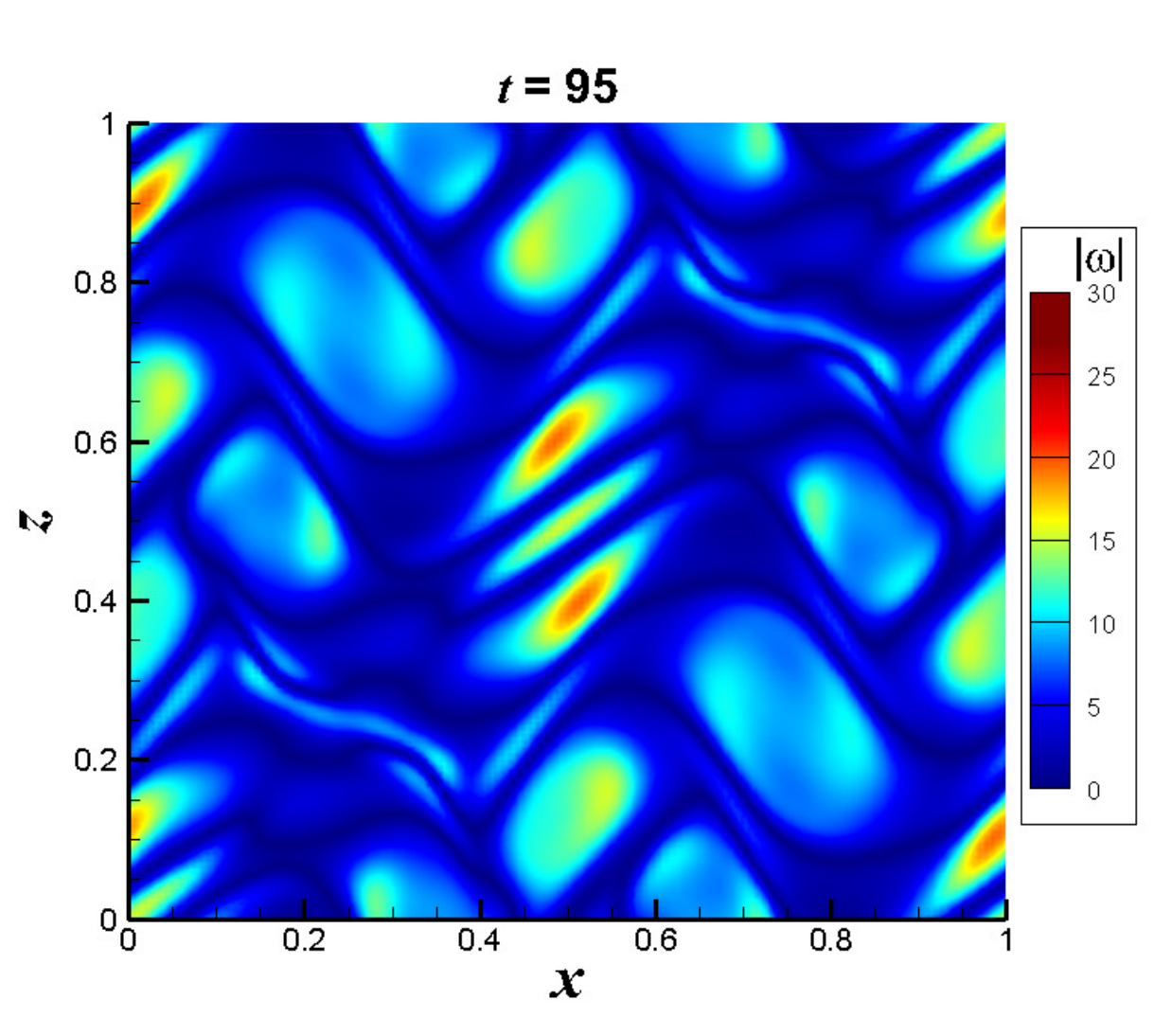}}
             \subfigure[]{\includegraphics[width=1.7in]{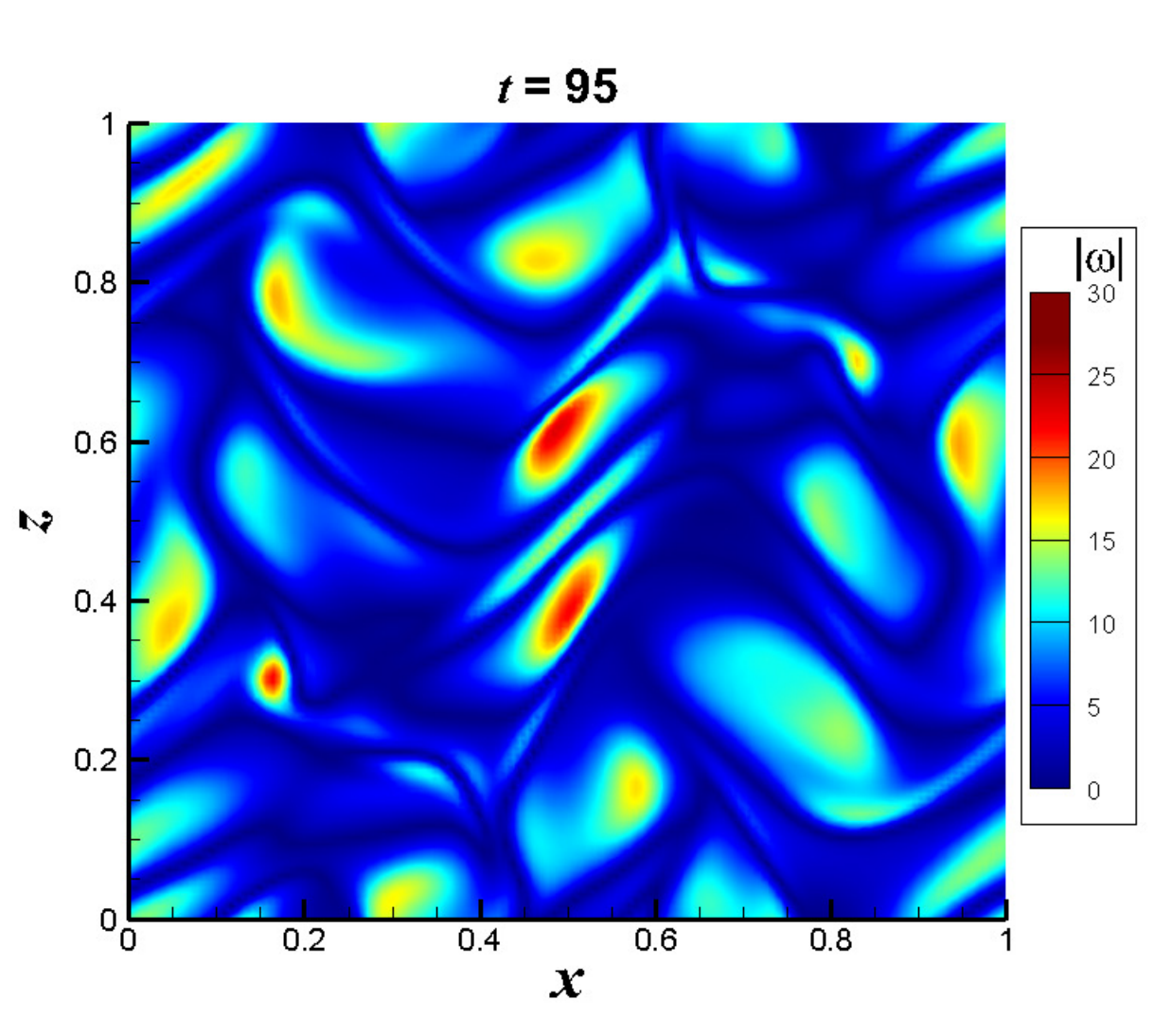}}\\
             \subfigure[]{\includegraphics[width=1.7in]{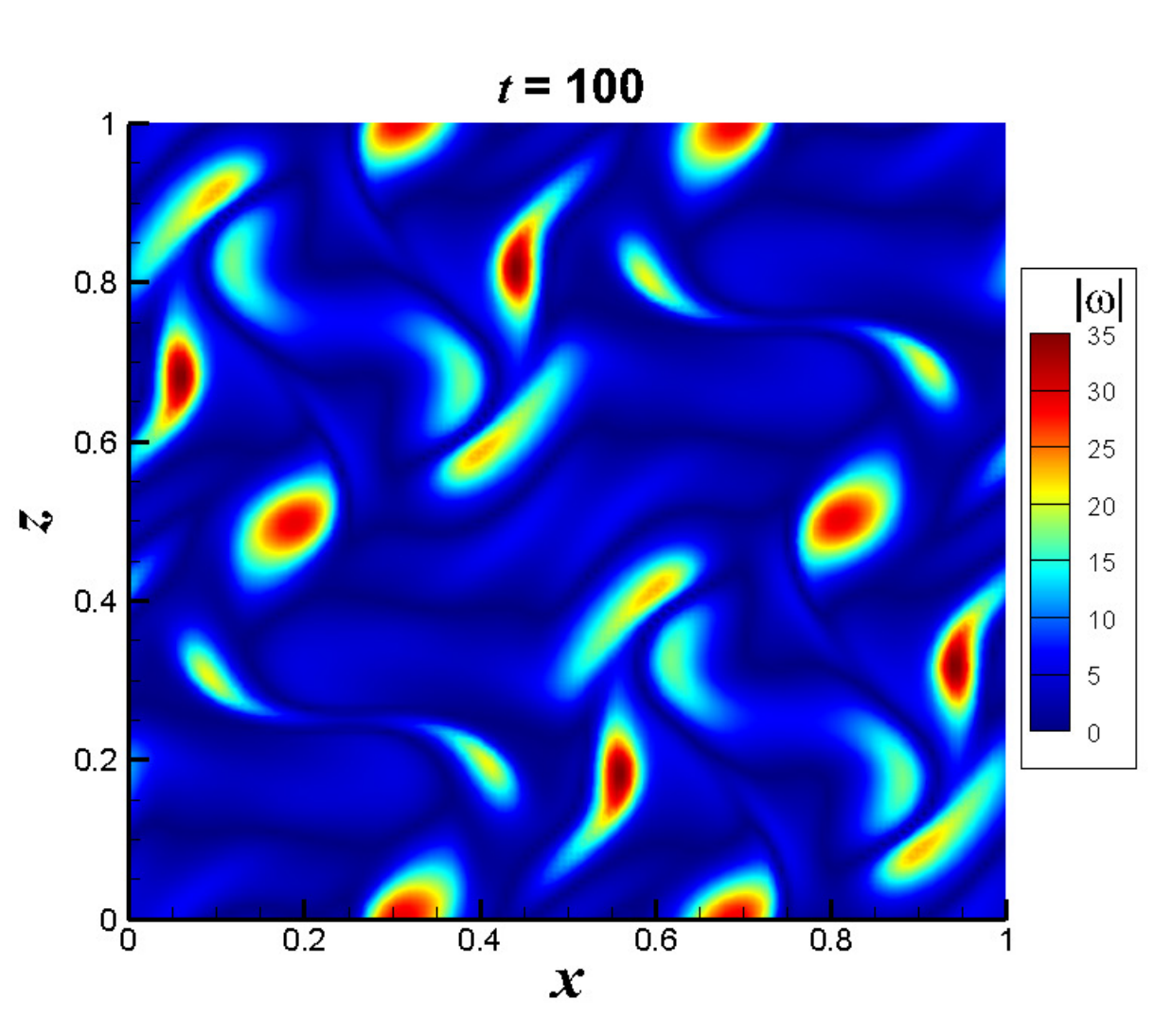}}
             \subfigure[]{\includegraphics[width=1.7in]{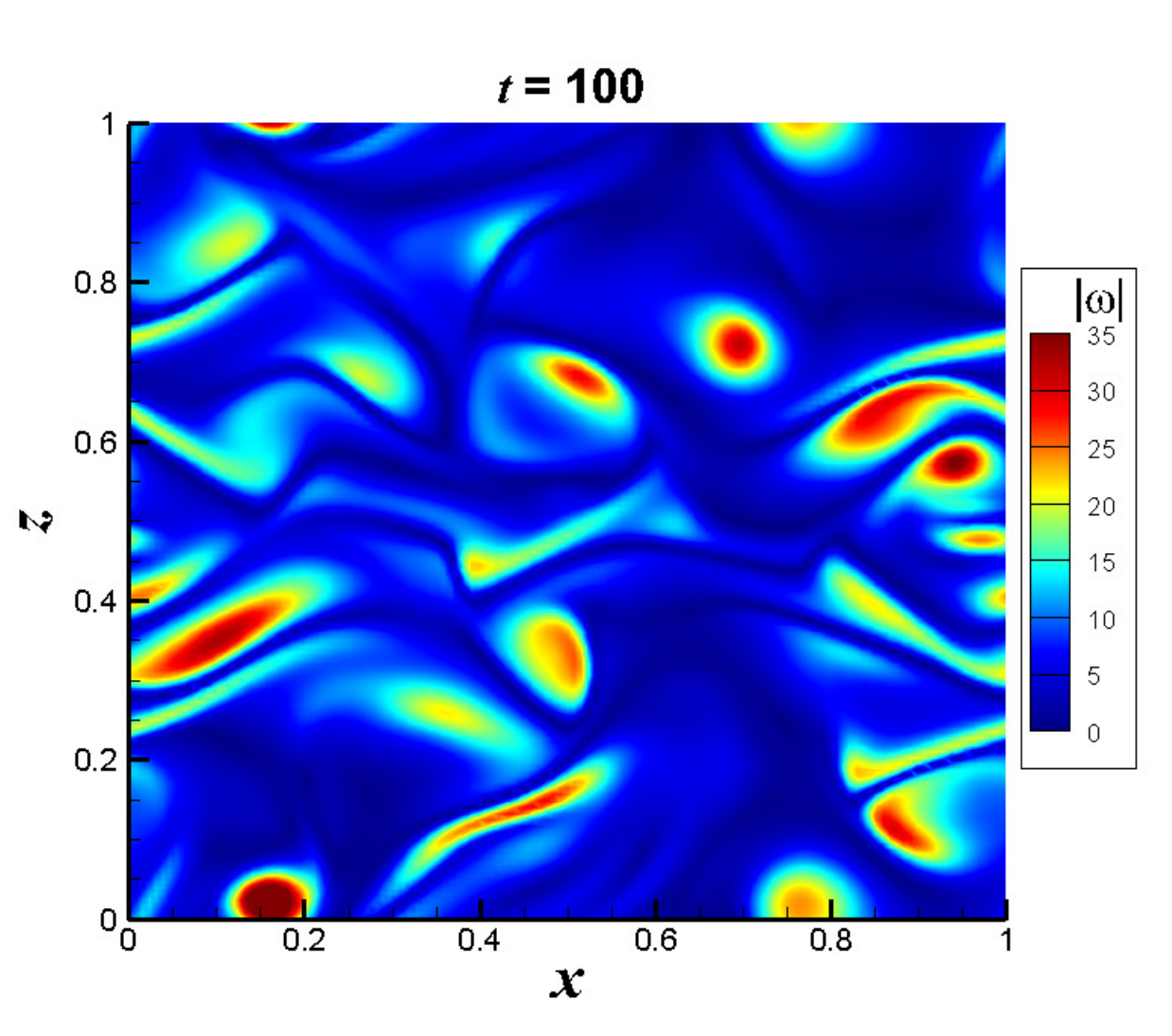}}\\
             \subfigure[]{\includegraphics[width=1.7in]{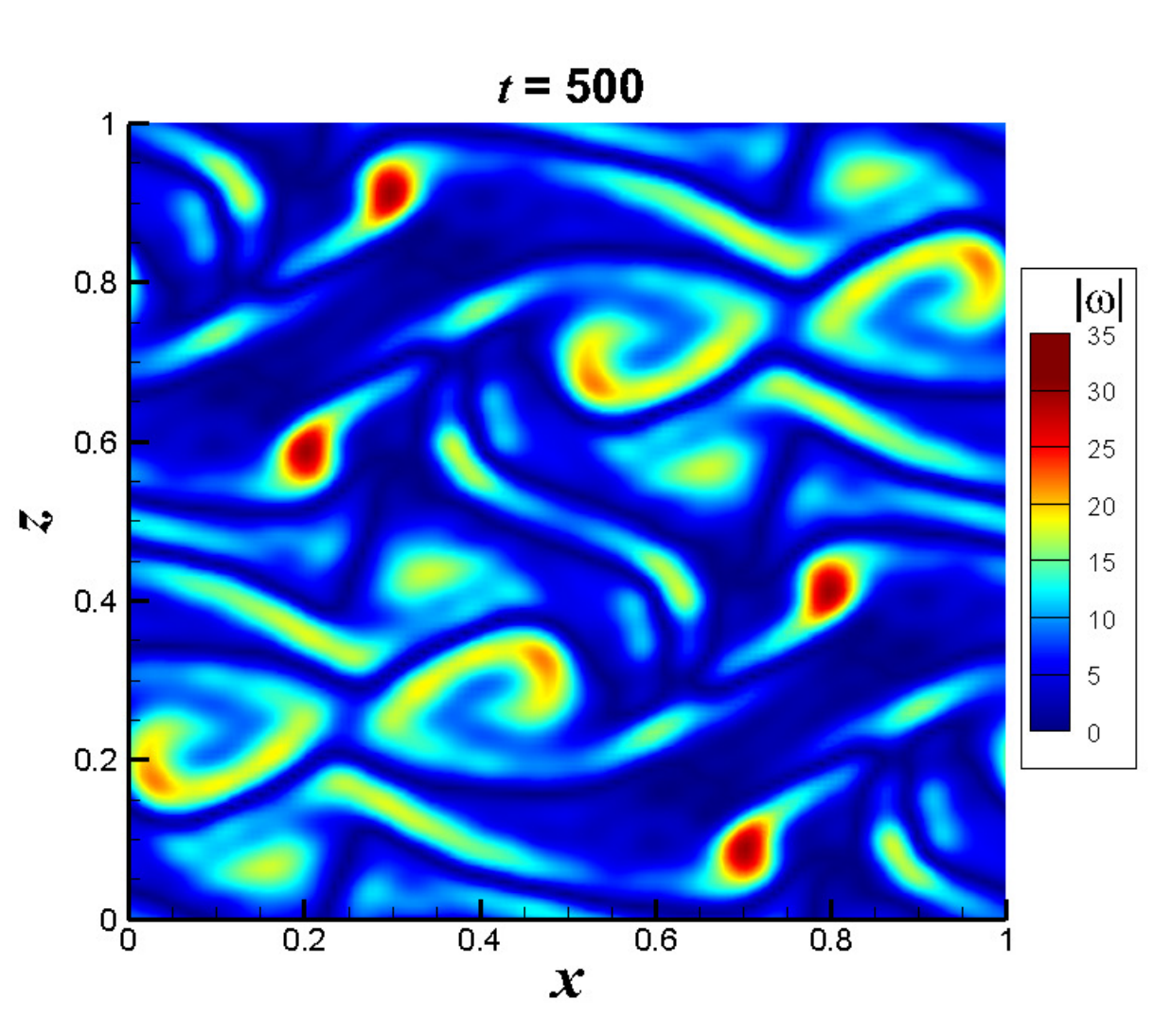}}
             \subfigure[]{\includegraphics[width=1.7in]{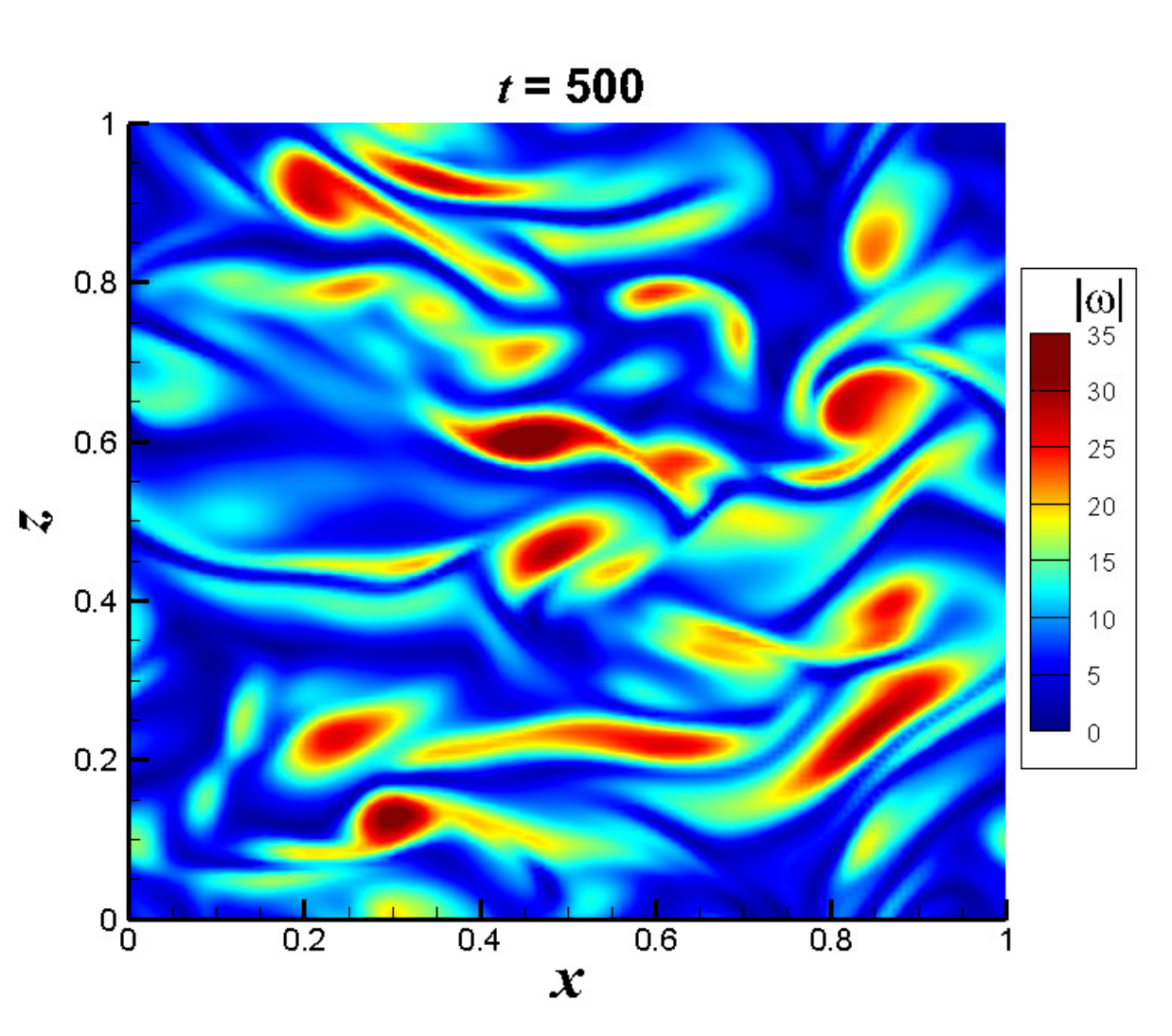}}\\
        \end{tabular}
    \caption{Distribution of the vorticity modulus at $y=0$, i.e. $|\bm{\omega}(x,0,z,t)|$, of the 3D turbulent Kolmogorov flow governed by Eq.~(1) subject to and the initial condition (2) with the spatial symmetry (3) in the case of $n_K=4$ and $Re=1211.5$ given by the CNS benchmark solution (left) and the DNS result (right), respectively, at some typical times, i.e. (a)-(b) $t=85$, (c)-(d) $t=95$, (e)-(f) $t=100$, and (g)-(h) $t=500$.}     \label{Vor_2D-1}
    \end{center}
\end{figure*}

\begin{figure*}[!htb]
    \begin{center}
        \begin{tabular}{cc}
             \subfigure[]{\includegraphics[width=1.7in]{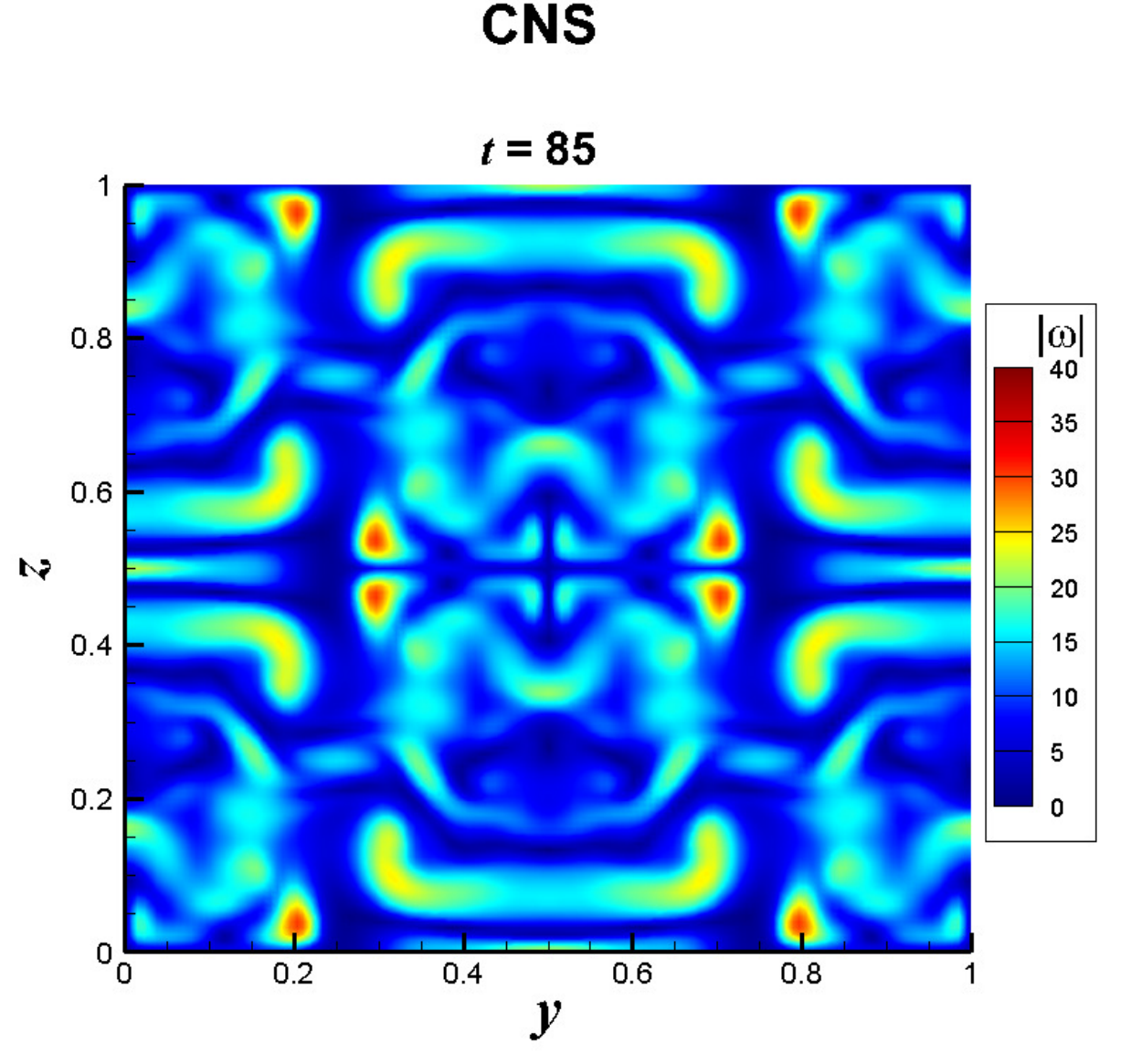}}
             \subfigure[]{\includegraphics[width=1.7in]{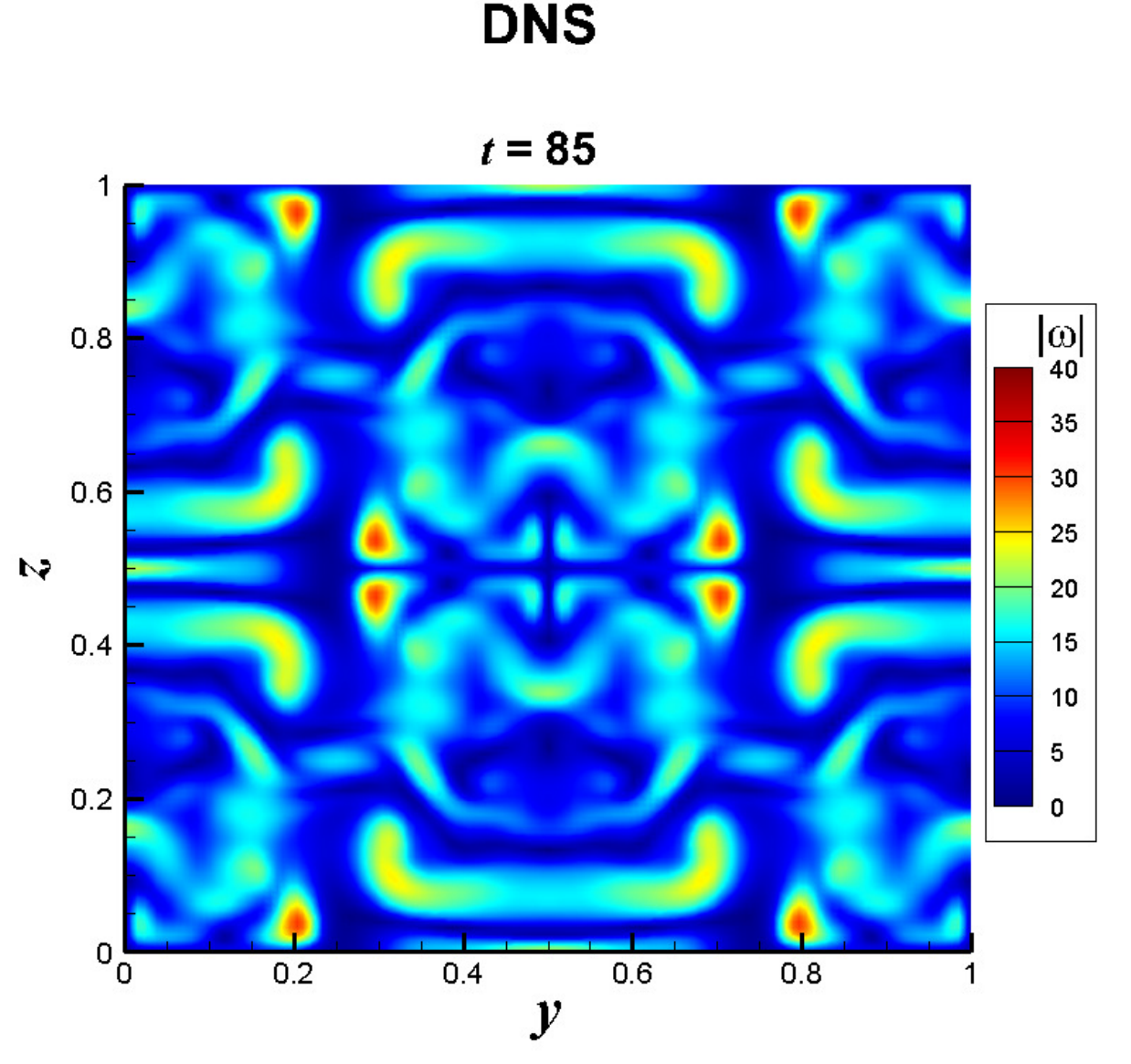}}\\
             \subfigure[]{\includegraphics[width=1.7in]{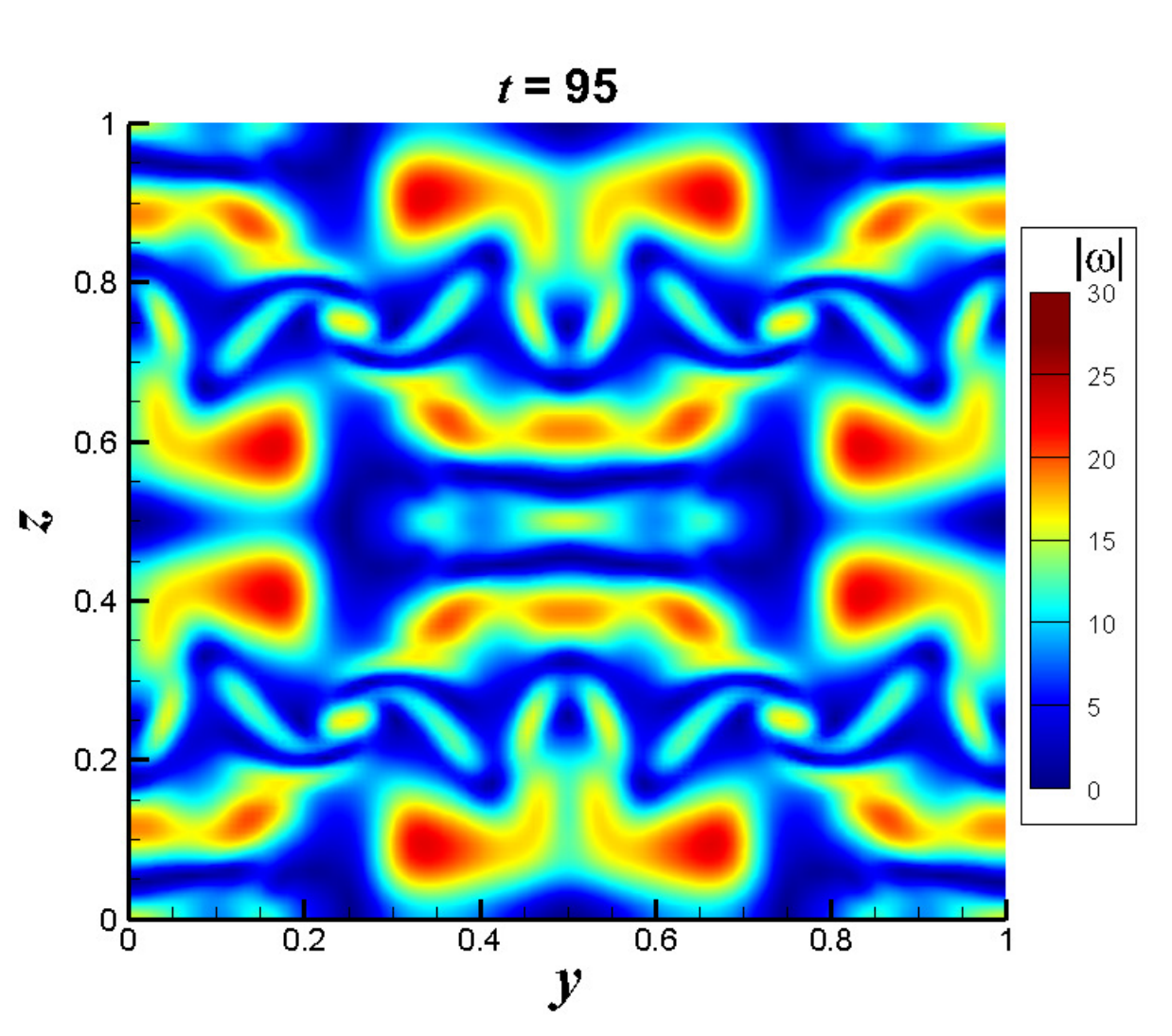}}
             \subfigure[]{\includegraphics[width=1.7in]{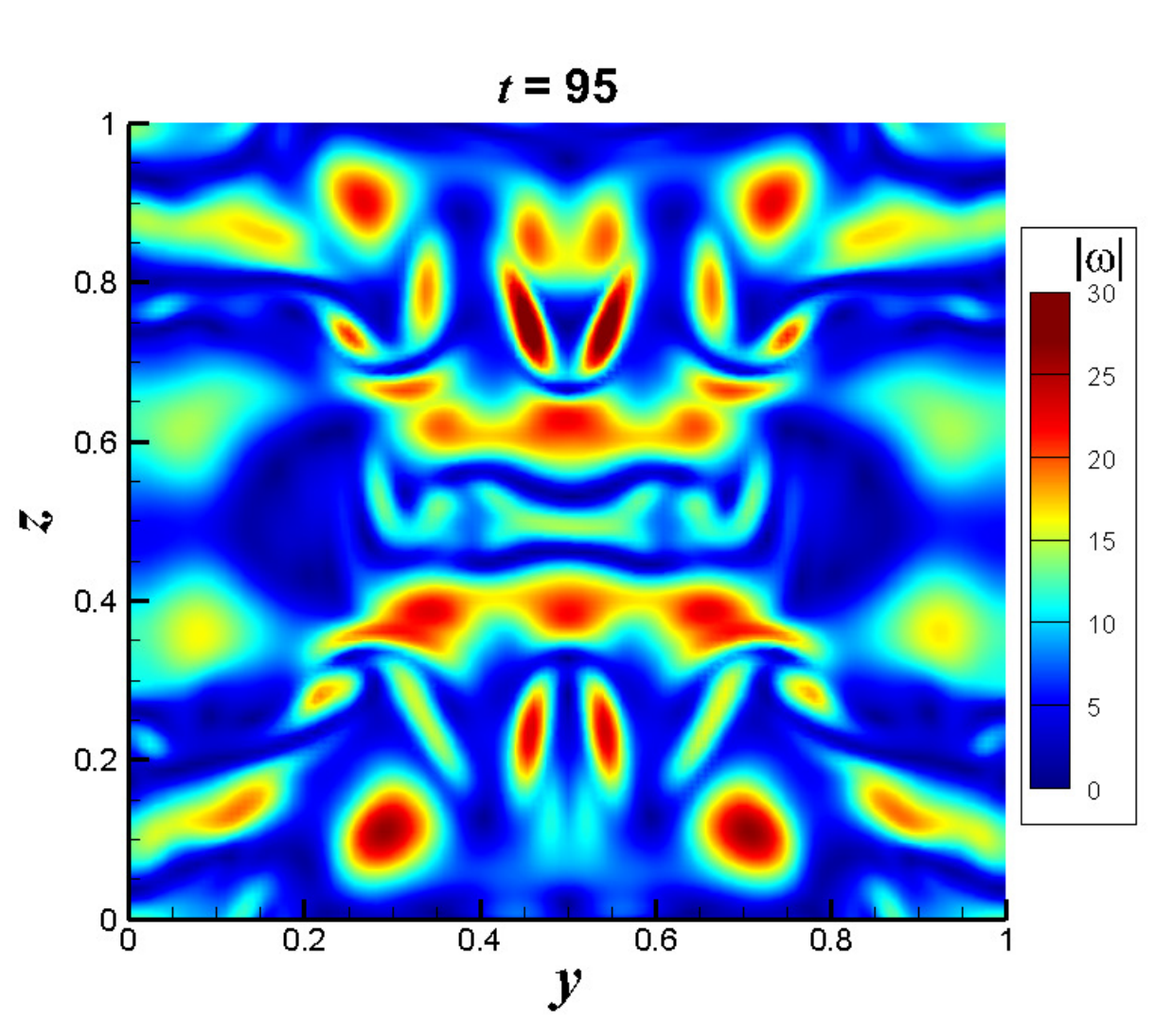}}\\
             \subfigure[]{\includegraphics[width=1.7in]{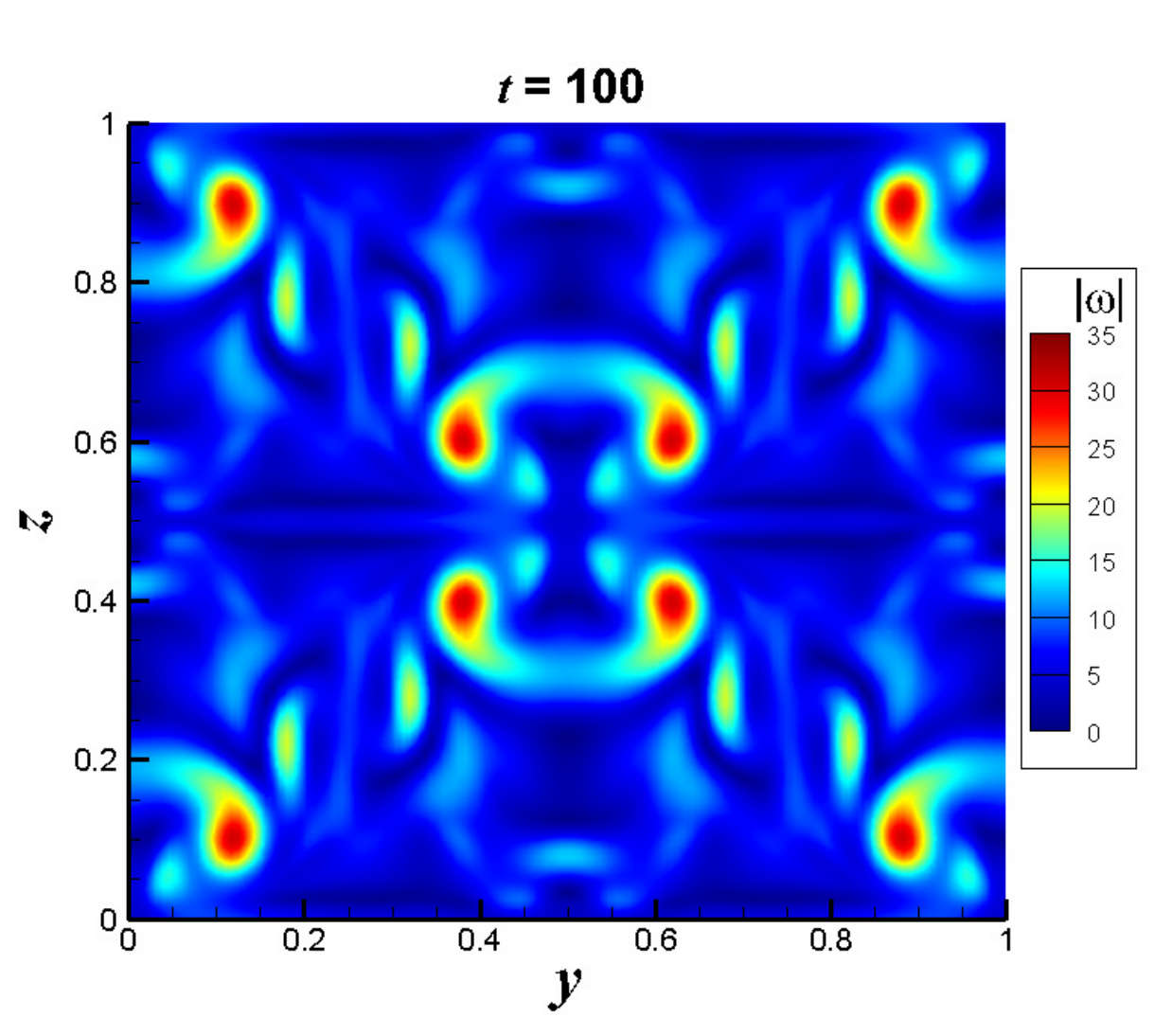}}
             \subfigure[]{\includegraphics[width=1.7in]{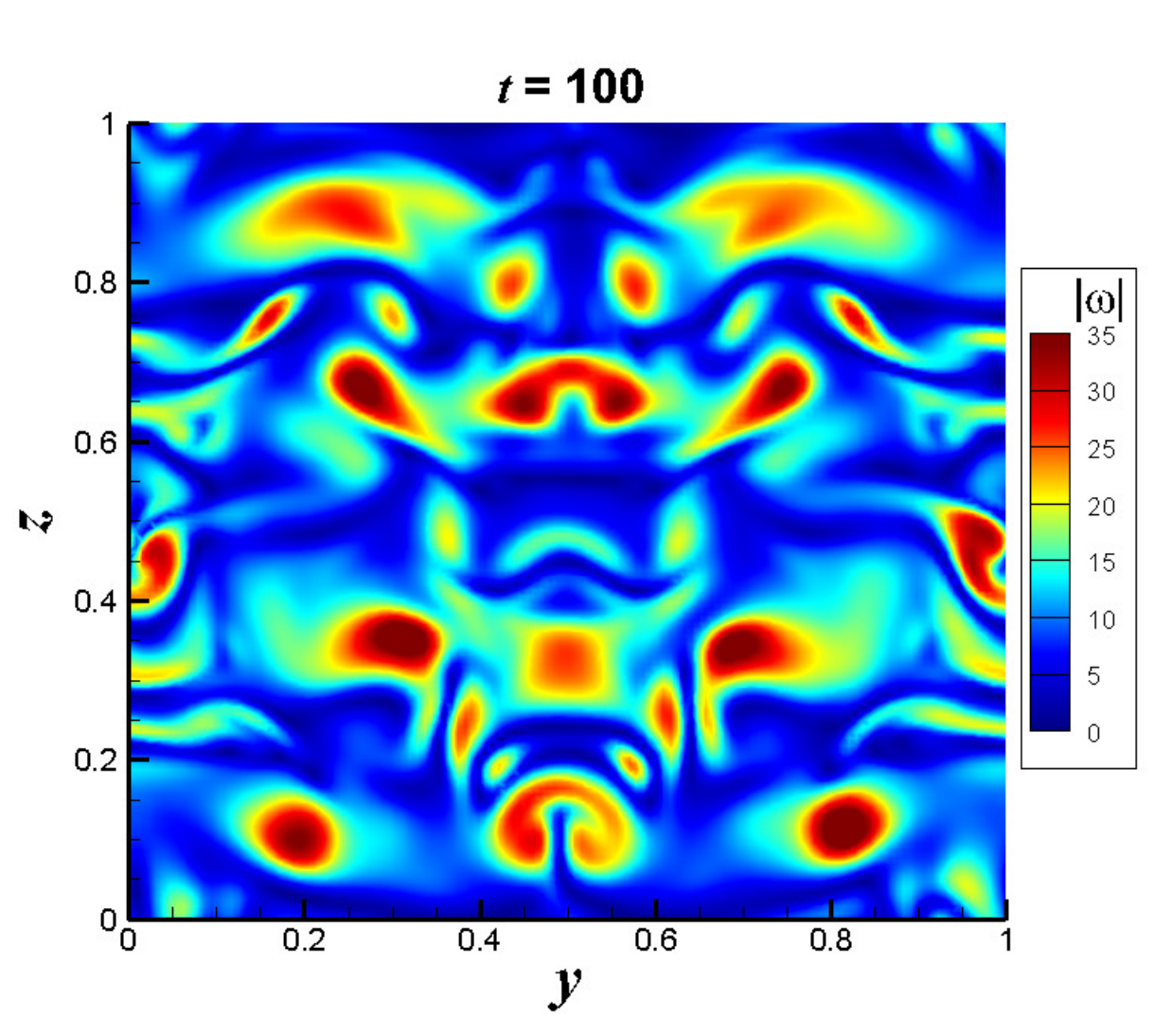}}\\
             \subfigure[]{\includegraphics[width=1.7in]{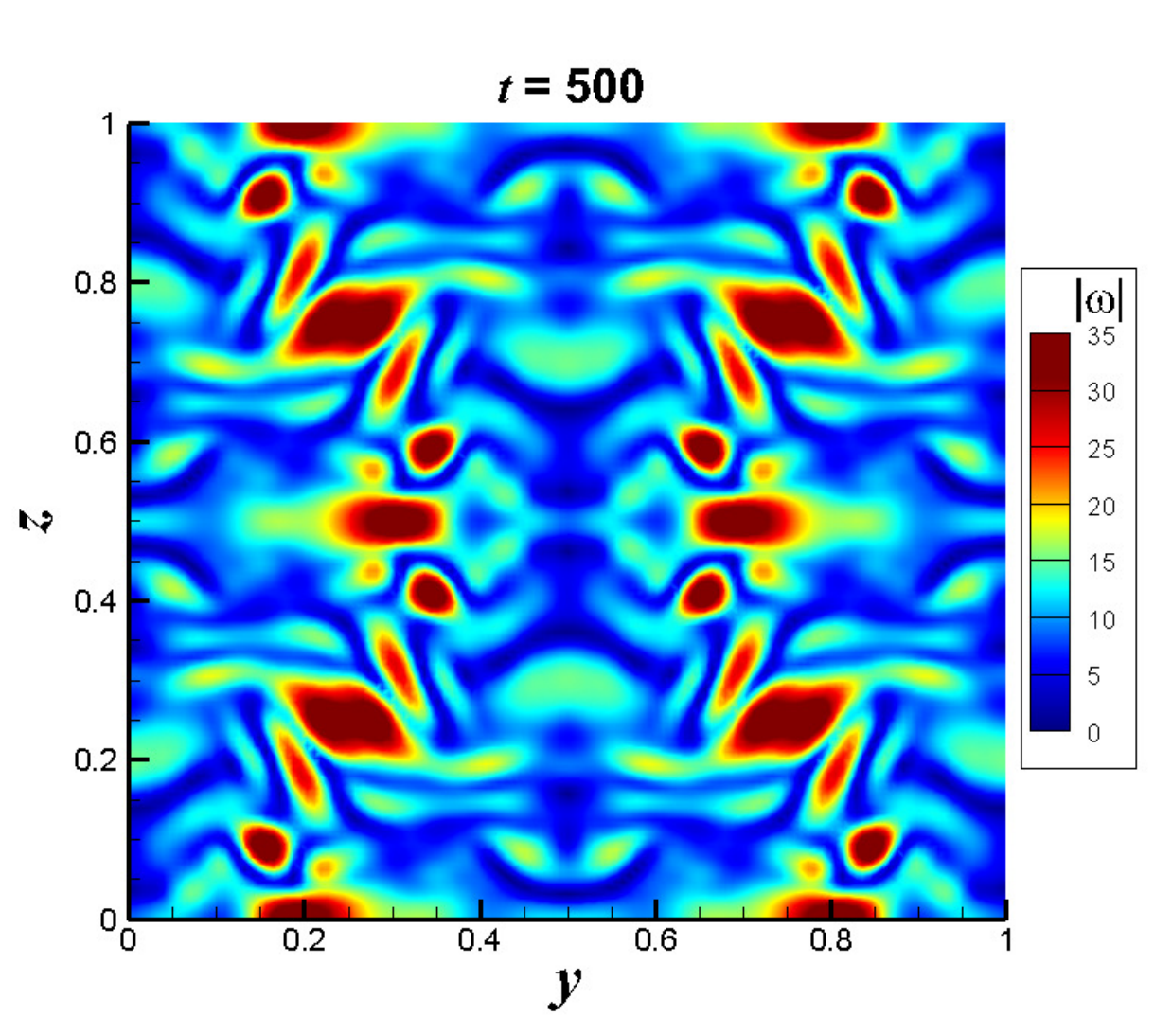}}
             \subfigure[]{\includegraphics[width=1.7in]{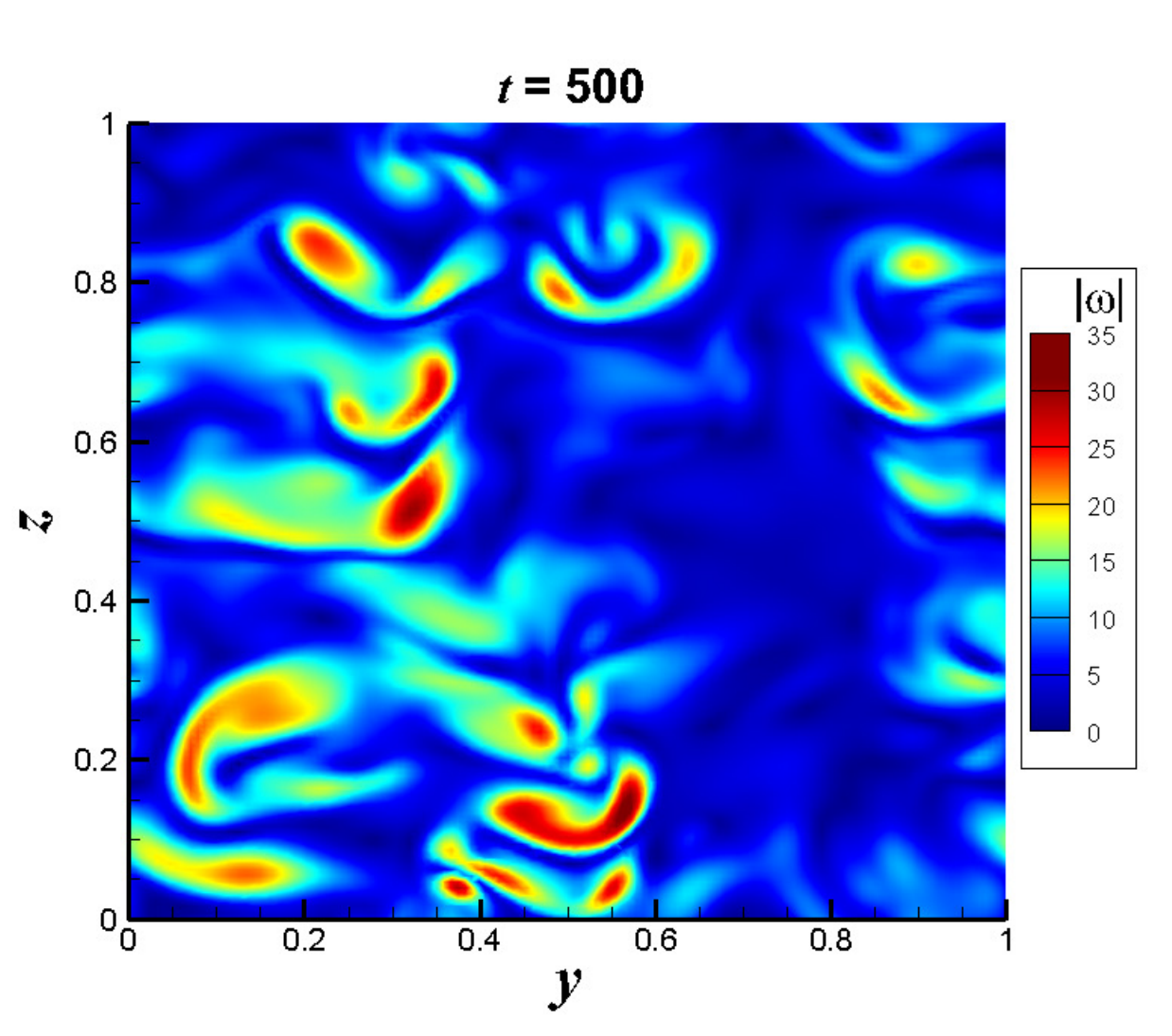}}\\
        \end{tabular}
    \caption{Distribution of the vorticity modulus at $x=0$, i.e. $|\bm{\omega}(0,y,z,t)|$, of the 3D turbulent Kolmogorov flow governed by Eq.~(1) subject the periodic boundary condition and the initial condition (2) with the spatial symmetry (3) in the case of $n_K=4$ and $Re=1211.5$ given by the CNS benchmark solution (left) and the DNS result (right), respectively, at some typical times, i.e. (a)-(b) $t=85$, (c)-(d) $t=95$, (e)-(f) $t=100$, and (g)-(h) $t=500$.}     \label{Vor_2D-2}
    \end{center}
\end{figure*}

\begin{figure*}[!htb]
    \begin{center}
        \begin{tabular}{cc}
             \subfigure[]{\includegraphics[width=1.7in]{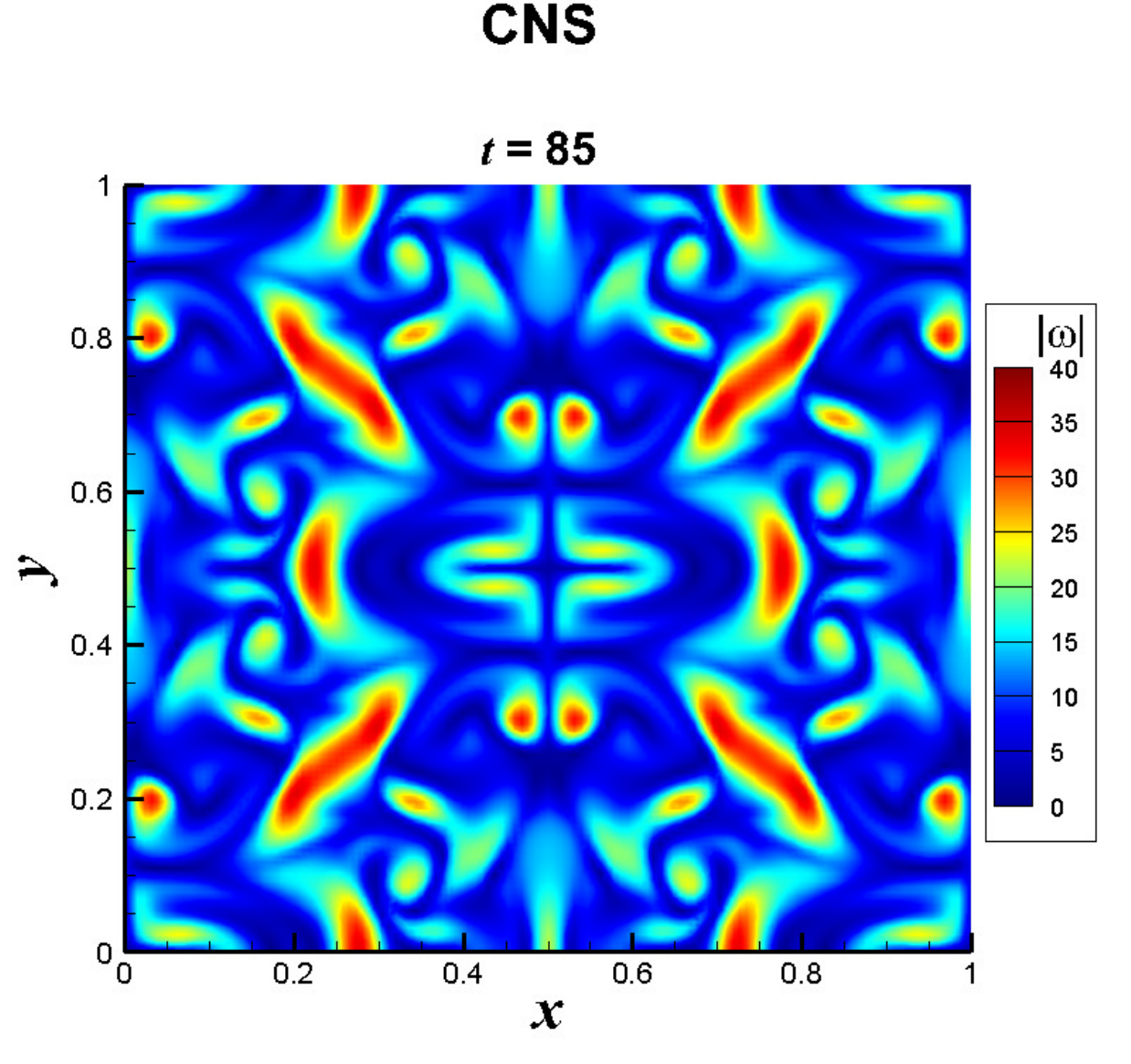}}
             \subfigure[]{\includegraphics[width=1.7in]{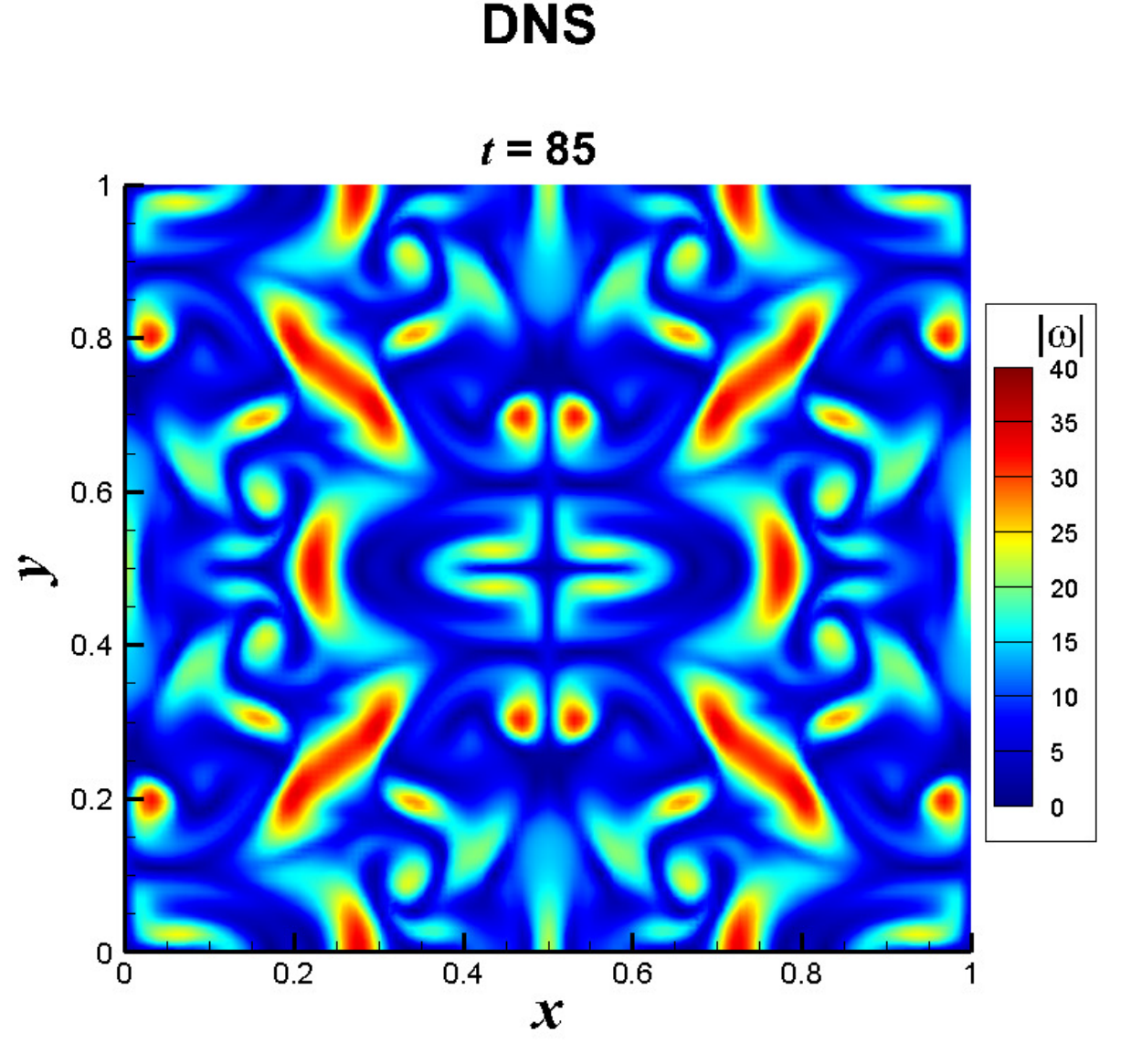}}\\
             \subfigure[]{\includegraphics[width=1.7in]{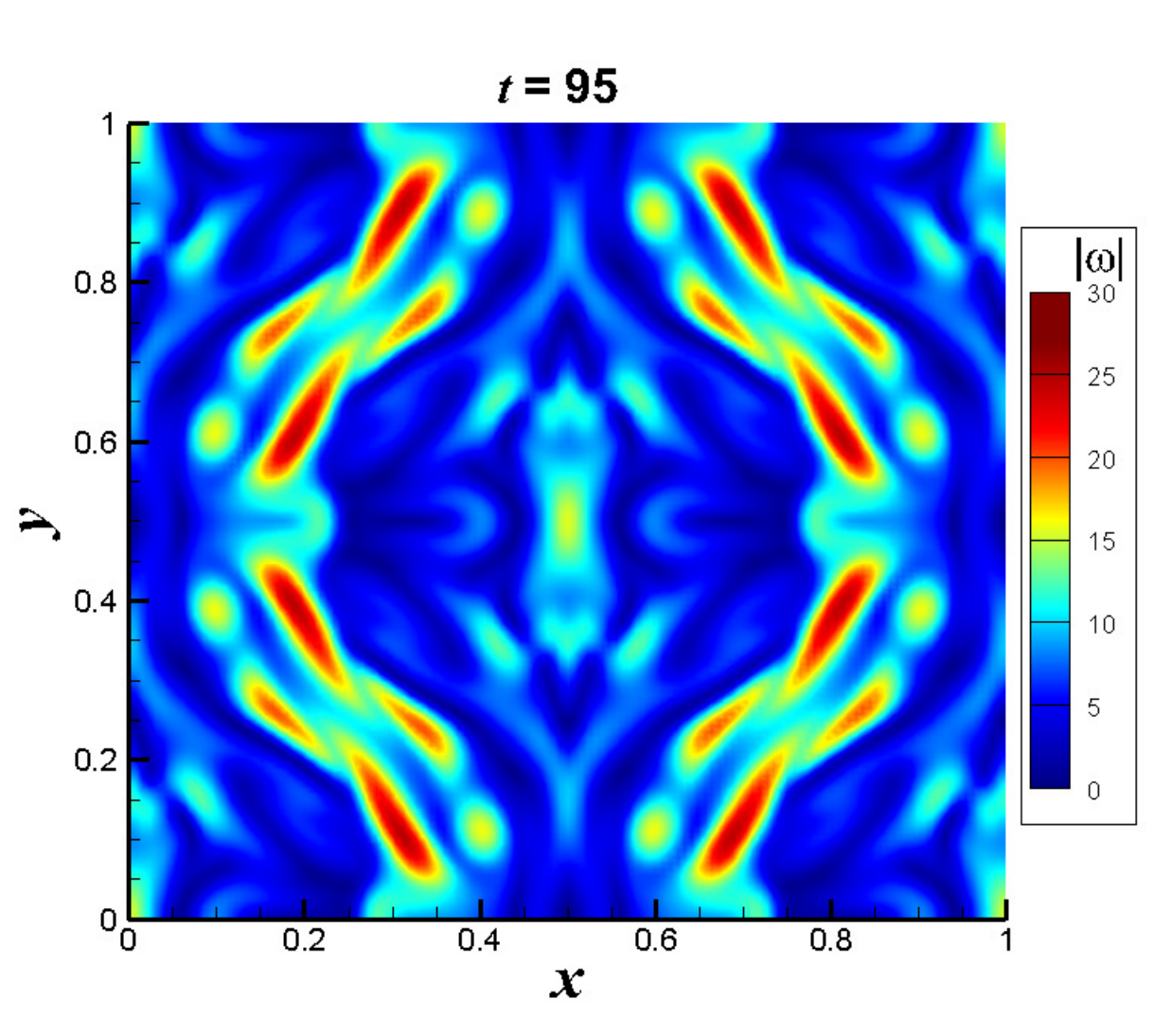}}
             \subfigure[]{\includegraphics[width=1.7in]{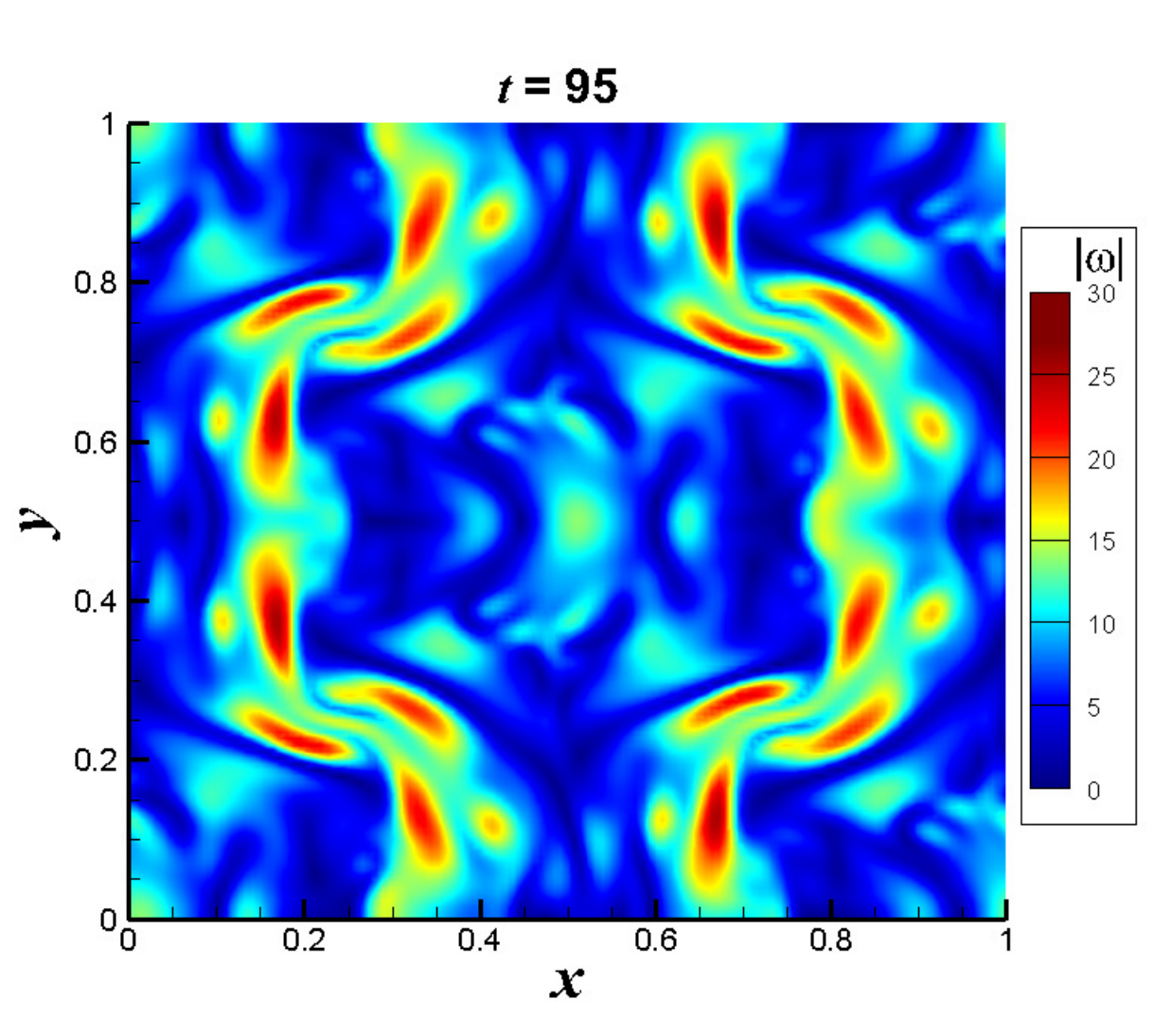}}\\
             \subfigure[]{\includegraphics[width=1.7in]{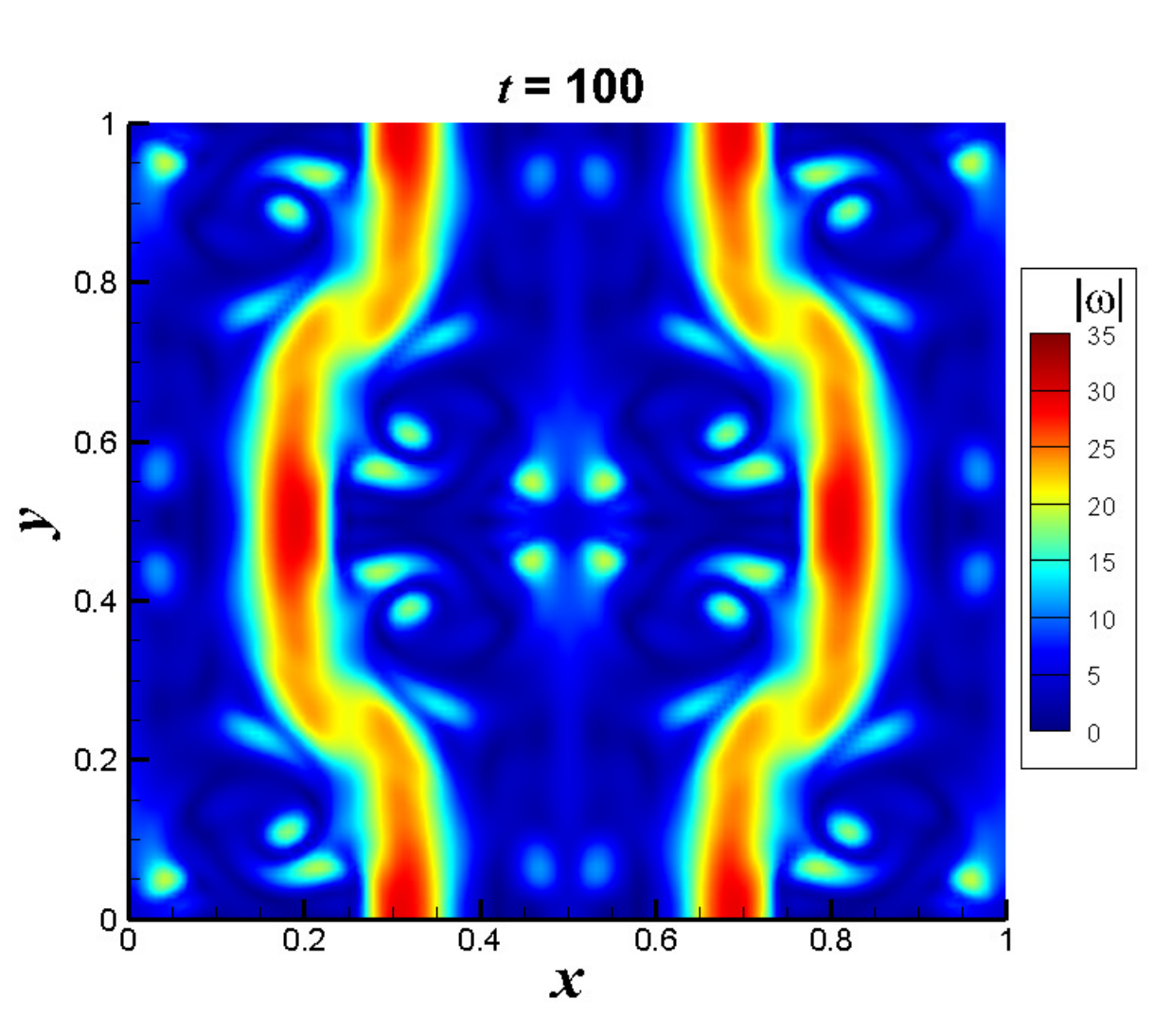}}
             \subfigure[]{\includegraphics[width=1.7in]{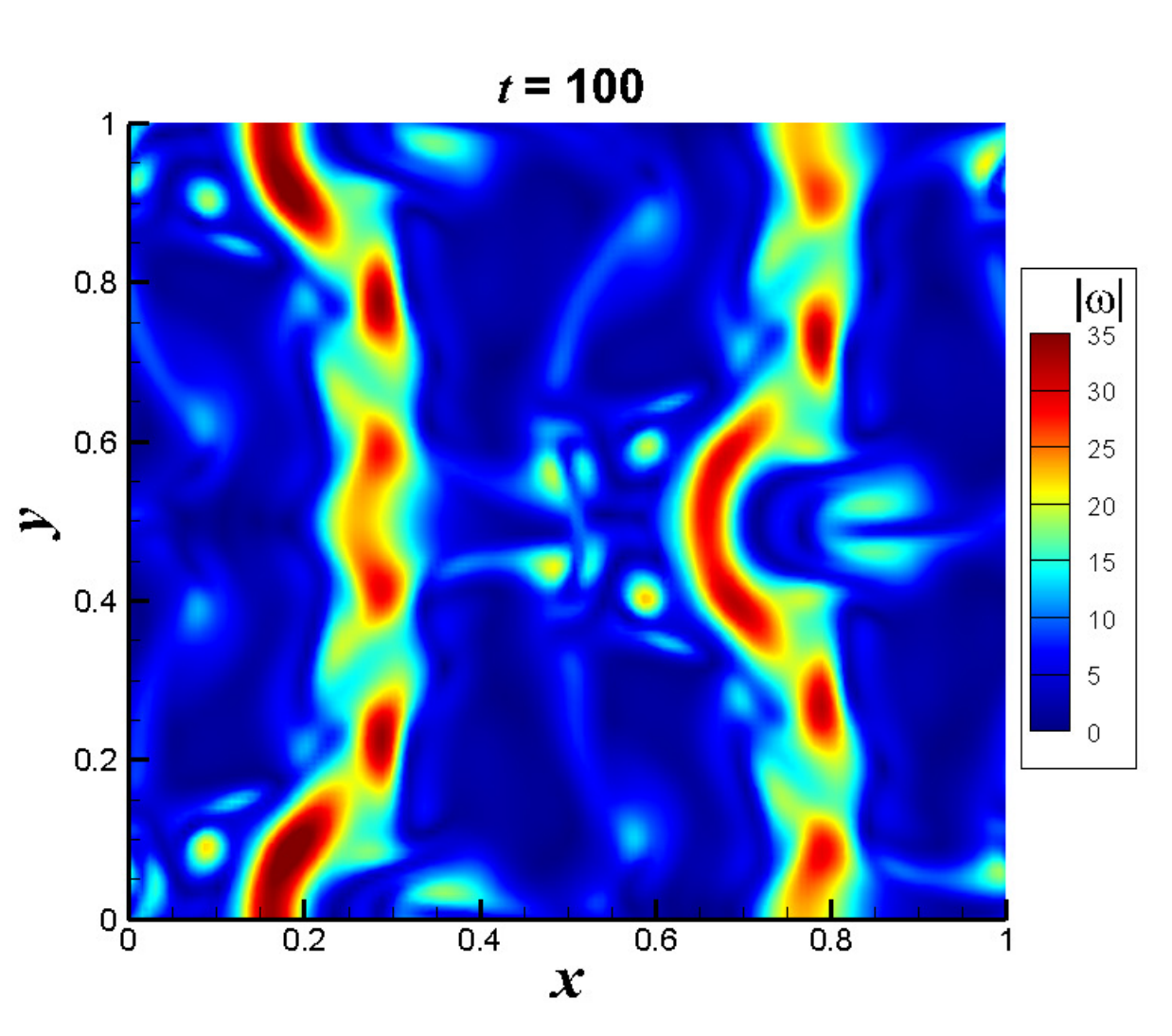}}\\
             \subfigure[]{\includegraphics[width=1.7in]{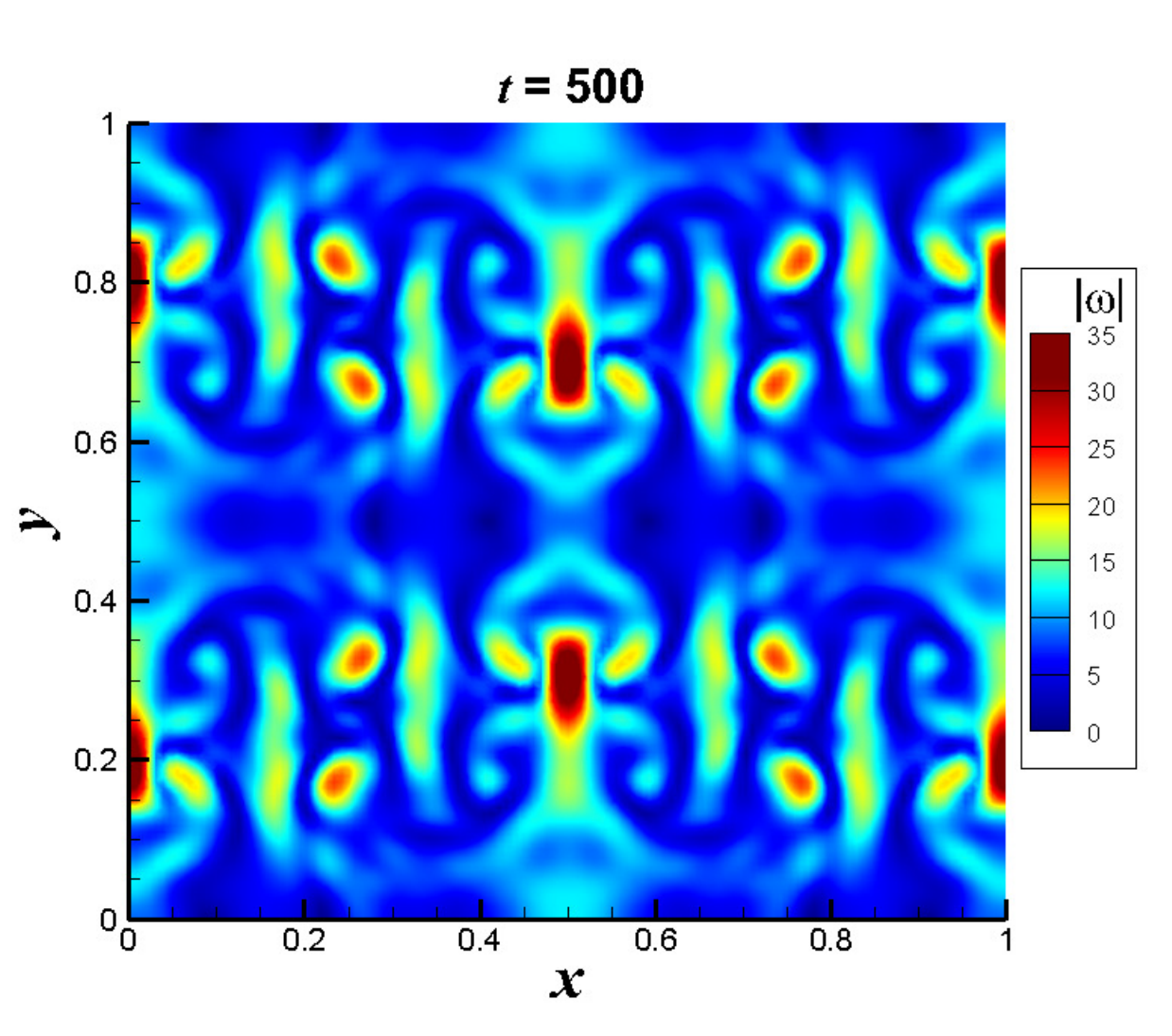}}
             \subfigure[]{\includegraphics[width=1.7in]{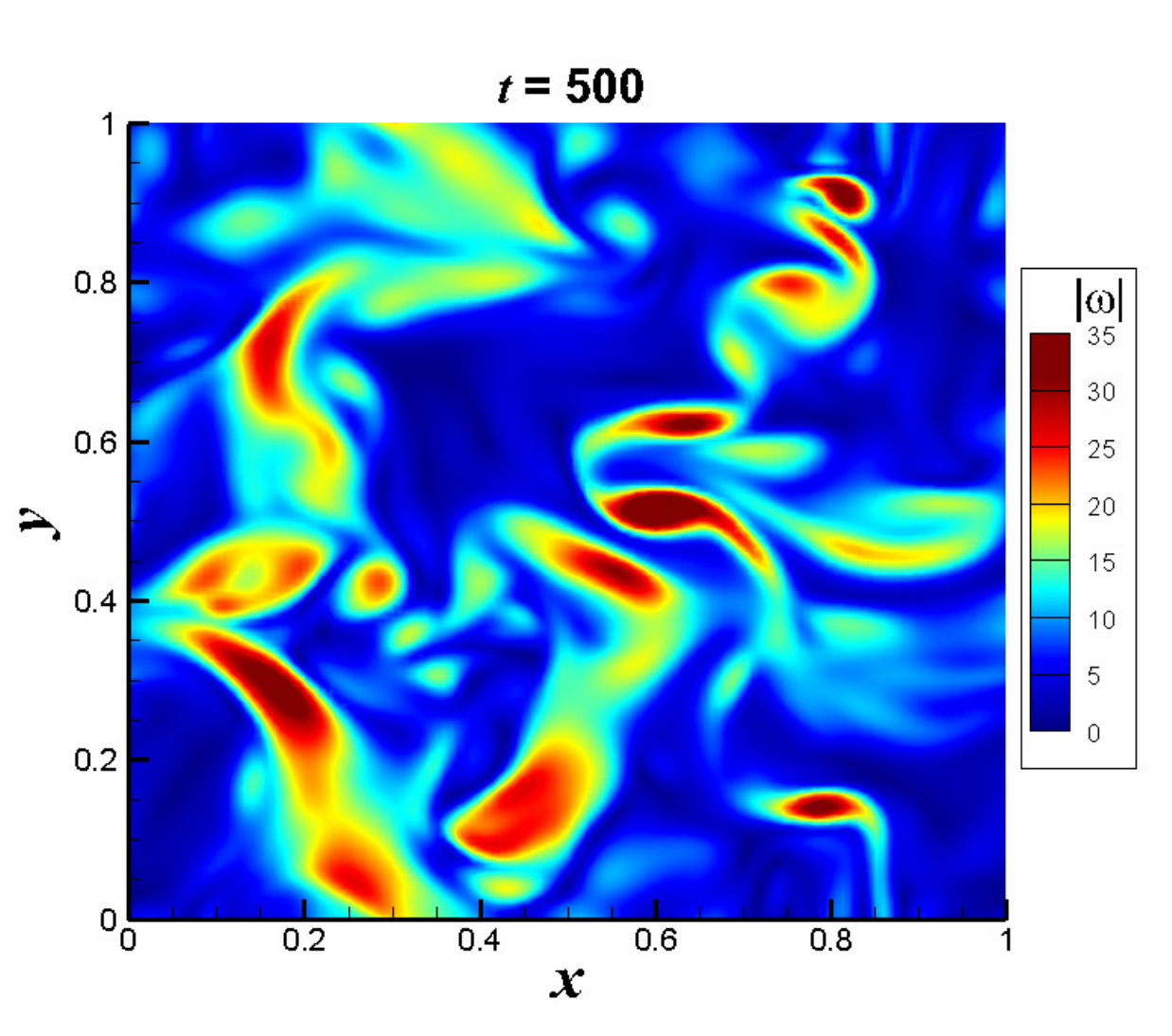}}\\
        \end{tabular}
    \caption{Distribution of the vorticity modulus at $z=0$, i.e. $\Large|\bm{\omega}(x,y,0,t)\Large|$, of the 3D turbulent Kolmogorov flow governed by Eq.~(1) subject the periodic boundary condition and the initial condition (2) with the spatial symmetry (3) in the case of $n_K=4$ and $Re=1211.5$ given by the CNS benchmark solution (left) and the DNS result (right), respectively, at some typical times, i.e. (a)-(b) $t=85$, (c)-(d) $t=95$, (e)-(f) $t=100$, and (g)-(h) $t=500$.}     \label{Vor_2D-3}
    \end{center}
\end{figure*}

\begin{figure*}[!htb]
    \begin{center}
        \begin{tabular}{cc}
             \includegraphics[width=2.3in]{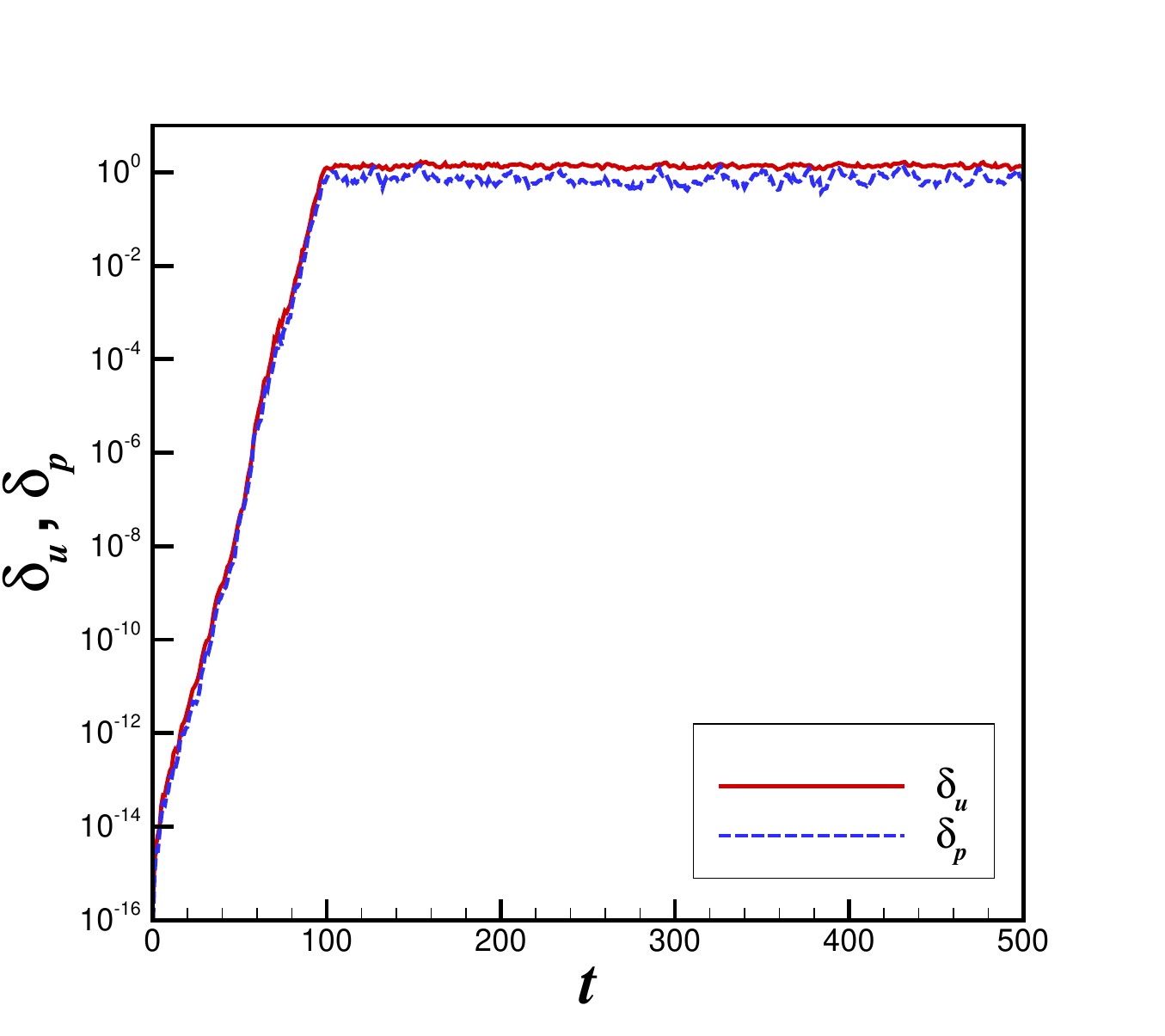}
        \end{tabular}
    \caption{Time histories of the mean squared deviations (between the DNS result and the CNS benchmark solution) of  $\delta_u$ defined by (\ref{delta_u}) (for the velocity) and $\delta_p$ defined by (\ref{delta_p}) (for the pressure) in Appendix~C of the 3D turbulent Kolmogorov flow governed by Eq.~(1) subject the periodic boundary condition and the initial condition (2) with the spatial symmetry (3) in the case of $n_K=4$ and $Re=1211.5$. Red solid line: $\delta_u$; blue dashed line: $\delta_p$.}     \label{delta}
    \end{center}
\end{figure*}

For the 2D turbulent Kolmogorov flow, the flow fields of the CNS benchmark solution always remain the same spatial symmetry (for $t \geq 0$) as the corresponding initial condition, as illustrated in  \cite{Qin2024JOES, Liao-2025-JFM-NEC}.  
It is found that this is also {\em true} for the CNS benchmark solution of the 3D turbulent Kolmogorov flow, but unfortunately {\em false} for the DNS result: as shown in Figs.~\ref{Vor_Evolutions} to \ref{Vor_2D-3}, the spatio-temporal trajectory given by DNS agrees well with the CNS benchmark solution and indeed has the spatial symmetry (\ref{spatial-symmetry}) within $t\leq 90$, but loses the  spatial symmetry lightly at $t\approx 95$ and obviously at $t \approx 100$, however the spatio-temporal trajectory given by CNS {\em always} remains the spatial symmetry (\ref{spatial-symmetry}) throughout the {\em whole} interval of time $t\in[0,500]$.
 Note that, within $t\leq 90$, both of DNS and CNS results are  the same, for example as shown in (a) and (b) (when $t=85$) of Figs.~\ref{Vor_2D-1} to \ref{Vor_2D-3}.  The corresponding ``statistic''  results and  ``scale-to-scale energy flux'' given by DNS in $t\in[0,90]$ agree also quite well with those given by CNS in the same interval of time, as shown in Figs.~\ref{t90} and \ref{EF-t90} of Appendix~A.  The definitions of these statistic values are given in Appendix~C.  
From the viewpoint of CNS, the so-called ``critical predictable time'' $T_{c}$ (see (\ref{Tc-1}) and (\ref{Tc-2})) of our DNS result is about 90, i.e. $T_{c}\approx 90$, which is unfortunately too short to obtain reliable statistic results.  Even so,  they clearly indicate that the DNS and CNS indeed initially give the {\em same} solution.  It should be emphasized that the so-called ``critical predictable time'' $T_{c}$ is an important concept in the frame of CNS, beyond which numerical noise might be at the same level as the true solution so that the corresponding numerical simulation might be {\em far} away from the true solution and thus is {\em useless}.  However, DNS has {\em no} such kind of concept, say,  one could gain a DNS result in an interval of time as long as one would like (with a {\em linearly} increasing CPU time).  This is a fundamental difference between CNS and DNS.  It is found that the spatial-symmetry (\ref{spatial-symmetry}) of a CNS result always remain within $t \leq T_{c}$, where the so-called ``critical predictable time''  $T_{c}$ can be arbitrarily large (but with an {\em exponentially} increasing CPU time) from mathematical viewpoint,  as illustrated via Lorenz equation by Liao and Wang~\cite{LIAO2014On}.  
 
 The reason why the DNS result (when $t>100$) deviates greatly from the CNS and loses the spatial symmetry (\ref{spatial-symmetry}) is its consistently increasing random numerical noise without spatial symmetry, which is much larger than that of CNS and thus reaches at a macro-level much earlier (at $t \approx 100$) than the CNS result.               
This is very clear by considering the mean squared deviations $\delta_u(t)$ (for velocity) defined by (\ref{delta_u}) and $\delta_p(t)$ (for pressure) defined by (\ref{delta_p}) between DNS and CNS results: as shown in Fig.~\ref{delta}, both of $\delta_u(t)$ and $\delta_p(t)$ increase exponentially until $t \approx 100$ when they become macroscopic, say, the random artificial numerical noise is at the same order of magnitude as the exact solution $s_{exact}$ thereafter.    
This provides us a rigorous evidence that the spatio-temporal trajectory given by DNS (after $t > 100$) is indeed badly contaminated by the artificial numerical noise, but the CNS benchmark solution is always a good approximation of the exact solution $s_{exact}$ of the NS equations throughout the {\em whole} interval of time $t\in[0,500]$.   Even so, it would be nice if some rigorous  mathematical proofs about such kind of symmetry property of the NS equations can be given in future.

\begin{figure*}[!htb]
    \begin{center}
        \begin{tabular}{cc}
             \subfigure[]{\includegraphics[width=2.in]{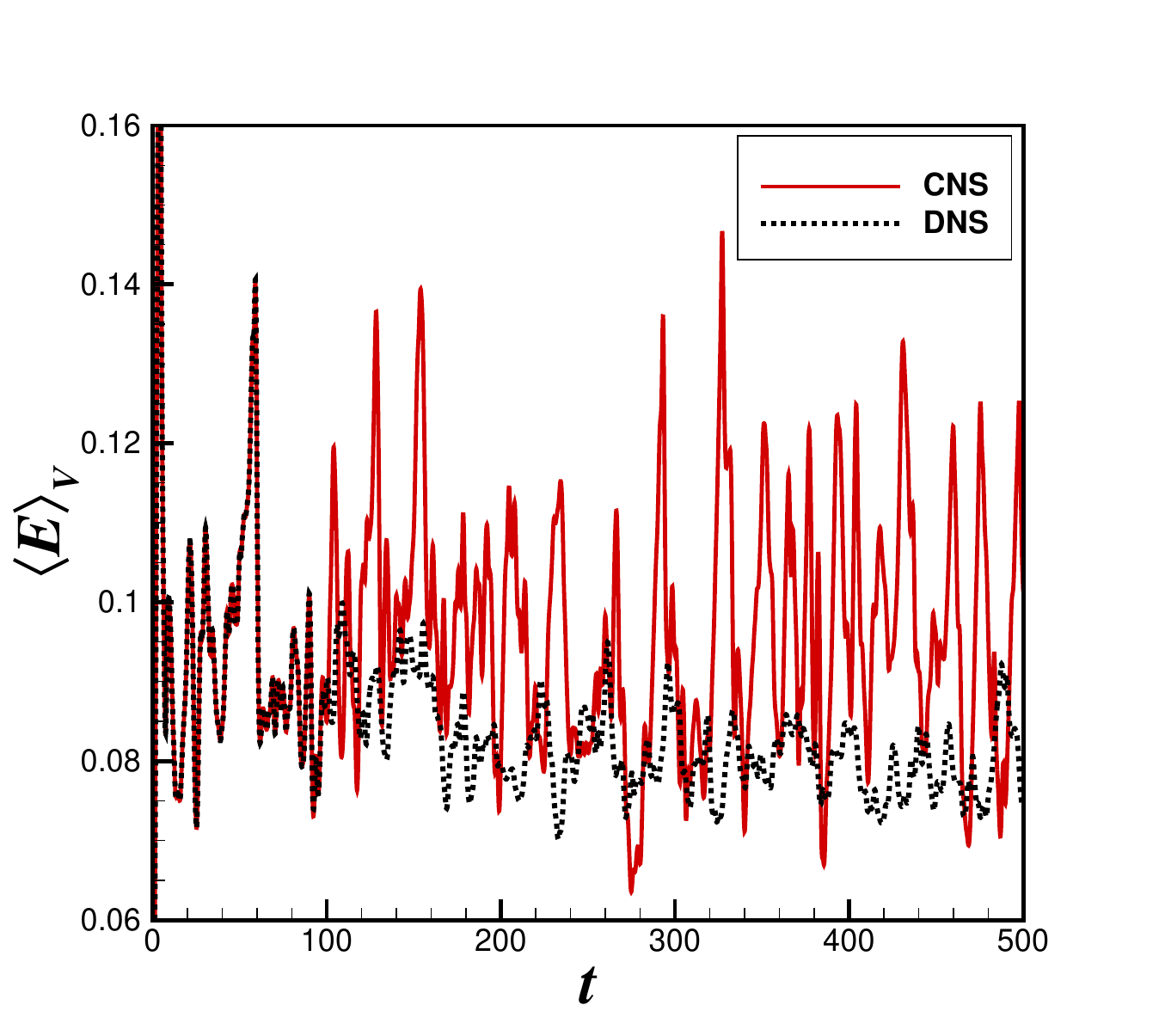}}
             \subfigure[]{\includegraphics[width=2.in]{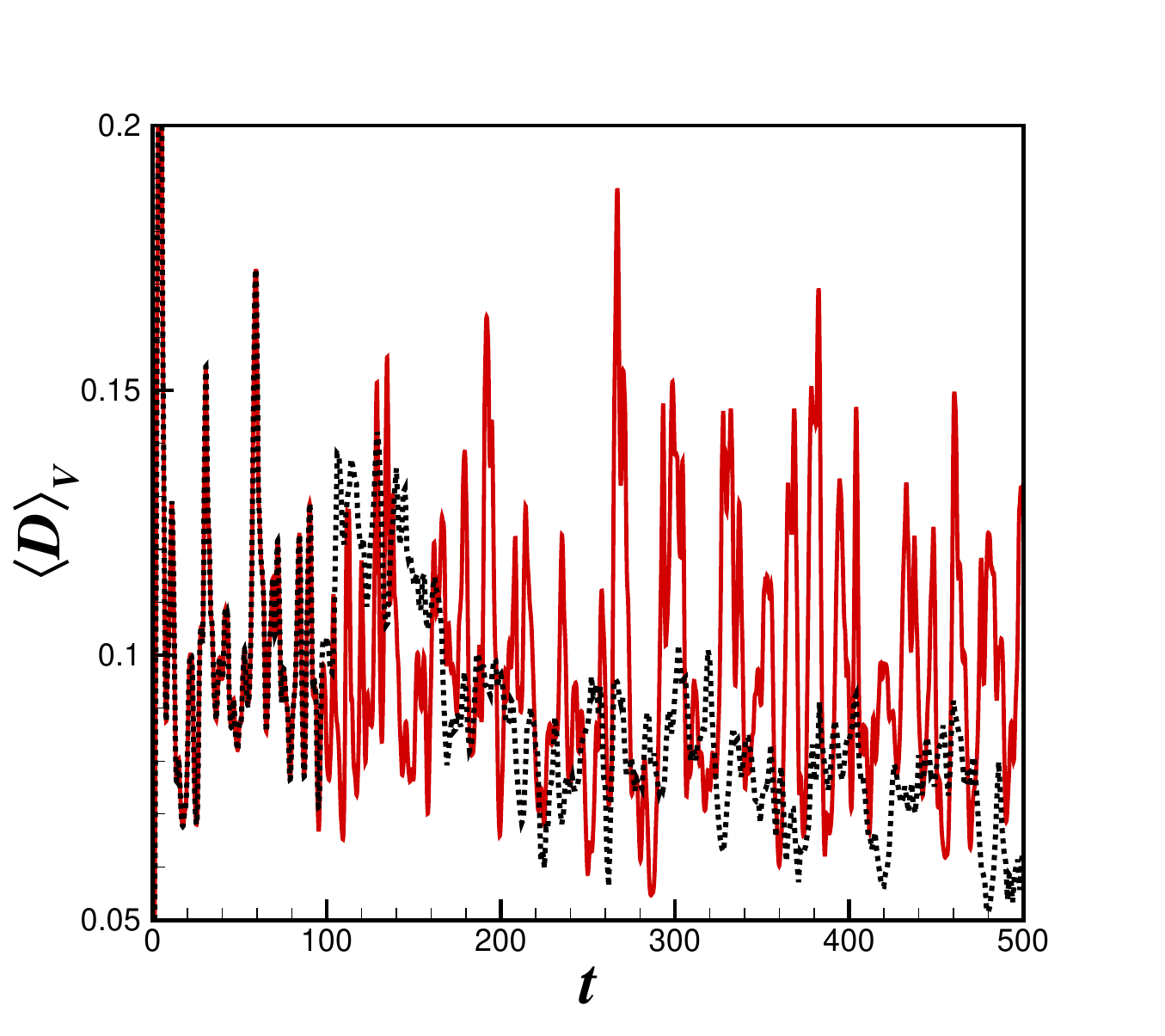}}
        \end{tabular}
    \caption{Comparisons of time histories of the spatially averaged (a) kinetic energy $\langle E\rangle_V$ and (b) kinetic energy dissipation rate $\langle D\rangle_V$ of the 3D turbulent Kolmogorov flow governed by Eq.~(1) subject the periodic boundary condition and the initial condition (2) with the spatial symmetry (3) in the case of $n_K=4$ and $Re=1211.5$, given by the CNS benchmark solution (solid line in red) and the DNS result (dashed line in black), respectively.}     \label{DE_t}
    \end{center}

    \begin{center}
        \begin{tabular}{cc}
             \subfigure[]{\includegraphics[width=2.in]{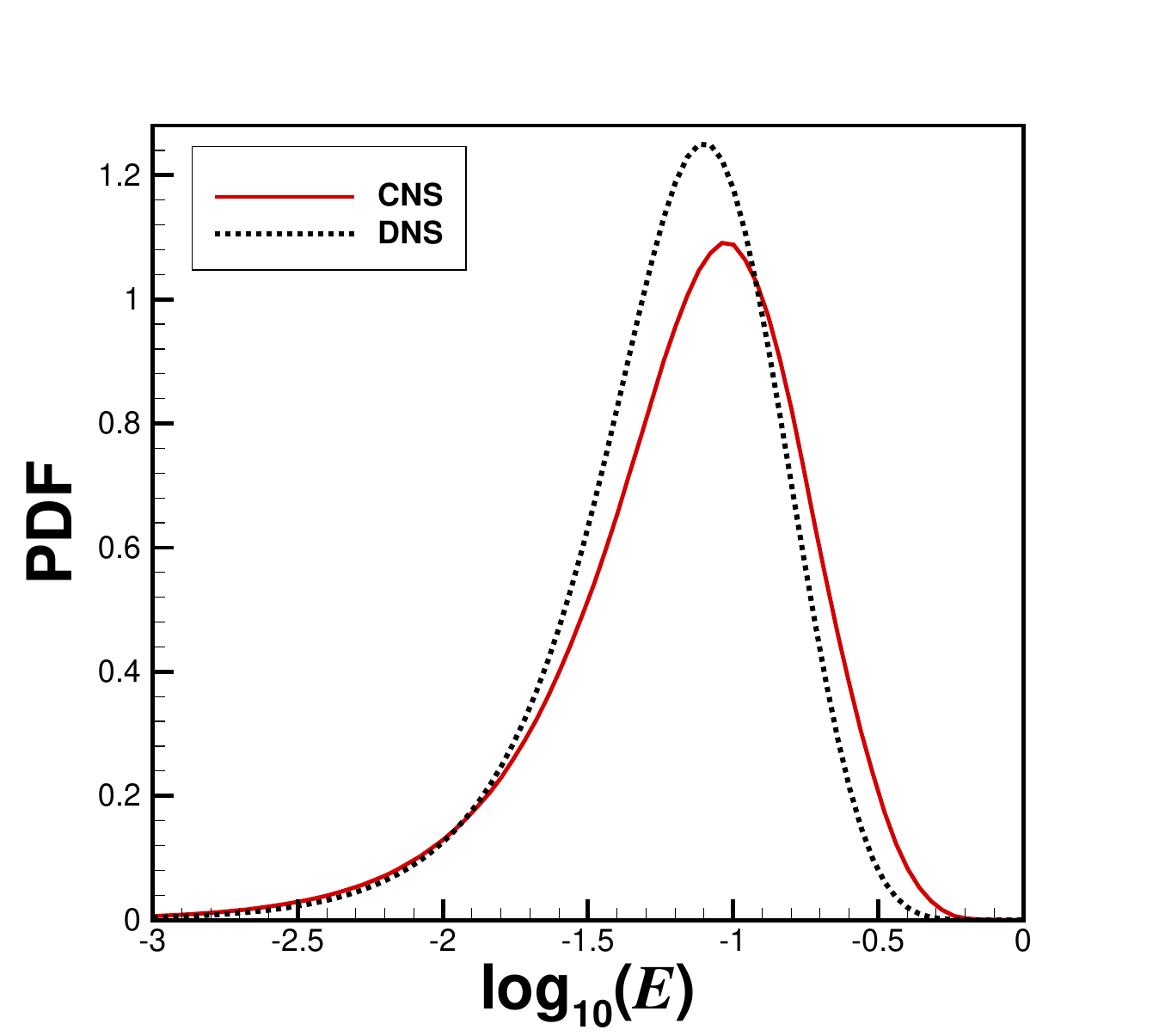}}
             \subfigure[]{\includegraphics[width=2.in]{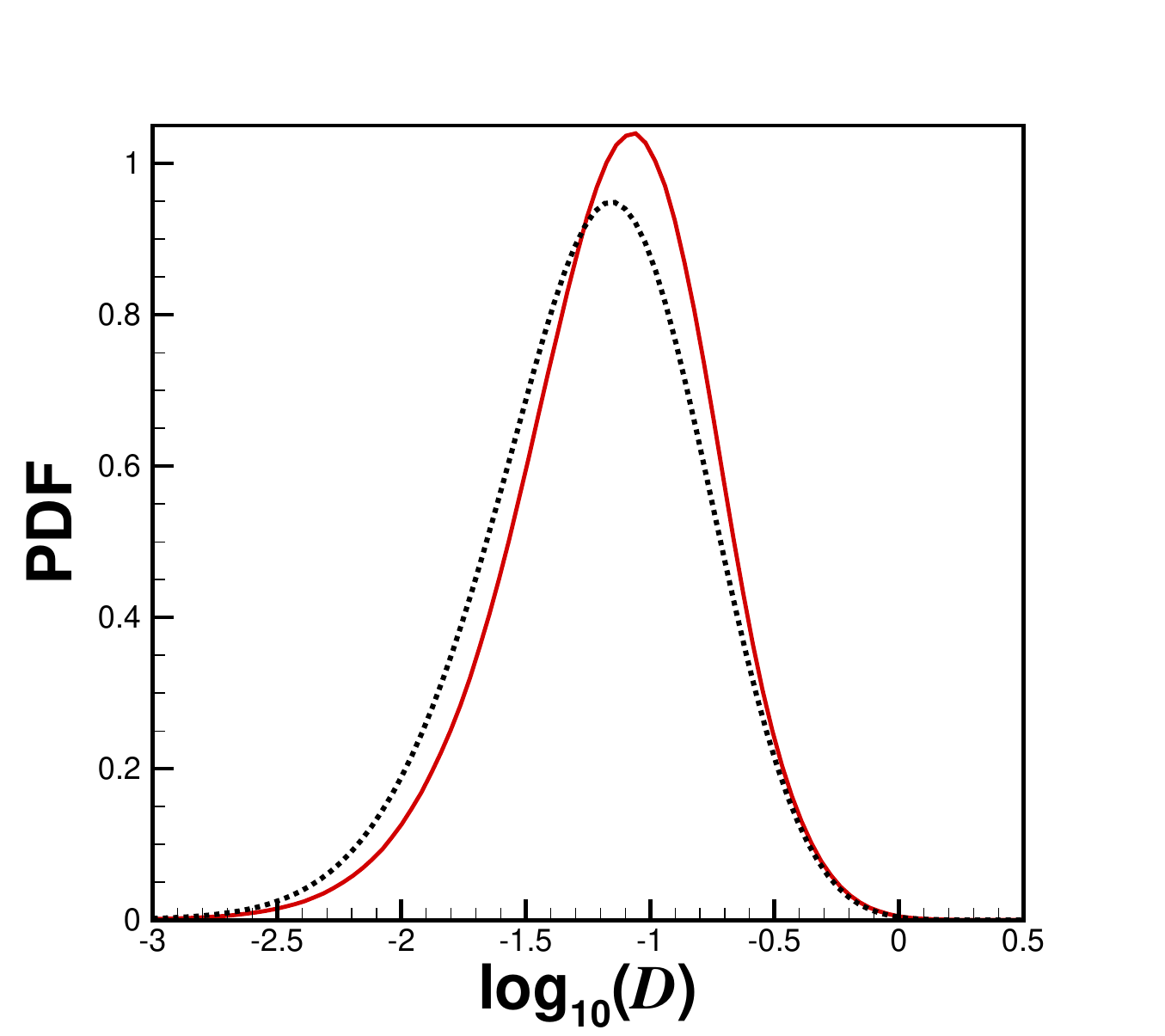}}
        \end{tabular}
    \caption{Comparisons of probability density functions (PDFs) of (a) the kinetic energy $E(\textbf{x},t)$ and (b) the kinetic energy dissipation rate $D(\textbf{x},t)$ of the 3D turbulent Kolmogorov flow governed by Eq.~(1) subject the periodic boundary condition and the initial condition (2) with the spatial symmetry (3) in the case of $n_K=4$ and $Re=1211.5$, given  by the CNS benchmark solution (solid line in red) and the DNS result (dashed line in black), respectively, where the PDFs are integrated in $(x,y,z)\in[0,1]^3$ and $t \in [100, 500]$.}     \label{ED-PDF}
    \end{center}
\end{figure*}

\begin{figure*}[!htb]
    \begin{center}
        \begin{tabular}{cc}
             \subfigure[]{\includegraphics[width=2.in]{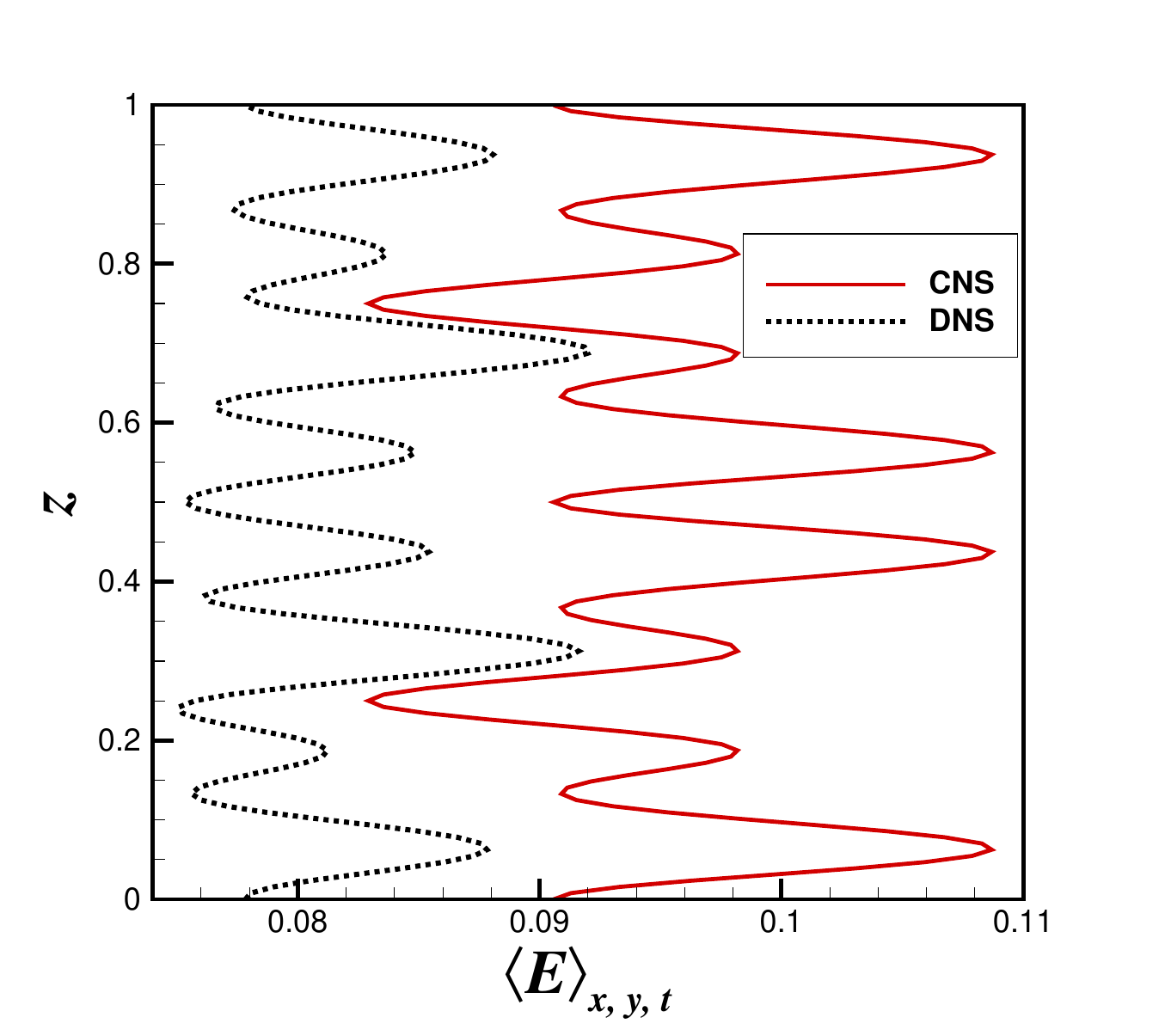}}
             \subfigure[]{\includegraphics[width=2.in]{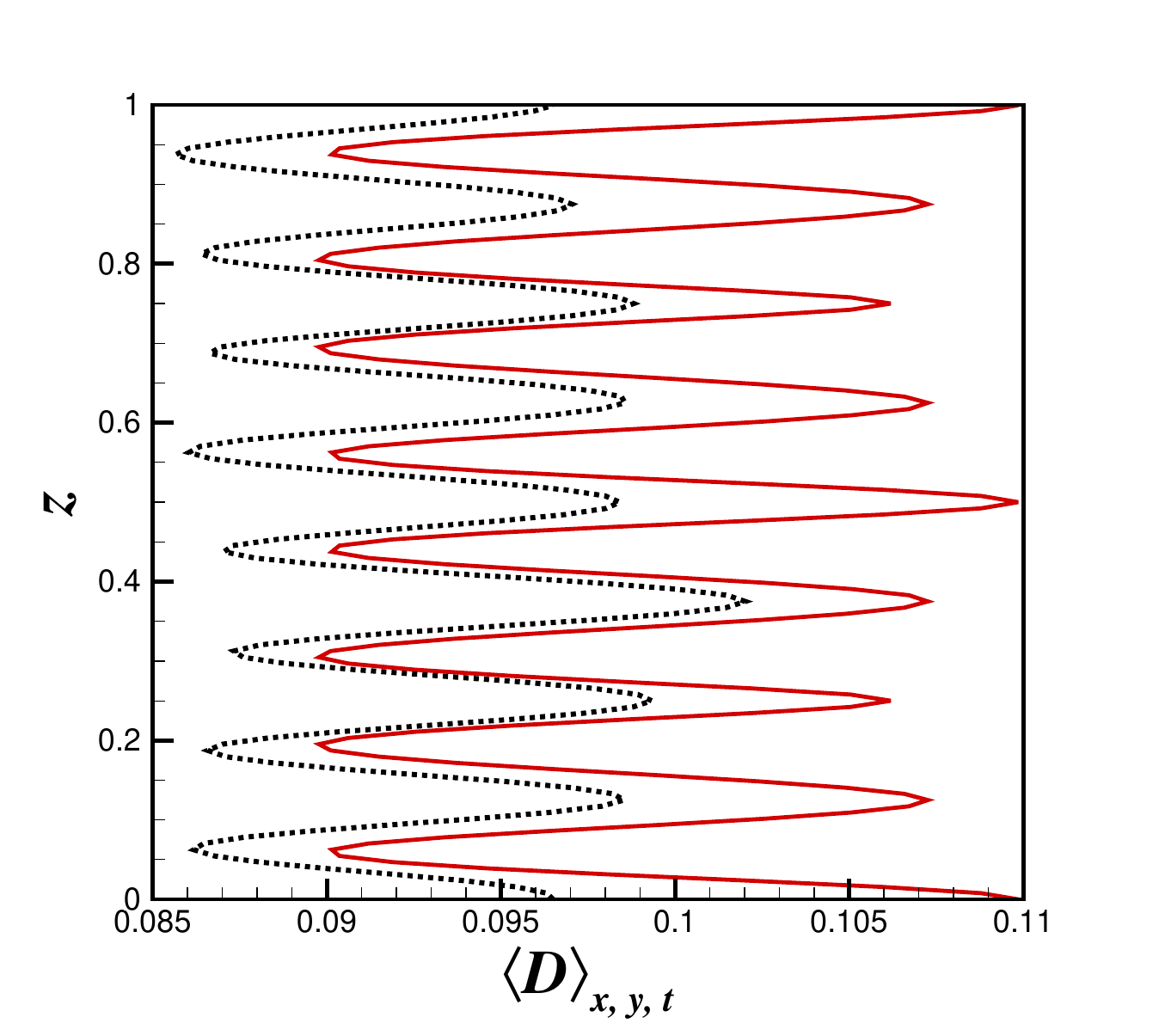}}
        \end{tabular}
    \caption{Comparisons of vertical distributions of the spatio-temporal averaged (a) kinetic energy $\langle E\rangle_{x,y,t}$ and (b) kinetic energy dissipation rate $\langle D\rangle_{x,y,t}$ of the 3D turbulent Kolmogorov flow governed by (\ref{NS}) subject to the periodic boundary condition and the initial condition (\ref{initial_condition}) with the spatial symmetry (\ref{spatial-symmetry}) in the case of $n_K=4$ and $Re=1211.5$, given by the CNS benchmark solution (solid line in red) and the DNS result (dashed line in black), respectively, where the spatio-temporal averages are integrated in $(x,y)\in[0,1]^2$ and $t \in [100, 500]$.}     \label{ED-mean_z}
    \end{center}

    \begin{center}
        \begin{tabular}{cc}
            \includegraphics[width=2.3in]{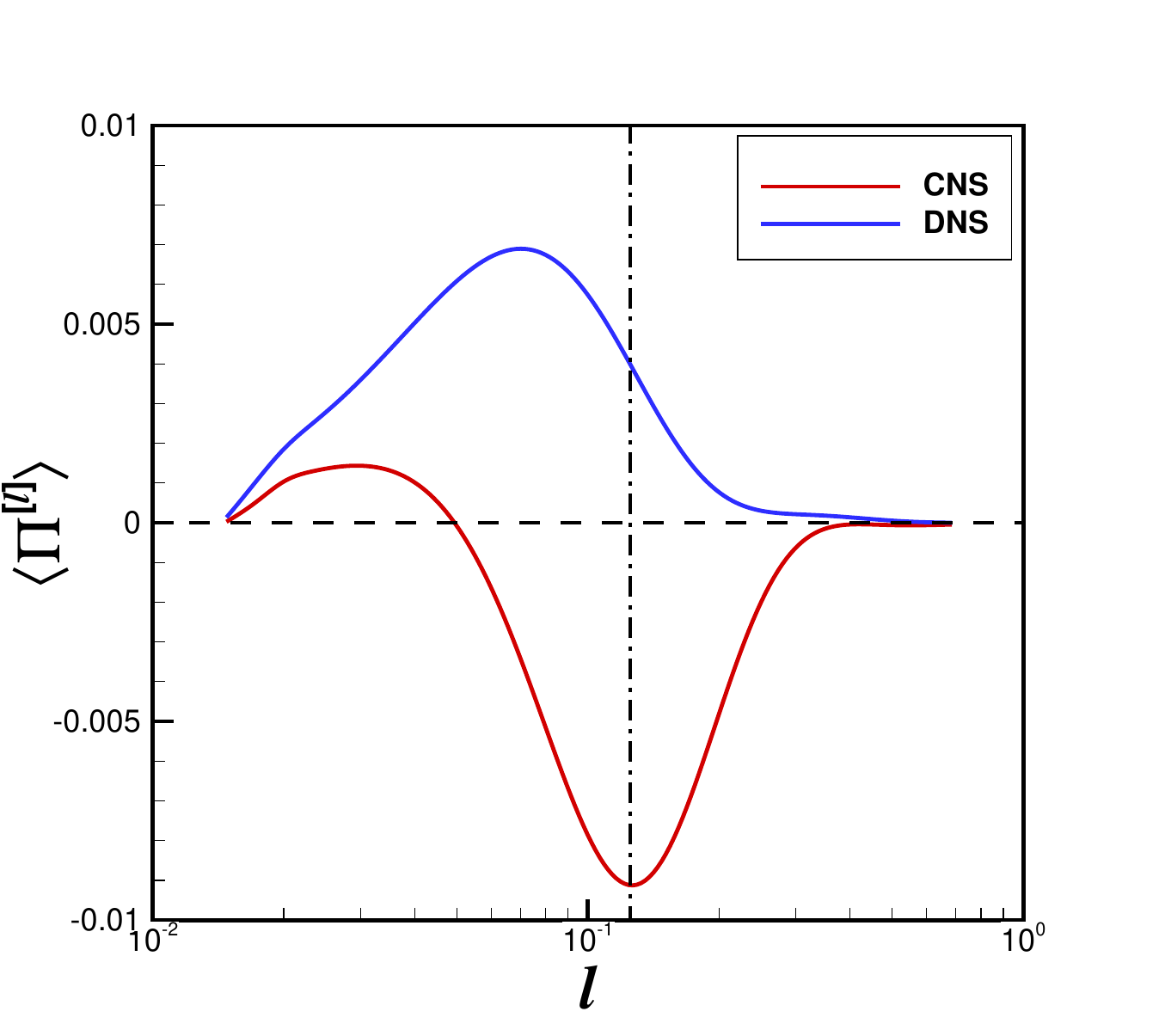}
        \end{tabular}
    \caption{Comparisons of  the spatio-temporal averaged scale-to-scale energy flux $\langle \Pi^{[l]} \rangle$ of the 3D turbulent Kolmogorov flow governed by (\ref{NS}) subject to the periodic boundary condition and the initial condition (\ref{initial_condition}) with the spatial symmetry (\ref{spatial-symmetry}) in the case of $n_K=4$, $Re=1211.5$, given by the CNS benchmark solution (solid line in red) and the DNS result (solid line in blue), respectively, where the spatio-temporal averages are integrated in $(x,y,z) \in [0,1]^3$ and $t \in [100, 500]$. The  black dash-dot line corresponds to the forcing scale $l_f=0.5/n_K=0.125$.}     \label{EF}
    \end{center}
\end{figure*}

\subsection{Statistics}

As mentioned above, the corresponding ``statistic''  results given by DNS in $t\in[0,90]$ agree quite well with those given by CNS in the same interval of time, as shown in Figs.~\ref{t90} of Appendix~A, indicating that DNS and CNS  initially give the same result.  Here, let us compare the statistic results given by CNS and DNS when $t\in[100,500]$.  

As shown in Fig.~\ref{DE_t}, the time histories (when $t>100$) of the spatially averaged kinetic energy $\langle E \rangle_V$ and kinetic energy dissipation rate $\langle D \rangle_V$ given by the DNS result are obviously different from those given by the CNS benchmark solution, respectively, where $E$ and $D$ are defined by (\ref{kinetic_energy}) and (\ref{dissipation_rate}) in Appendix~C, $\langle  \rangle_V$ is an operator defined by (\ref{average_V}).  
Besides, the comparisons of the probability density functions (PDFs) of the kinetic energy $ E $ and the kinetic energy dissipation rate $D$ are as shown in Fig.~\ref{ED-PDF}(a) and (b), respectively, where the PDFs are integrated in $(x,y,z)\in[0,1]^3$ and $t \in [100, 500]$. Obvious deviations  are also observed in the PDFs of the kinetic energy  $E$ and the kinetic energy dissipation rate $D$, indicating the macro-scale deviations of the fluid fields given by the DNS result and the CNS benchmark solution, respectively.    
For the spatio-temporally averaged kinetic energy $\langle E\rangle_{x,y,t}$ and kinetic energy dissipation rate $\langle D\rangle_{x,y,t}$, where $\langle \rangle_{x,y,t}$ is an operator defined by (\ref{average_xyt} ) in Appendix~C, Fig.~\ref{ED-mean_z} illustrates the comparison of their vertical distributions given by the DNS result and the CNS benchmark solution, respectively. Note that, for these two spatio-temporal averages, the results given by the CNS benchmark solution are apparently larger than those based on the DNS result. In addition, the variation ranges of $\langle E\rangle_{x,y,t}$ and $\langle D\rangle_{x,y,t}$ given by the CNS benchmark solution are also apparently larger than those by the DNS result, respectively. Furthermore, the distributions of these two spatio-temporal averages given by the CNS benchmark solution have the spatial symmetry with respect to $z=0.5$, but those by the DNS result have {\em no} such kind of spatial symmetry at all, because when $t\geq 100$ the spatiotemporal trajectory given by DNS  loses the spatial symmetry~(\ref{spatial-symmetry}).    

All of these illustrate that not only the spatio-temporal trajectories of the 3D turbulent Kolmogorov flow given by the DNS are badly polluted by artificial numerical noises  but also its statistics and energy cascade have obvious deviations from the CNS benchmark solution. Note that the above conclusions agree well qualitatively with our conclusions about some 2D turbulent flows \cite{qin_liao_2022, Liao-2025-JFM-NEC, Liao-2025-JFM-PS}.

\subsection{Energy cascade}

As mentioned in Section~3.2, the spatio-temporally averaged  ``scale-to-scale energy flux'' given by DNS in $t\in[0,90]$ agree quite well with those given by CNS in the same interval of time, as shown in Fig.~\ref{EF-t90} of Appendix~A, indicating that DNS and CNS  initially give the same result.  Here, let us compare the  spatio-temporally averaged  ``scale-to-scale energy flux''  given by CNS and DNS  when $t \in [100,500]$.  

The energy cascade is one of most important characteristic of turbulent flow \cite{pope2001turbulent}. The spatio-temporally averaged scale-to-scale energy flux $\langle \Pi^{[l]} \rangle$ of the 3D turbulent Kolmogorov is as shown in Fig.~\ref{EF}, where $\langle \rangle$ is an operator defined by (\ref{average_xyzt}) and $ \Pi^{[l]} $ is defined by (\ref{energy-flux}) in Appendix~C. Note that the CNS benchmark solution has the {\em inverse} energy cascade  near the forcing scale  $l = 0.05$, where kinetic energy transfers from smaller scale to larger ones so as to remain the large-scale spatial symmetry (\ref{spatial-symmetry}) of the 3D turbulent Kolmogorov flow,  but has the direct energy cascade in the small scales, respectively. However, on the contrary, the DNS result  {\em always} maintains the {\em direct} energy cascade, say, $\langle \Pi^{[l]} \rangle>0$ for {\em all} scales and the energy {\em always} transfers from larger to smaller ones. This is mainly because, unlike the CNS benchmark solution, the spatiotemporal trajectory given by DNS within $t>100$ loses the large-scale spatial symmetry (\ref{spatial-symmetry}) of the flow when the {\em random} numerical noise, which has {\em no} spatial symmetry at all, is exponentially enlarged to the {\em same} order of magnitude as the exact solution $s_{exact}$ at $t\approx 100$. Therefore, the energy cascade of the DNS result in $t\in[100,500]$ is {\em qualitatively} different from that of the CNS benchmark solution: this again indicates the significant deviation of the DNS result from the CNS benchmark solution of the NS turbulence considered in this paper.  
This is easy to understand from the mathematical viewpoint: the CNS result that is a very accurate solution of the Navier-Stokes turbulence is one thing,  but the DNS result (when $t >100$) that is {\em far away} from the true solution of the {\em same} Navier-Stokes turbulence is a completely different thing, so certainly their energy cascades are different.  
In other words, the CNS result is the solution of a dynamic system  {\em without  disturbances},  but the DNS result (when $t>100$) is the solution of a dynamic system {\em with stochastic  disturbances}: these two dynamic systems are essentially quite different.   It should be emphasized  that the energy cascade given by DNS result in $t\in[0,90]$ (see Fig.~\ref{EF-t90} of Appendix~A) qualitatively supports our CNS result shown in Fig.~\ref{EF}.    
Note that energy cascade is one of the most important property of turbulence.   Unfortunately, the numerical experiment based on DNS mentioned in this paper may lead to unreliable conclusions.

Note that, although the DNS result has {\em direct} energy cascade for all scales, say, energy always transfers from larger to smaller scale, its  numerical noise has been enlarged exponentially from a micro-level to a macro-level until $t \approx 100$,  as shown in Fig.~\ref{delta}. In other words, even the {\em direct} energy cascade {\em cannot} stop the expansion of the artificial numerical noise of DNS. This highly suggests that the noise-expansion should have {\em no} relationship with energy cascade: it is an essential property of turbulence, say, turbulence is chaotic in essence. This fundamental property of turbulence is called ``noise-expansion cascade'', which was currently discovered by Liao and Qin \cite{Liao-2025-JFM-NEC}: all disturbances at different orders of magnitude to the initial condition of the NS equations increase separately, say, one by one like an inverse cascade, to macro level, and each of them is capable of completely altering the characteristics (such as the vorticity spatial symmetry) of turbulent flow. Note that energy cascade is essentially different from ``noise-expansion cascade'': the former describes the energy distribution in {\em spatial} dimension, but the latter is about the disturbance expansion in {\em temporal} dimension.

\subsection{Necessary condition for validity of DNS}

As shown above, the spatiotemporal trajectories of turbulence given by DNS are quickly polluted badly by  numerical noises that are artificial. However, many numerical experiments based on DNS agree with physical experiments in statistics.  Why?  What is the necessary condition for the validity of DNS?       

Traditionally, DNS results of the Navier-Stokes equations were widely regarded as ``reliable'' benchmark solutions of turbulence as long as grid spacing is fine enough (i.e. less than the Kolmogorov scale \cite{pope2001turbulent}) and time-step is small enough, say, satisfying the Courant-Friedrichs-Lewy condition \cite{courant1928partiellen}, i.e. Courant number $<1$.
According to our CNS benchmark solution, we have the spatio-temporally averaged kinetic energy dissipation rate $\langle D \rangle=0.098$, corresponding to the non-dimensional Kolmogorov scale  \cite{pope2001turbulent}, say, 
\begin{equation}
\langle \eta \rangle \approx Re^{-3/4}\langle D \rangle^{-1/4}=0.0087.    \label{scale}  
\end{equation}
Since the grid spacing $h =1/N\approx0.0078$, the criterion on the grid spacing \cite{pope2001turbulent}, i.e.  
\begin{equation}
h < \langle \eta \rangle    \label{criterion}
\end{equation}
is satisfied for the CNS. The DNS result gives the spatio-temporally averaged kinetic energy dissipation rate $\langle D \rangle=0.085$, corresponding to the Kolmogorov scale $\langle \eta \rangle=0.009$, so that the grid spacing criterion (\ref{criterion}) is also satisfied. Therefore, the spatial resolution adopted in this paper is fine enough for {\em both} of the CNS and DNS of the 3D turbulent Kolmogorov flow considered in this paper, from the traditional view-point of DNS.
Besides, the corresponding Courant number is 0.09 for the CNS and 0.008 for the DNS respectively: {\em both} of them satisfy the Courant-Friedrichs-Lewy condition, i.e, Courant number $<1$. Therefore, the time-steps used here for {\em both} of the DNS and CNS are small enough from the traditional view-point of DNS. Unfortunately, even so, the DNS result has {\em distinct deviations} not only in spatio-temporal trajectory but also even in energy cascade and statistics from the CNS benchmark solution, clearly indicating that the small enough spacing and time-step {\em cannot} guarantee the validity of DNS for 3D turbulent flows: in other words, they are {\em not} the sufficient and necessary conditions for the validity of DNS.   
 
Note that spatiotemporal trajectory given by DNS is a mixture ($s_{exact}+\delta$)  of the exact solution $s_{exact}$ and the  deviation $\delta$ caused by the numerical noise, where $\delta$ is {\em random} and thus has {\em no} spatial symmetry at all so that it destroys the spatial symmetry (\ref{spatial-symmetry}) as it increases to the {\em same} order of magnitude as the exact solution $s_{exact}$.  However, the CNS benchmark solution always remains the spatial symmetry (\ref{spatial-symmetry}) in $t\in[0,500]$, indicating that the CNS result is a good approximation to the exact solution $s_{exact}$ at least in $t\in[0,500]$ since its numerical deviation $\delta$ is much small than the exact solution $s_{exact}$.    
For the 3D turbulent Kolmogorov flow considered in this paper, the DNS result (in $t>100$) has obvious deviations from the CNS solution not only in the trajectory, the spatial symmetry of flow field and the energy cascade but also in statistics, say, 
$ \langle s_{exact} + \delta \rangle \neq \langle s_{exact} \rangle$, where $\langle \rangle$ denotes an operator of statistics in general, suggesting that DNS result might be wrong sometimes even in statistics. This reveals a new {\em necessary} condition, i.e.        
\begin{equation}
 \langle s_{exact} + \delta \rangle = \langle s_{exact}  \rangle  \label{def:normal-chaos}
\end{equation}
for the validity of DNS, say, turbulent flow must has {\em statistic stability} under large disturbances. The left-hand side term $\langle s_{exact} + \delta \rangle$ of Eq.~(\ref{def:normal-chaos}) can be gained by DNS, whose spatiotemporal trajectory is badly polluted by artificial numerical noise quickly. The right-hand side term $\langle s_{exact}  \rangle $ can by obtained by clean numerical experiment based on CNS, respectively. Note that a chaotic system satisfying the criterion (\ref{def:normal-chaos}) is called ``normal-chaos'' whose statistics are stable to small disturbances, but otherwise ``ultra-chaos'' \cite{AAMM-14-799},  i.e. not only its spatiotemporal trajectories but also its statistics are unstable to small disturbances.
The necessary condition (\ref{def:normal-chaos}) can well explain why many numerical experiments based on DNS, although their spatio-temporal trajectories are badly polluted by numerical noise, could agree with their corresponding physical experiment in {\em statistics}: the corresponding turbulent flows must belong to a normal-chaos so that DNS is valid in the meaning of statistics. 
The same phenomena has been discovered by CNS \cite{Liao-2025-JFM-PS}:  for some turbulent flows,  the statistic results  of  DNS on a rather sparse mesh are the same as those of DNS on a very fine mesh.   
But unfortunately, {\em not} all turbulent flows belong to normal-chaos, as illustrated by Liao and Qin \cite{Liao-2025-JFM-PS}.   So, it is very important whether a turbulent flow has statistic stability  (\ref{def:normal-chaos}) or not. 

\section{Concluding remarks and discussions}

In this paper, for the first time,  we solve a 3D turbulent Kolmogorov flow by means of clean numerical simulation (CNS), and compare our CNS result with that given by DNS in details. It is found that the spatiotemporal trajectories of the 3D Kolmogorov turbulent flow given by DNS are badly polluted by  numerical noise rather quickly, and  the DNS results have obvious  deviations from the CNS benchmark solution {\em not only} in the flow field, the spatial symmetry of vorticity,  and the energy cascade {\em but also} even in statistics.   

As shown in Figs.~\ref{Vor_Evolutions} to \ref{Vor_2D-3}, the spatio-temporal trajectory given by DNS agrees well with the CNS benchmark solution and has the spatial symmetry (\ref{spatial-symmetry}) within $t\leq 90$, but loses the  spatial symmetry lightly at $t\approx 95$ and obviously at $t \approx 100$.  However,  the spatio-temporal trajectory given by CNS {\em always} remains the spatial symmetry (\ref{spatial-symmetry}) throughout the {\em whole} interval of time $t\in[0,500]$.  The reason why the DNS result deviates greatly from the CNS  benchmark solution and loses the spatial symmetry (\ref{spatial-symmetry}) is its consistently increasing random numerical noise, which has no  spatial symmetry but is much larger than that of CNS and thus reaches at a macro-level much earlier (at $t \approx 100$) than the CNS benchmark solution. It should be emphasized that the loss of the spatial symmetry (\ref{spatial-symmetry}) is an obvious sign and a criterion for numerical noise to be enlarged to the same order of magnitude as the true solution.  This is the reason why the initial condition (\ref{initial_condition}) with the spatial symmetry (\ref{spatial-symmetry}) is used in this paper.  

Note that the numerical noise of the CNS benchmark solution does not reach at the macro-level throughout the {\em whole} interval of time $t\in[0,500]$, clearly indicating that its ``critical predictable time'' $T_{c}$ must be greater than 500.  As pointed out in \S~2,  DNS can be regarded as a special case of the CNS, although its ``critical predictable time'' $T_{c}\approx 90$ is much smaller than that of CNS, indicating that the DNS is ``clean'' (i.e. not badly polluted by numerical noise) only in an interval of time $t\in[0,90]$, which is unfortunately too short for calculating statistics.  It is a pity that, in the frame of DNS, people have no concept of  ``critical predictable time'' and believe that simulation can be done in an arbitrary interval of time, which could be as long as one would like.     

As shown in \S~3.3,  the energy cascade given by the DNS for the 3D turbulent Kolmogorov flow is {\em always direct}, say, energy always transfers from larger to smaller scale. However, according to our clean numerical experiment based on CNS, the flow has the {\em inverse} energy cascade within the spatial scale $l >0.05$,  where kinetic energy transfers from small scale to large ones so as to remain the large-scale spatial symmetry (\ref{spatial-symmetry}) of the 3D turbulent Kolmogorov flow,  but has the {\em direct} energy cascade within the small scale $l < 0.05$,  respectively.   This is true in physics, since the forcing scale $l_f=0.5/n_K=0.125$ is in the middle  but the symmetric flow in a larger scale needs energy to remain, so that the energy absorbed by the flow near the forcing scale $l_f$ should transfer from small to large scales. Thus, from the viewpoint of energy transfer of the 3D turbulent Kolmogorov flow, the numerical experiment based on DNS may lead to unreliable conclusions.     
Note that, although the DNS result has {\em direct} energy cascade for all scales, i.e. kinetic energy transfers from large to small scales,  its  numerical noise has been enlarged exponentially from a micro-level to a macro-level.  In other words, even the {\em direct} energy cascade (in spatial dimension) {\em cannot} stop the expansion of the numerical noise of DNS, because small disturbances of turbulence are always enlarged (in temporal dimension) due to the butterfly-effect of chaos. This confirms once again that the so-called ``noise-expansion cascade''  discovered by Liao and Qin \cite{Liao-2025-JFM-NEC} is indeed a fundamental property of turbulence.         

As shown in \S~3.4,  the small enough spacing and time-step {\em cannot} guarantee the validity of DNS for the 3D turbulent flows under consideration: in other words, they are {\em not} the sufficient/necessary conditions for the validity of DNS.  Instead,  we proposed the necessary condition (\ref{def:normal-chaos}) for the validity of DNS:  turbulent flow must have {\em statistic stability} under disturbance.  This well explains why many numerical experiments based on DNS agree well with their corresponding physical experiments, because all of these turbulent flows must have statistic stability. Unfortunately, {\em not} every turbulence has such kind of statistic stability, as illustrated in this paper and also by some reported  2D turbulent flows \cite{qin_liao_2022, Liao-2025-JFM-NEC, Liao-2025-JFM-PS}. 
Note that, given a turbulent flow, it is nowadays unknown in mathematics whether its statistics is stable or not.  This is exactly the reason why the so-called ``modified Fourth Millennium Problem'' was proposed in [22] since statistic stability of turbulence is very important.

\begin{figure*}[!h]
    \begin{center}
        \begin{tabular}{cc}
             \subfigure[]{\includegraphics[width=2.in]{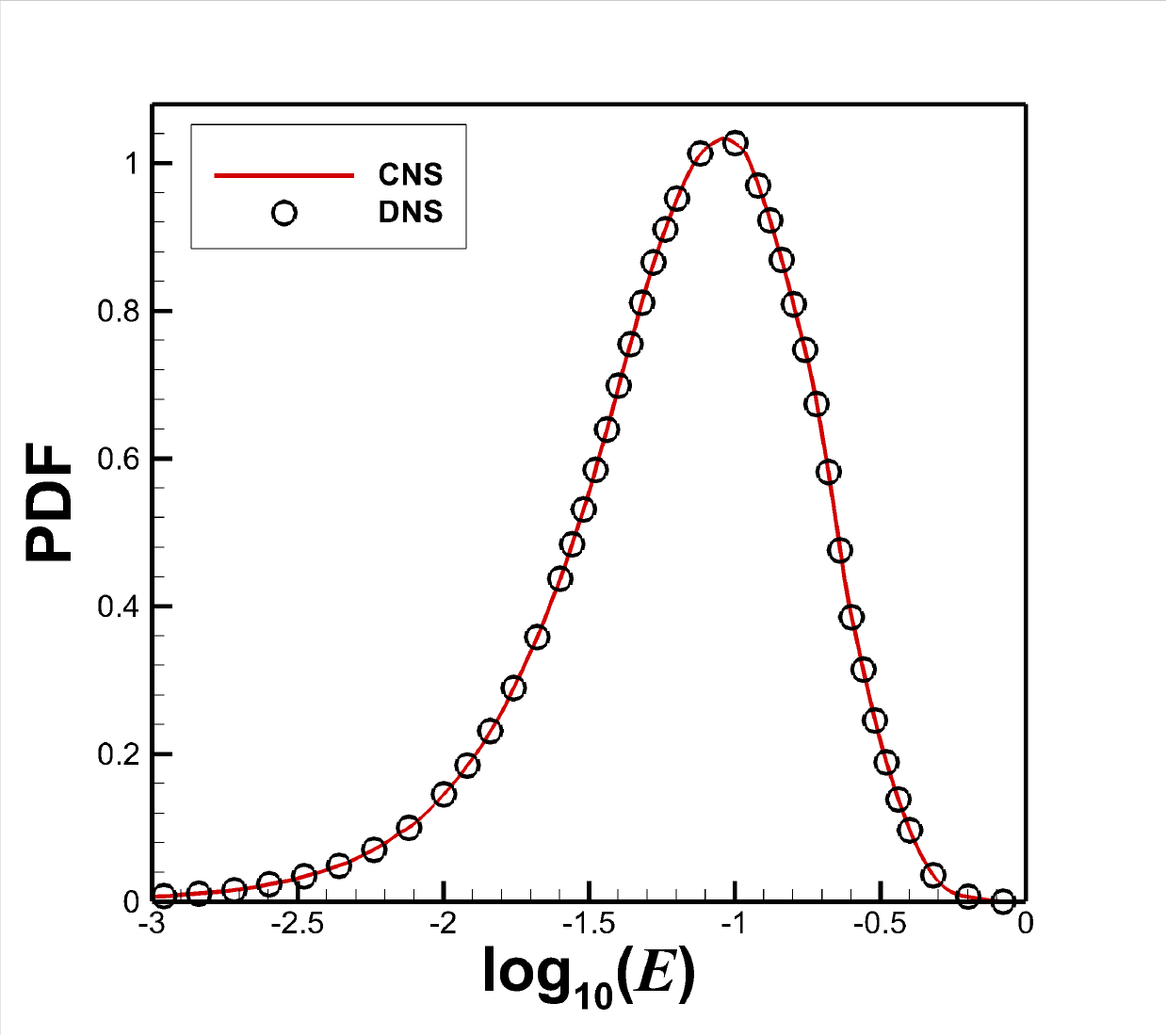}}
             \subfigure[]{\includegraphics[width=2.in]{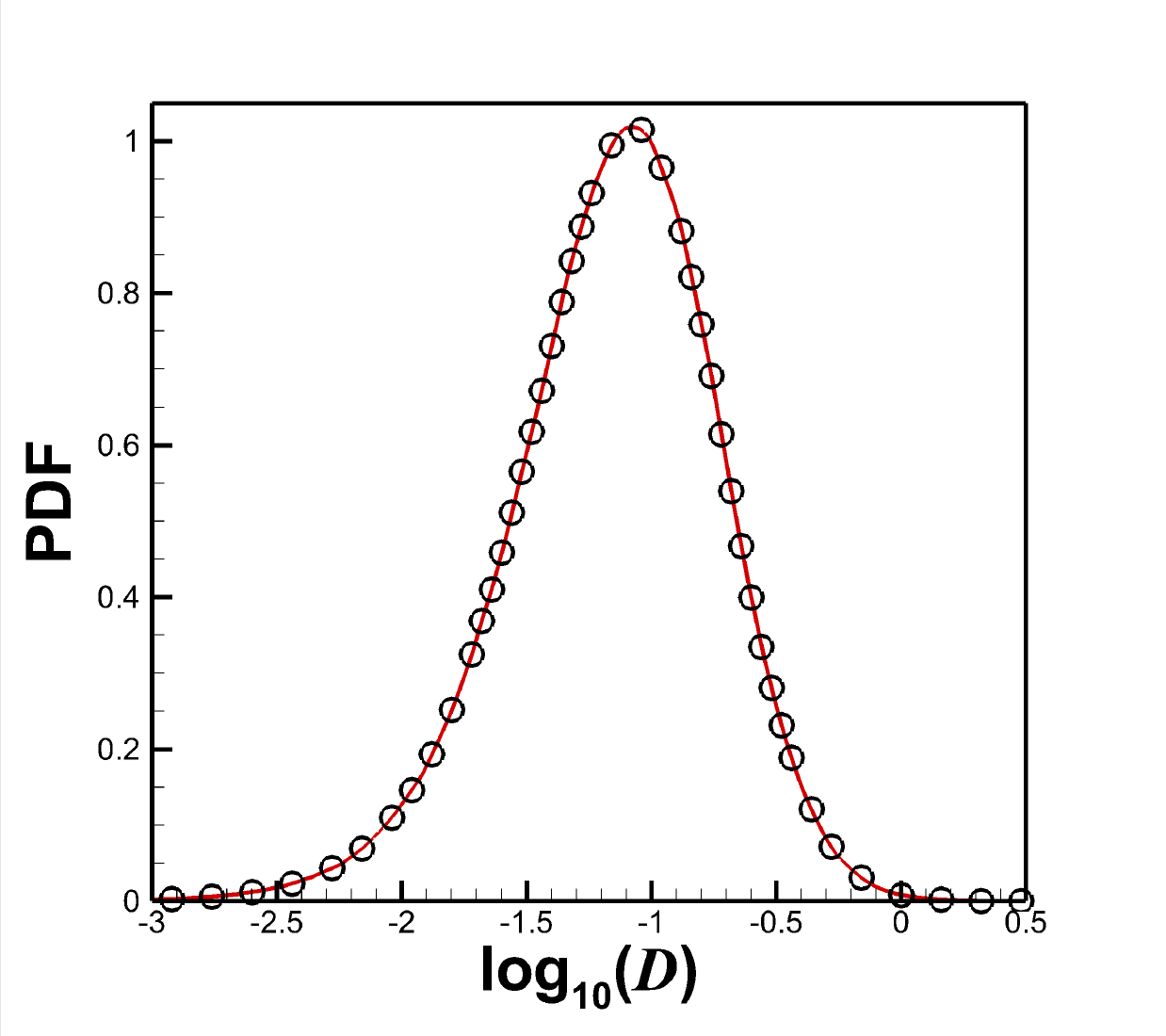}}\\
             \subfigure[]{\includegraphics[width=2.in]{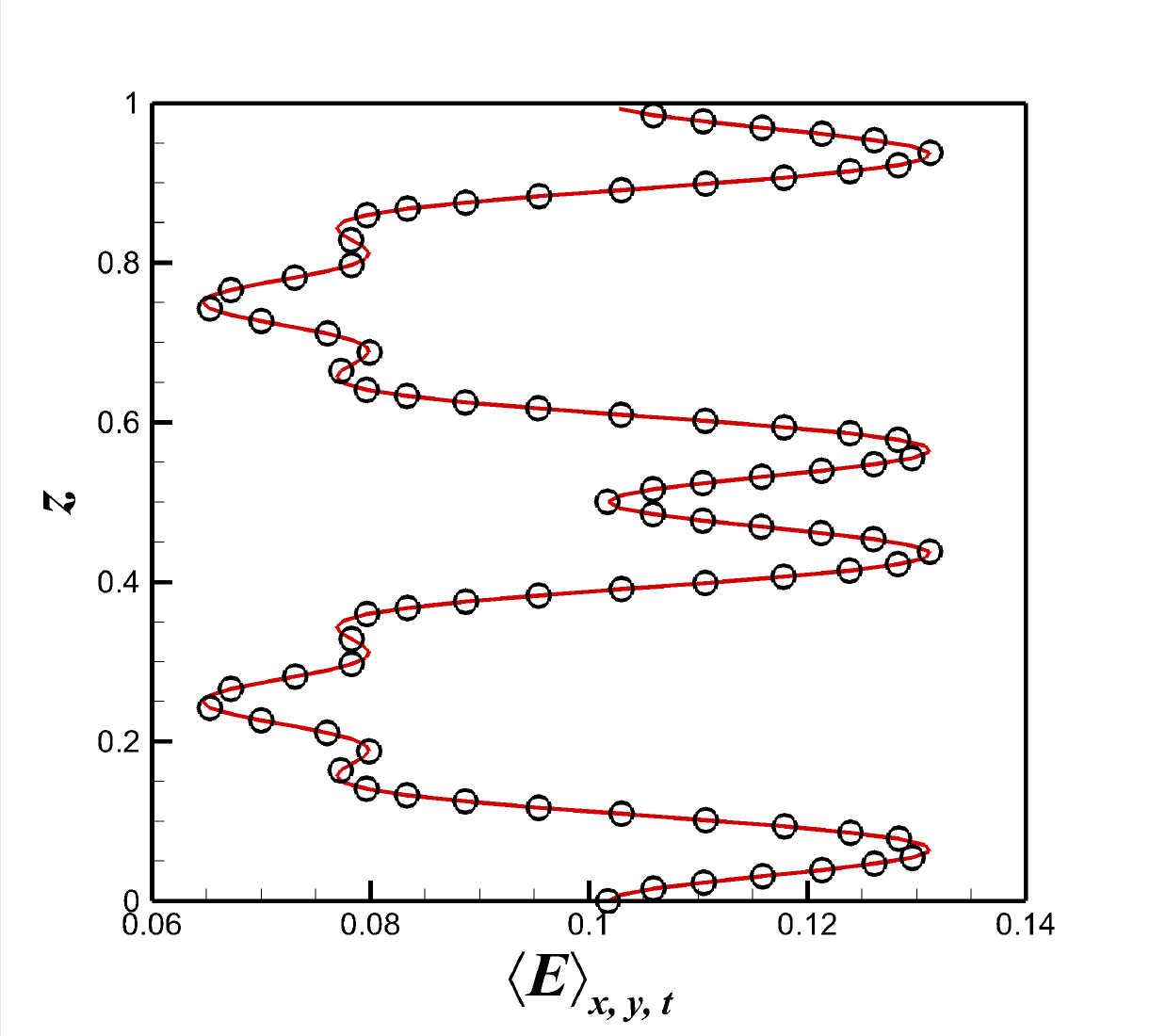}}
             \subfigure[]{\includegraphics[width=2.in]{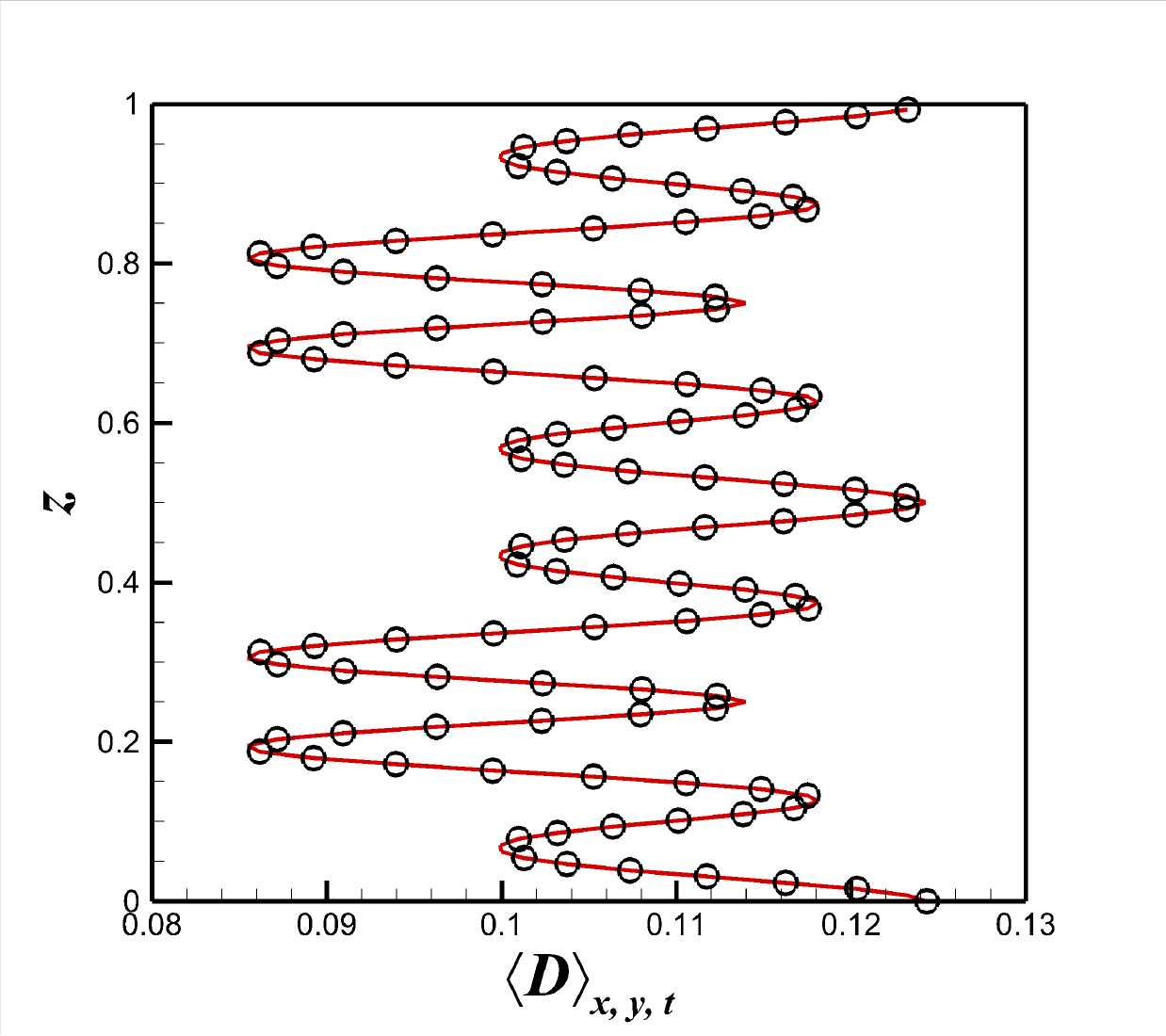}}
        \end{tabular}
 \caption{Comparisons of (a) PDFs of $E(\textbf{x},t)$, (b) PDFs of $D(\textbf{x},t)$, (c) vertical distributions of $\langle E\rangle_{x,y,t}$, and (d) vertical distributions of $\langle D\rangle_{x,y,t}$ of the 3D turbulent Kolmogorov flow governed by Eq.~(1) subject the periodic boundary condition and the initial condition (2) with the spatial symmetry (3) in the case of $n_K=4$ and $Re=1211.5$, given by the CNS benchmark solution (red line) and the DNS result (black circle) in $t \in [0, 90]$, respectively.}     \label{t90}
    \end{center}
\end{figure*}

\begin{figure*}[!htb]
    \begin{center}
        \begin{tabular}{cc}
            \includegraphics[width=2.3in]{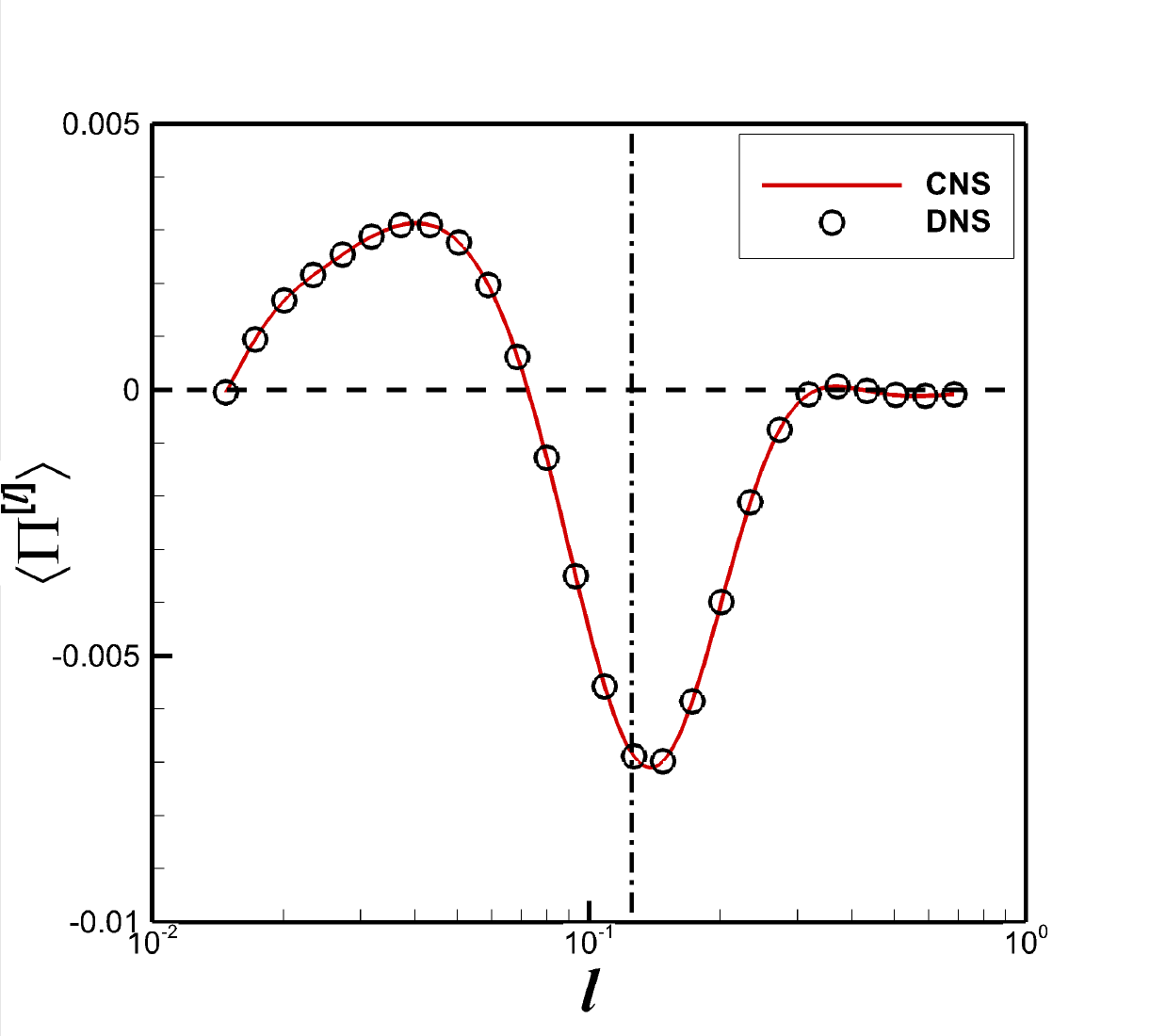}
        \end{tabular}
    \caption{Comparison of the spatio-temporal averaged scale-to-scale energy flux $\langle \Pi^{[l]} \rangle$ of the 3D turbulent Kolmogorov flow governed by (1) subject to the periodic boundary condition and the initial condition (2) with the spatial symmetry (3) in the case of $n_K=4$, $Re=1211.5$, given by the CNS benchmark solution (red line) and the DNS result (black circle), respectively, where the spatio-temporal averages are integrated in $(x,y,z) \in [0,1]^3$ and $t \in [0, 90]$. The black dash-dot line corresponds to the forcing scale $l_f=0.5/n_K=0.125$.}     \label{EF-t90}
    \end{center}
\end{figure*}

From mathematical viewpoint, it is clear that the DNS result is the solution of the NS turbulence only in a short interval of time $t\in [0,90]$ (as shown in Figs.~~\ref{t90} and \ref{EF-t90}), but is far away from the true solution of the NS turbulence when $t > 100$ (see Figs.~\ref{DE_t} tpo \ref{EF}), say, badly polluted by numerical noise, i.e. the numerical noise is at the same order of magnitude as its true solution.  In other words, mathematically speaking, the DNS result when $t > 100$ might have a rather weak relationship with the NS equations considered in this paper. 
Even so, it clearly indicates the significant influences of small disturbances to the Navier-Stokes turbulence in mathematics.  Note that the Navier-Stokes turbulence as a mathematical model completely neglects the influence of  ``physical'' disturbances when $t > 0$, which unfortunately is {\em unavoidable} from physical viewpoint.  So, the Navier-Stokes equations as a turbulence model might lead to a paradox in logic: it neglects the unavoidable small physical stochastic disturbances in the real physical world, which however in mathematics has significant influences even in statistics.  To avoid this paradox, a stochastic turbulence model (such as Landau-Lifshitz-Navier-Stokes equations \cite{LLNS1959} that consider the influence of thermal fluctuation) should be suggested.  Note that the Navier-Stokes turbulence is one thing in mathematics, but real turbulence is another thing in the physical world.  As a noise-negligible numerical realization, the CNS results report in this paper reveal such kind of paradox of the Navier-Stokes turbulence, indicating that CNS should be a powerful tool in theoretical investigation of turbulence.

Actually, although the computational efficiency of CNS has been increased several orders of magnitude since 2009 due to the algorithmic optimizations and a self-adaptive strategy  \cite{hu2020risks, qin2020influence, AAMM-15-1191}, it still requires substantial computer resources and is time-consuming at present, just like the initial DNS when Orszag \cite{Orszag1970} proposed it. For the 3D turbulent Kolmogorov flow under consideration, large-scale parallel computing of the CNS takes 533 hours (i.e. about 22 days) using 4096 CPUs (Intels CPU: Xeon Gold 6348, 2.60GHz) of the Tian-He New Generation Supercomputer at the National Supercomputer Center in Tianjin, China. By contrast, the DNS only needs 198 hours (i.e. about 8 days) using 512 CPUs at the same computing platform: this number of CPUs corresponds to the optimal computing efficiency, say, more CPU time is needed when using the same number (i.e. 4096) CPUs as the CNS.  A relatively small time step $\Delta t=10^{-4}$ is adopted for the DNS so that the truncation error and round-off error are at the approximately same order of magnitude. 
Even so, CNS can provide us a powerful tool to theoretically investigate some turbulent flows via clean numerical experiment (with negligible numerical noise), especially about influence of small (physical or/and artificial) disturbances to turbulence.  Without doubt,  given that even DNS is considered prohibitively expensive for engineering applications (where RANS and LES dominate),   CNS is nowadays just at its initial stage and thus quite far from practical applications. 

Finally, we emphasize once again that turbulent Kolmogorov flow is primarily studied as a {\em theoretical and idealized model} to understand the fundamental aspects of turbulence. While Kolmogorov flow itself is not directly used in practical applications, the insights gained from studying Kolmogorov flow have significant implications for various fields,  and the knowledge derived from its study enhances our ability to model, predict, and control turbulence in real world.

\newpage

  \begin{center}

{\bf\Large Appendix A \\
 Comparison of CNS and DNS in $t\in[0,90]$}

\end{center}
 
Our DNS result is the same as the CNS one only in a short interval of time $t\in[0,90]$, as shown in (a) and (b) (when $t=85$) of Figures~\ref{Vor_2D-1} to \ref{Vor_2D-3}.   The corresponding ``statistic''  results and  spatiotemporal averaged scale-to-scale energy flux   given by DNS in $t\in[0,90]$ agree also quite well with those given by CNS in the {\em same} interval of time, as shown in Figs.~\ref{t90} and \ref{EF-t90}.   From the viewpoint of CNS, the so-called ``critical predictable time'' of our DNS result is about 90, i.e. $T_{c}\approx 90$, which is unfortunately too short to obtain reliable statistic results.  Although these DNS results in Figs.~\ref{t90} and \ref{EF-t90} are unreliable,  they clearly indicate that the DNS and CNS indeed initially give the {\em same} solution.  It should be emphasized  that the DNS result in $t\in[0,90]$ qualitatively supports our CNS result in $t\in[0,500]$ about energy cascade.             

\newpage

\setcounter{equation}{0} 
\renewcommand{\theequation}{B\arabic{equation}}

\begin{center}
{\bf\Large Appendix B \\
 CNS algorithm for 3D Kolmogorov flow}
\end{center}

Let $(u,v,w)$ denote the velocity field and $p$ stand for the pressure. For the 3D turbulent Kolmogorov flow under consideration, continuity equation and momentum equations are as follows:
\begin{eqnarray}
u_{x}+v_{y}+w_{z}=0,~~~~~~~~~~~~~~~    \label{NS0}\\
u_{t}+uu_{x}+vu_{y}+wu_{z}=-p_{x}+\frac{\Delta u}{Re}+f,    \label{NS1} \\
v_{t}+uv_{x}+vv_{y}+wv_{z}=-p_{y}+\frac{\Delta v}{Re},~~~~~    \label{NS2} \\
w_{t}+uw_{x}+vw_{y}+ww_{z}=-p_{z}+\frac{\Delta w}{Re},~~~~    \label{NS3}
\end{eqnarray}
subject to the periodic boundary condition and the initial condition (2) with the spatial symmetry (3) mentioned in the main text, where $x$, $y$, $z$ as subscripts denote the spatial derivatives, $\Delta $ is the Laplace operator, and $f=\sin(2\pi n_K z)$ is the external force.  
Based on (\ref{NS0})-(\ref{NS3}), we have the Poisson equation of the pressure $p$ as follows: 
\begin{eqnarray}
\Delta p &=& -\,\frac{\partial}{\partial t} \Big(  u_{x}+ v_{y} + w_{z}\Big)  +\frac{1}{Re} \Delta\Big(  u_{x}+ v_{y} + w_{z}\Big) \nonumber\\
&&-\,\Big( 2 u_{y} v_{x}+2 v_{z} w_{y}+ 2 u_{z} w_{x} + u_{x} u_{x}+v_{y} v_{y} + w_{z} w_{z}\Big).    \label{poisson0} 
\end{eqnarray}
According to the continuity equation (\ref{NS0}), the above equation becomes 
\begin{eqnarray}
\Delta p &=&-\,\Big( 2 u_{y} v_{x}+2 v_{z} w_{y}+ 2 u_{z} w_{x} + u_{x} u_{x}+v_{y} v_{y} + w_{z} w_{z}\Big).    \label{poisson}
\end{eqnarray}
Note that the pressure $p$ satisfies the periodic condition on the boundary, too. So, as long as the velocity field is known, one can gain the corresponding pressure field by solving the above Laplace equation with the periodic boundary condition (see \cite{Rogallo1981NASA}).     

Like DNS, the spatial domain $(x, y, z)\in[0,1]^3$ is discretized by means of $N^3$ equidistant points, i.e.
\begin{equation}
x_{i}=\frac{i}{N},  \hspace{1.0cm} y_{j}=\frac{j}{N}, \hspace{1.0cm} z_{k}=\frac{k}{N},    \label{points} \nonumber
\end{equation}
where $i,j,k=0, \,1, \,2, \,..., \,N-1$.
To reduce truncation errors in the dimension of time, we, {\em unlike} DNS, use the following high-order Taylor expansions
\begin{equation}
u(x_{i},y_{j},z_{k},t+\Delta t)\approx\sum^{M}_{m=0}u^{[m]}(x_{i},y_{j},z_{k},t)(\Delta t)^{m},    \label{Taylor_u}
\end{equation}
\begin{equation}
v(x_{i},y_{j},z_{k},t+\Delta t)\approx\sum^{M}_{m=0}v^{[m]}(x_{i},y_{j},z_{k},t)(\Delta t)^{m},    \label{Taylor_v}
\end{equation}
\begin{equation}
w(x_{i},y_{j},z_{k},t+\Delta t)\approx\sum^{M}_{m=0}w^{[m]}(x_{i},y_{j},z_{k},t)(\Delta t)^{m},    \label{Taylor_w}
\end{equation}
where $\Delta t$ is the time-step and $M$ is the order of Taylor expansion, with the following definitions
\begin{equation}
u^{[m]}=\frac{1}{m!}\frac{\partial^{m}u}{\partial t^{m}},\hspace{1.0cm}
v^{[m]}=\frac{1}{m!}\frac{\partial^{m}v}{\partial t^{m}},\hspace{1.0cm}
w^{[m]}=\frac{1}{m!}\frac{\partial^{m}w}{\partial t^{m}}.  \nonumber
\end{equation}
Here, the order $M$ should be large enough so as to reduce the temporal truncation errors to a required tiny level.

Differentiating  both sides of Eqs.~(\ref{NS1})-(\ref{NS3}) $(m-1)$ times with respect to $t$ and then dividing them by $m!$, we obtain the governing equations of $u^{[m]}$, $v^{[m]}$ and $w^{[m]}$ at $(x_{i},y_{j},z_{k},t)$ as follows
\begin{eqnarray}
u^{[m]}&=&\frac{1}{m}\Big[-\sum^{m-1}_{s=0}\Big(u^{[s]}u_{x}^{[m-1-s]}+v^{[s]}u_{y}^{[m-1-s]}+w^{[s]}u_{z}^{[m-1-s]}\Big)- p_{x}^{[m-1]}\nonumber\\
&&+\,\frac{\Delta u^{[m-1]}}{Re}+F_m\Big],    \label{NS1_m} \\
v^{[m]}&=&\frac{1}{m}\Big[-\sum^{m-1}_{s=0}\Big(u^{[s]}v_{x}^{[m-1-s]}+v^{[s]}v_{y}^{[m-1-s]}+w^{[s]}v_{z}^{[m-1-s]}\Big)- p_{y}^{[m-1]}\nonumber\\
&&+\,\frac{\Delta v^{[m-1]}}{Re}\Big],    \label{NS2_m} \\
w^{[m]}&=&\frac{1}{m}\Big[-\sum^{m-1}_{s=0}\Big(u^{[s]}w_{x}^{[m-1-s]}+v^{[s]}w_{y}^{[m-1-s]}+w^{[s]}w_{z}^{[m-1-s]}\Big)- p_{z}^{[m-1]}\nonumber\\
&&+\,\frac{\Delta w^{[m-1]}}{Re}\Big],    \label{NS3_m}
\end{eqnarray}
where $m\geq1$, and
\begin{equation}
F_m=\left\{
\begin{array}{l}
\sin(n_K2\pi z_k),    \hspace{1.0cm}    m=1,\\
0,    \hspace{3cm}    m>1.
\end{array}
\right.  \label{Fm}
\end{equation}

Besides, differentiating both sides of Eq.~(\ref{poisson}) $(m-1)$ times with respect to $t$ and then dividing them by $(m-1)!$, the governing equation of $p^{[m-1]}$ at $(x_{i},y_{j},z_{k},t)$, whose value is needed in Eqs.~(\ref{NS1_m})-(\ref{NS3_m}), is obtained as follows:
\begin{eqnarray}
\Delta p^{[m-1]} &=&-\,2\sum^{m-1}_{s=0}\Big( u_{y}^{[s]}v_{x}^{[m-1-s]}+v_{z}^{[s]}w_{y}^{[m-1-s]}
+u_{z}^{[s]}w_{x}^{[m-1-s]}\Big) \nonumber\\
&&-\,\sum^{m-1}_{s=0}\Big( u_{x}^{[s]}u_{x}^{[m-1-s]}+v_{y}^{[s]}v_{y}^{[m-1-s]}
+w_{z}^{[s]}w_{z}^{[m-1-s]}\Big).    \label{poisson_m}
\end{eqnarray}
The above Laplace equation of $p^{[m-1]}$ can be easily solved by means of Fourier transforms, as long as $u^{[m-1]}$, $v^{[m-1]}$ and $w^{[m-1]}$ are known. Thereafter, one can further gain $u^{[m]}$, $v^{[m]}$ and $w^{[m]}$ by means of Eqs.~(\ref{NS1_m})-(\ref{NS3_m}), respectively, and then $p^{[m]}$ by means of (\ref{poisson_m}), and so on.    

Note that there exist some spatial partial derivatives (denoted by subscripts) in Eqs.~(\ref{NS1_m})-(\ref{poisson_m}).
Without loss of generality, we take $u_{xx}^{[m-1]}(x_{i},y_{j},z_{k},t)$ as an example. In order to approximate this spatial partial derivative with  high precision from the known discrete variable $u^{[m-1]}(x_{i},y_{j},z_{k},t)$, we, like DNS,  adopt the spatial Fourier series
\begin{equation}
u^{[m-1]}(x,y,z,t)\approx\sum_{\mid \textbf{n} \mid \leq N/3}
U^{[m-1]}(n_x,n_y,n_z,t)\exp[2\pi\mathbf{i}\hspace{0.02cm}(n_xx+n_yy+n_zz)],   \label{Appro_u}
\end{equation}
with $\textbf{n}=(n_x, n_y, n_z)$,
where $n_x$, $n_y$, $n_z$ are integers, $\mathbf{i}=\sqrt{-1}$ denotes the imaginary unit, for dealiasing $U^{[m-1]}(n_x,n_y,n_z,t)=0$ is imposed for wave-numbers outside the above domain $\sum$, and 
\begin{eqnarray}
&& U^{[m-1]}(n_x,n_y,n_z,t) \nonumber\\
&=&\frac{1}{N^3}\sum^{N-1}_{i=0}\sum^{N-1}_{j=0}\sum^{N-1}_{k=0}u^{[m-1]}(x_{i},y_{j},z_{k},t)
\exp[-2\pi\mathbf{i}\hspace{0.02cm}(n_xx_i+n_yy_j+n_zz_k)]   \label{Coe_u}
\end{eqnarray}
is determined by the known $u^{[m-1]}(x_{i},y_{j},z_{k},t)$. Then, we have the rather accurate approximation of the spatial partial derivative
\begin{eqnarray}
&& u_{xx}^{[m-1]}(x_{i},y_{j},z_{k},t) \nonumber\\
&\approx &\sum_{\mid \textbf{n} \mid \leq N/3}
(2\pi\mathbf{i}\hspace{0.03cm}n_x)^{2}\hspace{0.03cm}
U^{[m-1]}(n_x,n_y,n_z,t)\exp[2\pi\mathbf{i}\hspace{0.02cm}(n_xx_i+n_yy_j+n_zz_k)].        \label{Appro_uxx}
\end{eqnarray}
Here, the fast Fourier transform (FFT) is used to increase computational efficiency. In this way, the spatial truncation error can be decreased to any required tiny level, as long as the mode number $N$ is large enough.  

It should be emphasized that, if the order $M$ of the Taylor expansions (\ref{Taylor_u})-(\ref{Taylor_w}) in temporal dimension is large enough, temporal truncation error can be decreased to {\em any} required tiny level. Besides, if the spatial discretization is fine enough, i.e. the mode number $N$ is large enough, spatial truncation error in Eqs.~(\ref{Appro_u})-(\ref{Appro_uxx}) can be reduced to {\em any}  required tiny level. However, these  are {\em not} enough, since there always exists round-off error that might be larger than temporal and/or spatial truncation errors. Thus, in the frame of CNS, we, {\em unlike} DNS, express all parameters and variables  in {\em multiple precision} (MP) \cite{oyanarte1990mp} using a large enough number $N_s$ of significant digits so that round-off error can be reduced to {\em any}   required tiny level. In this way, background numerical noise, i.e. the maximum of truncation error and round-off error, can be decreased to any  required tiny level. This is fundamentally different from DNS that uses the {\em double precision} in general.
Note that result given by CNS is useful only in a {\em limited} temporal interval $t\in[0,T_{c}]$, where $T_{c}$ is the so-called ``critical predictable time'', in which numerical noise is negligible compared to the exact solution $s_{exact}$ of the NS equations. For more details about the CNS algorithm, please refer to \cite{qin_liao_2022, Liao2023book}. 
  
We emphasize here that DNS \cite{Orszag1970, Rogallo1981NASA, She1990Nature, Nelkin1992Science, FeracoScience2024, MoinARFM1998, Scardovelli1999ARFM, Coleman2010DNS, Huang2022JFM} can be regarded as a special case of CNS \cite{Liao2009, Liao2023book, hu2020risks, qin2020influence, qin_liao_2022}, since both of CNS and DNS are based on pseudo-spectral method in spatial dimension. However, in temporal dimension, DNS generally uses either a 2nd-order or 4th-order Runge-Kutta method, but CNS applies a much higher Taylor expansion so as to decrease temporal truncation error to any required tiny level. More importantly, {\em unlike} DNS that normally uses double-precision, CNS uses multiple-precision with much more significant digits so as to decrease round-off error to any required tiny level. Therefore, results given by CNS are much more accurate than those by DNS.

\setcounter{equation}{0} 
\renewcommand{\theequation}{C\arabic{equation}}

\begin{center}
{\bf\Large Appendix C \\
Definitions of some measures}
\end{center}     

For the sake of simplicity, the definitions of some statistic operators are briefly described below. The spatial average is defined by
\begin{equation}
 \langle\,\,\rangle_V=\int^{1}_0\int^{1}_0\int^{1}_0 dxdydz,       \label{average_V} 
\end{equation}
the spatio-temporal average varying along $z$ direction is defined by
\begin{equation}
 \langle\,\,\rangle_{x,y,t}=\frac{1}{T_2-T_1}\int^{1}_0\int^{1}_0\int^{T_2}_{T_1} dxdydt,  \label{average_xyt} 
\end{equation}
and the spatio-temporal average over the whole field is defined by
\begin{equation}
 \langle\,\,\rangle=\frac{1}{T_2-T_1}\int^{1}_0\int^{1}_0\int^{1}_0\int^{T_2}_{T_1} dxdydzdt,       \label{average_xyzt} 
\end{equation}
respectively, where $T_1=100$ and $T_2=500$ are chosen in this paper for the 3D turbulent Kolmogorov flow under consideration.

The vorticity $\bm{\omega}(\textbf{x},t)$ can be expanded as the Fourier series
\begin{equation}
 \bm{\omega}(\textbf{x},t)=\nabla\times\textbf{u}(\textbf{x},t)=\sum_{\mid \textbf{n} \mid \leq N/3}\bm{\Omega}(\textbf{n},t) \exp(2\pi\mathbf{i}\,\textbf{n}\cdot\textbf{x})       \label{vorticity}  
\end{equation}
with $\bm{\omega}=(\omega_x, \omega_y, \omega_z)$, $\bm{\Omega}=(\Omega_x, \Omega_y, \Omega_z)$, and $\textbf{n}=(n_x, n_y, n_z)$,
where $n_x$, $n_y$, $n_z$ are integers, $\mathbf{i}=\sqrt{-1}$ denotes the imaginary unit, and for dealiasing $\bm{\Omega}(\textbf{n},t)=0$ is imposed for wavenumbers outside the above domain $\sum$.
The enstrophy spectrum is defined as
\begin{equation}
 B_k(t)=\sum_{k-1/2 \leq \mid \textbf{n} \mid < k+1/2} \Big| \bm{\Omega}(\textbf{n},t) \Big|^2,       \label{enstrophy_spectrum} 
\end{equation}
where the wave number $k$ is a non-negative integer.
For two different numerical simulations that correspond to $B_k(t)$ and $B'_k(t)$, respectively, where the latter has smaller numerical noise and thus has higher numerical precision, we define the so-called ``spectrum-deviation''
\begin{equation}
 \delta_s(t) = \frac{\sum\limits_{k=0}\big|B_k(t)-B'_k(t)\big|}{\sum\limits_{k=0}B_k'(t)}    \label{delta_s}
\end{equation}
to measure the deviation of $B_k(t)$ from $B'_k(t)$ at the given time $t$.
Besides, to clearly indicate deviation of the velocity $\textbf{u}$ and the pressure $p$ between the DNS result and the CNS benchmark solution, we define the mean squared deviation of the velocity and pressure as follows:   
\begin{align}
& \delta_u(t)=\sqrt{\langle|\textbf{u}_{\mathrm{\,DNS}}(\textbf{x},t)-\textbf{u}_{\mathrm{\,CNS}}(\textbf{x},t)|^2\rangle_V},       \label{delta_u}
\end{align}
\begin{align}
& \delta_p(t)=\sqrt{\langle[p_{\mathrm{\,DNS}}(\textbf{x},t)-p_{\mathrm{\,CNS}}(\textbf{x},t)]^2\rangle_V},       \label{delta_p}
\end{align}
where $\textbf{u}_{\mathrm{\,DNS}}(\textbf{x},t)$, $p_{\mathrm{\,DNS}}(\textbf{x},t)$,  $\textbf{u}_{\mathrm{\,CNS}}(\textbf{x},t)$ and $p_{\mathrm{\,CNS}}(\textbf{x},t)$ are velocity and pressure fields given by DNS and CNS, respectively, $\langle \; \rangle_{V}$ denotes a spatial average defined by (\ref{average_V}).    

We also focus on the kinetic energy
\begin{equation}
E(\textbf{x},t) = \frac{1}{2} \big| \textbf{u}(\textbf{x},t) \big|^2    \label{kinetic_energy} 
\end{equation}
and
the kinetic energy dissipation rate
\begin{equation}
D(\textbf{x},t)=\frac{1}{2Re}\sum_{ij}\Big[ \partial_iu_j(\textbf{x},t)+\partial_ju_i(\textbf{x},t) \Big]^2,    \label{dissipation_rate} 
\end{equation}
where $i,j=1,2,3$, $u_1=u$, $u_2=v$, $u_3=w$, $\partial_1=\partial /\partial x$, $\partial_2=\partial /\partial y$, and $\partial_3=\partial /\partial z$.

Filter-Space-Technique (FST) is employed here to extract the scale-to-scale energy flux, denoted as $\Pi^{[l]}$. FST, initially developed for large eddy simulation in the 1970s \cite{Leonard1975AG}, involves applying a low-pass filter to the velocity field. Mathematically, it is expressed as:
\begin{equation}
f^{[l]}(\mathbf{x},t) = \int G^{[l]}(\mathbf{x}-\mathbf{x}') f(\mathbf{x}',t) d \mathbf{x}',
\end{equation}
where $f$ represents a 3D function, $\mathbf{x}=(x, y, z)$ denotes the coordinate vector, and $G^{[l]}$ is chosen to be a spherical Gaussian filter for the scale $l$ (see \cite{Chen2003PRL,Boffetta2012ARFM}).
For the incompressible Navier-Stokes equations, the scale-to-scale energy flux can be derived analytically as
\begin{equation}
\Pi^{[l]}= -\sum_{i,j=1,2,3} \left[ \left( u_iu_j \right)^{[l]} -u_i^{[l]}u_j^{[l]}\right]\frac{\partial u_i^{[l]}}{\partial x_j}. \label{energy-flux}
\end{equation}
Note that the sign of $\Pi^{[l]}$ reveals the direction of energy transfer: a positive value indicates a direct cascade from larger scale ($>l$) to smaller scale ($<l$), while a negative value signifies the inverse cascade.

\vspace{0.5cm}

\section*{Acknowledgements}
Sincere thanks to anonymous reviewers for their valuable comments and suggestions.  The calculations were performed on ``Tianhe New Generation Supercomputer'', National Supercomputer Center in Tianjin, China. This work is supported by National Natural Science Foundation of China (no. 12302288; 91752104),  Shanghai Pilot Program for Basic Research of Shanghai Jiao Tong University (no. 21TQ1400202), and State Key Laboratory of Ocean Engineering. 
	
\section*{Data and materials availability}
All data are available by sending requirement to the corresponding author.

\section*{Competing interests}
The authors declare no competing interests.

\bibliographystyle{elsarticle-num}

\bibliography{Kolmogorov3D-BIB}

\begin{thebibliography}{10}
\expandafter\ifx\csname url\endcsname\relax
  \def\url#1{\texttt{#1}}\fi
\expandafter\ifx\csname urlprefix\endcsname\relax\def\urlprefix{URL }\fi
\expandafter\ifx\csname href\endcsname\relax
  \def\href#1#2{#2} \def\path#1{#1}\fi

\bibitem{Orszag1970}
S.~A. Orszag, Analytical theories of turbulence, J. Fluid Mech. 41~(2) (1970)
  363--386.

\bibitem{Rogallo1981NASA}
R.~S. Rogallo, Numerical experiments in homogeneous turbulence, Tech. Rep.
  NASA-TM-81315, NASA, USA (September 1981).

\bibitem{She1990Nature}
Z.-S. She, E.~Jackson, S.~A. Orszag, Intermittent vortex structures in
  homogeneous isotropic turbulence, Nature 344 (1990) 226--228.

\bibitem{Kuhnen2018}
J.~K\"{u}hnen, B.~Song, D.~Scarselli, N.~B. Budanur, M.~Ried, A.~P. Willis,
  M.~Avila, B.~Hof, Destabilizing turbulence in pipe flow, Nature Physics 14
  (2018) 386--390.

\bibitem{Nelkin1992Science}
M.~Nelkin, In what sense is turbulence an unsolved problem?, Science 255 (1992)
  566--570.

\bibitem{FeracoScience2024}
A.~Alexakis, R.~Marino, P.~D. Mininni, A.~van Kan, R.~Foldes, F.~Feraco,
  Large-scale self-organization in dry turbulent atmospheres, Science 383
  (2024) 1005--1009.

\bibitem{MoinARFM1998}
P.~Moin, K.~Mahesh, Direct numerical simulation: a tool in turbulence research,
  Annu. Rev. Fluid Mech. 30 (1998) 539--578.

\bibitem{Scardovelli1999ARFM}
R.~Scardovelli, S.~Zaleski, Direct numerical simulation of free-surface and
  interfacial flow, Annu. Rev. Fluid Mech. 31 (1999) 567--603.

\bibitem{Coleman2010DNS}
G.~N. Coleman, R.~D. Sandberg, A primer on direct numerical simulation of
  turbulence --methods, procedures and guidelines, Tech. Rep. AFM-09/01a,
  Aerodynamics \& Flight Mechanics Research Group, University of Southampton,
  UK (March 2010).

\bibitem{Huang2022JFM}
J.~Huang, L.~Duan, M.~M. Choudhari, Direct numerical simulation of hypersonic
  turbulent boundary layers: effect of spatial evolution and {Reynolds} number,
  J. Fluid Mech. 937 (2022) A3.

\bibitem{lorenz1963deterministic}
E.~N. Lorenz, Deterministic nonperiodic flow, J. Atmos. Sci. 20~(2) (1963)
  130--141.

\bibitem{Deissler1986PoF}
R.~G. Deissler, Is {N}avier-{S}tokes turbulence chaotic?, Phys. Fluids 29
  (1986) 1453--1457.

\bibitem{boffetta2017chaos}
G.~Boffetta, S.~Musacchio, Chaos and predictability of homogeneous-isotropic
  turbulence, Phys. Rev. Lett. 119~(5) (2017) 054102.

\bibitem{berera2018chaotic}
A.~Berera, R.~D. J.~G. Ho, Chaotic properties of a turbulent isotropic fluid,
  Phys. Rev. Lett. 120~(2) (2018) 024101.

\bibitem{Vassilicos2023JFM}
J.~Ge, J.~Rolland, J.~C. Vassilicos, The production of uncertainty in
  three-dimensional {N}avier-{S}tokes turbulence, J. Fluid Mech. 977 (2023)
  A17.

\bibitem{Lorenz2006Tellus}
E.~N. Lorenz, Computational periodicity as observed in a simple system, Tellus
  A: Dynamic Meteorology and Oceanography 58A (2006) 549 -- 557.

\bibitem{Liao2009}
S.~Liao, On the reliability of computed chaotic solutions of non-linear
  differential equations, Tellus Ser. A-Dyn. Meteorol. Oceanol. 61~(4) (2009)
  550--564.

\bibitem{Liao2023book}
S.~Liao, Clean Numerical Simulation, Chapman and Hall/CRC, 2023.

\bibitem{LIAO2014On}
S.~Liao, P.~Wang, On the mathematically reliable long-term simulation of
  chaotic solutions of {Lorenz} equation in the interval [0, 10000], Sci.
  China-Phys. Mech. Astron. 57~(2) (2014) 330--335.

\bibitem{hu2020risks}
T.~Hu, S.~Liao, On the risks of using double precision in numerical simulations
  of spatio-temporal chaos, J. Comput. Phys. 418 (2020) 109629.

\bibitem{LiaoNA2022}
S.~Liao, X.~Li, Y.~Yang, Three-body problem -- from {Newton} to supercomputer
  plus machine learning, New Astronomy 96 (2022) 101850.

\bibitem{qin2020influence}
S.~Qin, S.~Liao, Influence of numerical noises on computer-generated simulation
  of spatio-temporal chaos, Chaos Solitons Fractals 136 (2020) 109790.

\bibitem{qin_liao_2022}
S.~Qin, S.~Liao, Large-scale influence of numerical noises as artificial
  stochastic disturbances on a sustained turbulence, J. Fluid Mech. 948 (2022)
  A7.

\bibitem{Qin2024JOES}
S.~Qin, Y.~Yang, Y.~Huang, X.~Mei, L.~Wang, S.~Liao, Is a direct numerical
  simulation {(DNS)} of {Navier-Stokes} equations with small enough grid
  spacing and time-step definitely reliable/correct?, J. Ocean Eng. Sci. 9
  (2024) 293--310.

\bibitem{Liao-2025-JFM-NEC}
S.~Liao, S.~Qin, Noise-expansion cascade: an origin of randomness of
  turbulence, J. Fluid Mech. 1009 (2025) A2.

\bibitem{Liao-2025-JFM-PS}
S.~Liao, S.~Qin, Physical significance of artificial numerical noise in direct
  numerical simulation of turbulence, J. Fluid Mech. 1008 (2025) R2.

\bibitem{Zhang2025CPC}
B.~Zhang, S.~Liao, \href{https://doi.org/10.1016/j.cpc.2025.109855}{An
  automated parallel program of clean numerical simulation for chaotic systems
  governed by {ODE}s}, (accepted).
\newline\urlprefix\url{https://doi.org/10.1016/j.cpc.2025.109855}

\bibitem{Xiaoming2025-3D-3body}
X.~Li, S.~Liao, \href{https://arxiv.org/abs/2508.08568}{Discovery of 10,059 new
  three-dimensional periodic orbits of general three-body problem} (2025).
\newblock \href {http://arxiv.org/abs/2508.08568} {\path{arXiv:2508.08568}}.
\newline\urlprefix\url{https://arxiv.org/abs/2508.08568}

\bibitem{Lee1951}
T.~D. Lee, Difference between turbulence in a two-dimensional fluid and in a
  three-dimensional fluid, J. Appl. Phys. 22 (1951) 524.

\bibitem{wu2021quadratic}
W.~Wu, F.~G. Schmitt, E.~Calzavarini, L.~Wang, A quadratic {Reynolds} stress
  development for the turbulent {Kolmogorov} flow, Phys. Fluids 33 (2021)
  125129.

\bibitem{oyanarte1990mp}
P.~Oyanarte, {MP}-a multiple precision package, Comput. Phys. Commun. 59~(2)
  (1990) 345--358.

\bibitem{pope2001turbulent}
S.~B. Pope, Turbulent Flows, IOP Publishing, 2001.

\bibitem{courant1928partiellen}
R.~Courant, K.~Friedrichs, H.~Lewy, {\"U}ber die partiellen
  {Differenzengleichungen} der mathematischen {Physik}, Math. Ann. 100~(1)
  (1928) 32--74.

\bibitem{AAMM-14-799}
S.~Liao, S.~Qin, Ultra-chaos: an insurmountable objective obstacle of
  reproducibility and replication, Adv. Appl. Math. Mech. 14~(4) (2022)
  799--815.

\bibitem{LLNS1959}
L.~D. Landau, E.~M. Lifshitz, Course of Theoretical Physics: Fluid Mechanics
  (Vol. 6), Addision-Wesley, Reading, 1959.

\bibitem{AAMM-15-1191}
S.~Qin, S.~Liao, A self-adaptive algorithm of the clean numerical simulation
  {(CNS)} for chaos, Adv. Appl. Math. Mech. 15~(5) (2023) 1191--1215.

\bibitem{Leonard1975AG}
A.~Leonard, Energy cascade in large-eddy simulations of turbulent fluid flows,
  Adv. Geophys. 18 (1975) 237--248.

\bibitem{Chen2003PRL}
S.~Chen, R.~E. Ecke, G.~L. Eyink, X.~Wang, Z.~Xiao, Physical mechanism of the
  two-dimensional enstrophy cascade, Phys. Rev. Lett. 91~(21) (2003) 214501.

\bibitem{Boffetta2012ARFM}
G.~Boffetta, R.~E. Ecke, Two-dimensional turbulence, Annu. Rev. Fluid Mech. 44
  (2012) 427--51.

\end{thebibliography}

\renewcommand{\figurename}{Extended Data Fig.}  
\renewcommand{\tablename}{Extended Data Table} 





\end{document}